\documentclass[prb,aps,amsmath,amssymb,twocolumn,floatfix,longbibliography]{revtex4-2}
\usepackage{graphicx,amsmath,amssymb, color,xcolor,orcidlink,bm}
\usepackage{overpic}
\usepackage[normalem]{ulem}
\usepackage[utf8]{inputenc}
\usepackage{tikz}
\usetikzlibrary{arrows.meta,positioning,fit,backgrounds}

\newcommand{\LLL}{\mathrm{P}_{\mathrm{LLL}}}

\newcommand{\ellstar}{\ell^{\star}}
\newcommand{\lB}{\ell_{\mathrm{B}}} 
\newcommand{\cm}{\mathrm{cm}}
\newcommand{\rel}{\mathrm{rel}}
\newcommand{\eff}{\mathrm{eff}}
\newcommand{\fd}{\mathrm{fd}}
\newcommand{\AB}{\mathrm{AB}}

\newcommand{\QHsym}{\mathrm{QH}}
\newcommand{\QPsym}{\mathrm{QP}}
\renewcommand{\vec}[1]{\mbox{\boldmath$#1$}}
\newcommand{\conj}[1]{\bar{#1}}
\def\beq{
\begin{eqnarray}}
  \def\eeq{
\end{eqnarray}}

\newcommand{\abs}[1]{\vert #1 \vert}

\newcommand{\ket}[1]{\vert #1 \rangle}
\newcommand{\bra}[1]{\langle #1 \vert}
\newcommand{\overlap}[2]{\langle #1 \vert #2 \rangle}

\newcommand{\QP}[1]{\QPsym_{#1}}
\newcommand{\QH}[1]{\QHsym_{#1}}

\newlength{\figwidth}
\begin{document}
\title{Shape Deformation and Braid Statistics of Fractional Quantum Hall Quasiparticles}
\author{Mytraya Gattu$^{1,2}$~\orcidlink{0000-0001-6994-389X} and J. K. Jain$^{1,2,3}$~\orcidlink{0000-0003-0082-5881}}
\affiliation{$^{1}$Department of Physics, 104 Davey Lab, Pennsylvania State University, University Park, Pennsylvania 16802, USA}
\affiliation{$^{2}$Center for Theory of Emergent Quantum Matter, Pennsylvania State University, University Park, Pennsylvania 16802, USA}
\affiliation{$^{3}$Lodha Theoretical Physics Institute, 17th Floor, Lodha NCP Supremus, Wadala (E), Mumbai 400037, India}
\date{\today}
\begin{abstract}

For ``ideal anyons,'' the Berry phase associated with a closed loop of an anyon around another is robust, i.e., independent of the size and shape of the loop, and directly yields the braid statistics. This is not the case for fractional quantum Hall (FQH) quasiparticles (QPs), which are charged and have finite size. The Berry phase $\Theta$ associated with a loop of a fractionally charged QP around another contains two contributions: $\Theta=\Theta_{\rm AB}+\Theta_{\rm braid}$, where $\Theta_{\rm AB}$, the Aharonov-Bohm phase, is proportional to the enclosed area, and $\Theta_{\rm braid}$ is the order-one contribution from braid statistics. The latter is obtained by determining $\Theta$ for a closed loop twice, once with and once without the other QP inside, and taking the difference. It is known that even small uncertainties in the enclosed areas can produce corrections to $\Theta_{\rm AB}$ larger than $\Theta_{\rm braid}$. We consider how the Berry phase depends on the shape of the QP, which, unlike its charge, is not topological and varies along the path in response to the local potential. Our calculations show that practically imperceptible differences in the shape of the outer QP in the two experiments can produce a non-universal value for the braid statistics, revealing an additional challenge for a precise determination of the braid statistics. We discuss these results in the context of interference experiments. We also note that the fractional phase jumps in these experiments can be understood without assuming QPs with sharply quantized fractional charges at the edges of the FQH system; instead, they directly measure the fractionally quantized vorticity of bulk QPs.
\end{abstract}
\maketitle
\section{Introduction}
\label{sec:introduction}

\begin{figure}[t!]
  \centering
  \tikzset{
    pnl/.style={inner sep=0pt, outer sep=0pt},
    txt/.style={align=center, font=\small, inner sep=1pt},
    arr/.style={-{Stealth[length=4.5pt,width=3.5pt]}, draw=gray!45!black, line width=0.8pt},
    lbl/.style={font=\small, inner sep=2pt},
    frame/.style={draw=gray!55, line width=0.6pt, inner sep=2.5pt},
  }
  \begin{tikzpicture}
    \node[pnl] (aL) at (0,0) {\includegraphics[width=0.40\columnwidth]{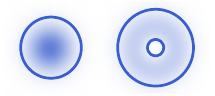}};
    \node[txt] (aT) at (0.57\columnwidth,0) {non-universal\\braid statistics};
    \node[pnl] (aB) at (0.57\columnwidth,-1.85cm) {\includegraphics[width=0.40\columnwidth]{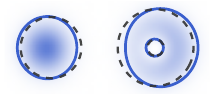}};
    \draw[arr, shorten >=4pt] (aL.east) -- node[lbl,above]{CF-ization} (aT.west);
    \draw[arr] (aL.south) |- (aB.west);
    \draw[arr] (aB.north) -- (aT.south);
    \begin{scope}[on background layer]
      \node[frame, fit=(aL)(aT)(aB)] (aF) {};
    \end{scope}
    \node[anchor=south west, font=\small\bfseries, inner sep=3pt] at (aF.south west) {(a)};
  \end{tikzpicture}\\[3pt]
  \begin{tikzpicture}
    \node[pnl,opacity=0] (bREF) at (0,0) {\includegraphics[width=0.40\columnwidth]{sch_panel_ni.pdf}};
    \node[txt] (bL) at (0,0) {\Huge\textbf{?}};
    \node[txt] (bT) at (0.57\columnwidth,0) {universal\\braid statistics?};
    \node[pnl] (bB) at (0.57\columnwidth,-1.85cm) {\includegraphics[width=0.40\columnwidth]{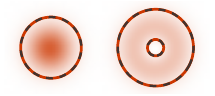}};
    \draw[arr, shorten >=4pt] (bREF.east) -- node[lbl,above]{CF-ization} (bT.west);
    \draw[arr] (bL.south) |- (bB.west);
    \draw[arr] (bB.north) -- (bT.south);
    \begin{scope}[on background layer]
      \node[frame, fit=(bREF)(bT)(bB)] (bF) {};
    \end{scope}
    \node[anchor=south west, font=\small\bfseries, inner sep=3pt] at (bF.south west) {(b)};
  \end{tikzpicture}
  \caption{(a) Composite-fermionization (CF-ization) of an integer quantum Hall wave function with two localized quasiparticles (QPs) yields fractional quantum Hall (FQH) QPs whose braid statistics, obtained by taking one QP around the other, is apparently shape-dependent, i.e., non-universal [Eq.~\eqref{eq:theta-standard-geometry}]. We notice that CF-ization deforms the shape of the QPs ever so slightly, with the deformation vanishing as the distance between the two QPs is increased. (b) We ask: Is it possible to write down IQH wave functions with two QPs that produce, upon CF-ization, undistorted FQH QPs? And if so, what is their braid statistics? We answer the first question in the affirmative by explicit construction, and find that these QPs indeed produce universal, shape-independent braid statistics.}
  \label{fig:hf-schematic}
\end{figure}

One of the remarkable aspects of the fractional quantum Hall effect (FQHE) is that its quasiparticles (QPs) and quasiholes (QHs) have a fractional charge relative to the uniform-density ground state, which follows from rather general principles when a gap opens at a fractional filling factor~\cite{Laughlin83}. (In the following, we will often refer only to QPs for convenience, with the understanding that all {\it generic} statements apply to both QPs and QHs.) The QP charge has been measured in shot-noise experiments~\cite{dePicciotto97,Saminadayar97,Reznikov99,Griffiths00,Comforti02,Chung03,Dolev08,Bid09,Dolev10,Hashisaka15,Biswas22,Veillon24}.

It was further predicted that the QPs also obey fractional braid statistics~\cite{Halperin84,Arovas84,Stern08,Feldman21}. In addition, a model even-denominator fractional quantum Hall state was proposed to support QPs satisfying non-Abelian braid statistics~\cite{Moore91,Stern04,Read09,Bonderson11}. This state has been proposed as a candidate description for the $5/2$ FQHE~\cite{Willett87,Morf98,Park98} and represents a topological $p$-wave paired state of composite fermions (CFs) analogous to a topological superconductor of electrons~\cite{Read00}. Non-Abelian anyons have inspired proposals for topological quantum computation~\cite{DasSarma05,DasSarma06,Stern06,Nayak08,Stern10,DasSarma15}. Many other even-denominator states~\cite{Suen92,Liu14,Shabani09,Shabani13,Singh24,Wang22,Wang25,Ki14,Kim19,Li17,Narayanan18,Shi20,Domaretskiy26} have also been modeled as paired CF states and therefore, based on general arguments, may support non-Abelian QPs~\cite{Scarola02,Moller08,Moller09,Balram18,Sharma21,Sharma23,Sharma24}.

Certain features of the fractional quantum Hall (FQH) QPs complicate a measurement of their braid statistics. The braid statistics of ``ideal anyons'' is given by the Berry phase associated with a closed loop of an anyon going around another anyon; it is robust, because it is independent of the size or shape of the loop. In contrast, the Berry phase of an FQH QP going around another is neither robust nor equal to the braid statistics. Because the FQH QPs are charged, the Berry phase of a closed loop also contains the contribution from the Aharonov-Bohm (AB) phase, which must be subtracted from the Berry phase to extract the braid statistics. What makes this difficult in practice is that the AB contribution, which is proportional to the area of the loop, is dominant, and even a minuscule error in it can overwhelm the order-one contribution arising from the braid statistics. Moreover, the FQH QPs are not point particles but are, in general, fairly large (6--10 $\lB$ or more, where $\lB=\sqrt{\hbar c/eB}$ is the magnetic length and $B$ is the strength of the applied magnetic field) cloud-like objects. As a result, neither the trajectory nor the area enclosed is sharply defined. These issues have made it rather tricky to extract the braid statistics even in computer calculations, as described below; some calculations yielded incorrect braid statistics, which were later attributed to an imperceptible shift in the location of the already present QPs when a new QP is added.

In this article, we investigate the role of QP shape, which has so far largely been overlooked. The QP shape, unlike its charge, is not a topological property. In fact, an FQH QP is, in general, a large, floppy object whose shape will depend on the local potential due to the disorder and other localized QPs in the vicinity, and will thus vary along a trajectory.

We briefly describe our findings here, with details to follow.

Let us first define our model. Our analysis below is based on the CF theory, which is known to provide an accurate microscopic account of the FQH QPs. The QPs of the $\nu=n/(2pn+1)$ state are analogous to the  QPs of the $\nu=n$ integer quantum Hall (IQH) state, and a wave function for the FQH QPs can be constructed by what is known as composite-fermionization (CF-ization) of the known wave function of the latter.  A single QP of an IQH state consists of $n$ fully occupied Landau levels (LLs) plus one electron in the $(n+1)^{\rm st}$ LL. A convenient basis for the IQH QP is constructed by placing the electron in different angular-momentum orbitals $k$ around a given point. A QP with an arbitrary shape can be produced by taking a linear superposition of the basis functions. States containing two or more QPs can similarly be constructed by placing two electrons at two different locations, each in an angular momentum orbital around its location. Composite-fermionization of these states produces FQH QPs at $\nu=n/(2pn+1)$.

To study the effect of the shape on the braiding properties, we begin by constructing two noninteracting IQH QPs in orbitals with angular momenta $k_1$ and $k_2$ around positions $\omega_1$ and $\omega_2$, and composite-fermionize this state to obtain the state of two FQH QPs. An earlier study of QPs with the smallest angular momenta had shown that composite-fermionization shifts the locations of the QPs, which ought to be accounted for to obtain the correct braid statistics~\cite{Jeon03,Jeon04}. We now find that, after correcting for the shift, the braid statistics calculated within this model depends on the angular momenta of the QPs, and is thus non-universal. If true, this would invalidate the notion of braid statistics, as one can obtain arbitrary braid statistics by constructing QPs that are a superposition of the different angular momentum states.

A closer inspection reveals, however, that composite-fermionization distorts the shape of the two QPs (except for the QP with the smallest value of $k$). In other words, for states consisting of more than one QP, composite-fermionization of IQH QPs that are rotationally symmetric about their centers does not produce FQH QPs that are rotationally symmetric about their centers. We ask if it is in principle possible to construct, within the basis specified by the CF theory, QPs that are rotationally symmetric about their centers. We demonstrate, by explicit construction, that IQH QPs with highly non-local correlations produce ``undistorted'' FQH QPs. Remarkably, the braid statistics of the undistorted FQH QPs is universal, i.e. does not depend on the angular momentum of the QP.

Our work thus demonstrates that the notion of braid statistics remains valid in principle, but at the same time, it also reveals the challenges in its measurement. Let us elaborate. When a QP is taken around another, the Berry phase contains contributions from the AB effect and the braid statistics:
\begin{equation}
  \label{eq:berry-decomposition}
  \Theta=\Theta_{\rm AB}+\Theta_{\rm braid}
\end{equation}
The first must be subtracted from the Berry phase to determine the braid statistics. In general, because of the finite size of the QP, we do not know the AB contribution with the accuracy required to extract the braid statistics. The AB contribution, being proportional to the area of the loop, is dominant, and even the tiniest uncertainty in the full Berry phase or the AB phase can wash out the order-one contribution from the braid statistics. Furthermore, as shown in this paper, even a practically immeasurable change in the shape can swamp the braid statistics. (In this context, it is interesting to note that the initial calculation was done for the Laughlin QH for which the shape dependence happens to be absent, and the AB contribution can be determined by defining the position of the QH as the location where the density vanishes. As we see below, this is not the case for general QP or QH.) 

In general, therefore, one needs to perform the experiment twice: once with and once without another QP enclosed by the loop, and then take the difference to tease out the braid statistics. An accurate determination of the braid statistics would require that the trajectory, as well as the shape of the looping QP in the two cases, be identical to a very high degree of precision. In particular, one must ensure that adding a new QP has no effect whatsoever on the already present QPs. 

We finally discuss the relevance of our work to interference experiments during the past few years~\cite{Nakamura20,Nakamura23,Kundu23,Kim24,Werkmeister25,Samuelson26,Ghosh25,Ghosh25A,Kim26}. These experiments observe fractional phase jumps, which have been interpreted as signatures of the braid statistics of FQH QPs. However, the interfering path mostly goes along the edges of the FQH system, which, due to the absence of a gap, do not support QPs with a sharp fractionally quantized charge. One may argue that the tunneling across the constrictions selects a fractional charge, effectively making it legitimate to view the entire trajectory as that of a fractionally charged QP~\cite{Chamon97,Halperin11,Feldman21,Feldman22,Heiblum20}, as needed for a measurement of the braid statistics of the QPs. (We note here that the measured phase jump in this geometry corresponds to two exchanges~\cite{Read24}; modifications have been proposed that will reveal the phase associated with a single exchange~\cite{Kivelson25}.) We suggest another interpretation of the experiments that does not require the existence of FQH QPs with sharply quantized charge at the edges of the FQH system. We show that the experimental observations can be understood if we assume the existence of CFs at the FQH edges -- unlike fractionally charged QPs, the CFs are known to be well defined in compressible states. From this perspective, a closed loop of a CF measures the effective flux enclosed, and a fractional phase jump arises because a QP added in the bulk of the FQH system adds a fractional effective flux equal to $2p\phi_0/(2pn\pm 1)$ for the $\nu=n/(2pn\pm 1)$ FQH state. Note that this interpretation relies fundamentally on the CF theory. The challenges noted above need to be overcome to make the measurement possible in either interpretation of the experiment, and we comment below on how experiments accomplish that. Going forward, it would be interesting to ask if any experiment can distinguish between the two interpretations, and also to design an experimental setup that implements braiding entirely within the bulk of the FQH system, without involving the edges (see Ref.~\cite{Gattu24} for such a proposal).

Earlier articles have also noted subtleties in measuring the fractional braid statistics. Coupling to electromagnetism (see, e.g., Ref.~\cite{Hansson26}), which causes Landau-level mixing,  produces a correction to the braid statistics~\cite{Hanna92,Sondhi92,Simon08}. This physics is neglected throughout our work. Ref.~\cite{Trung23} considered the effect of distortion of the QP wave function on its ``spin.'' Ref.~\cite{Kjall18} showed, in a matrix-product representation of the QP wave function, that screening of the QP operators must be treated carefully to get the right braid statistics. 

{\bf Plan of the rest of the paper:} The rest of the paper is organized as follows. In Sec.~\ref{sec:cf-theory} we review the CF theory of the QPs and QHs of an FQH liquid: a QP is an additional CF in the otherwise empty $\Lambda$L and a QH a missing one, and each CF experiences the reduced effective field $B^{\star}$. We show how their topological properties follow from this picture. The localized fractional charge $e^{\star}$ follows from the electron-flux (vortex to be more precise) binding involved in the formation of CFs, and, because every CF carries $2p$ vortices, that localized charge implies a localized fractional vorticity; the braid statistics is in turn a consequence of this vorticity. In Sec.~\ref{sec:first-look} we consider an aspect of the QPs of FQH liquids that has not been considered before in the context of a bulk measurement of their braid statistics: their shape. To ask whether the shape matters at all, we use a toy model of two anyons and show that the Berry phase acquired by the orbiting anyon carries a correction that depends on its angular momentum, and hence on its shape. This correction can exceed the braid statistics itself, and can therefore swamp any attempt to extract the latter. We trace it to a correlated shape deformation of the orbiting anyon: a deformation tied to the presence of the other. We then show that the deformation can be cured by building an inverse deformation into the two-anyon wave function, and that the resulting states give a shape-independent braid statistics. In Sec.~\ref{sec:localized-qh-qp-states} we review the CF construction of the incompressible states at $\nu=n/(2pn+1)$ and construct localized QH and QP wavepackets, whose shapes are specified by an angular-momentum label $k$ selected by the localizing potential; we also construct the two-QH and two-QP wave functions. In Sec.~\ref{sec:shape-dependence-braid-statistics} we evaluate the Berry phases of these wave functions and show that, while the Berry phase of a single QH or QP is independent of $k$, the extracted braid statistics depends on the shapes of the QHs or QPs; we trace this dependence to a correlated deformation of the two-QH and two-QP density profiles. In Sec.~\ref{sec:flux-dressed-cf-wavefunctions} we ask whether a new QP or QH can be added without distorting the shape of an already present one: we construct the first class (Class I) of one-parameter families of modified two-QH and two-QP wave functions and show that, at the physical value of the parameter, the deformation is removed and the expected shape-independent braid statistics is recovered. In Sec.~\ref{sec:hf-construction} we construct the second class (Class II), which achieves the same end with a single Slater determinant in which the localized QPs occupy orbitals different from those occupied by IQH QPs before composite-fermionization; these orbitals are inherently ``non-local''. In Sec.~\ref{sec:interference}, we turn to the interference experiments, in which the interfering path runs along the edges of the FQH system, where there is no gap and fractionally charged QPs are not well defined. We show that the observed period and the fractional phase jumps can be understood by assuming only that CFs exist at the edge: the period is set by the effective field $B^{\star}$ seen by the CF traversing the loop, and the phase jumps measure the fractional effective flux of an additional QP added to the interior. In Sec.~\ref{sec:discussion-outlook}, we summarize the challenges that the shape dependence poses for a measurement of the braid statistics and discuss open questions.

\section{CF theory of quasiparticles and quasiholes and their topological properties}
\label{sec:cf-theory}

While the existence of QPs with fractional charge and fractional braid statistics may be deduced from general principles and by making reasonable assumptions, a detailed microscopic understanding of the QPs now exists within the CF theory. Our analysis below will make extensive use of the CF theory.

The CFs are electrons carrying an even number of quantized vortices, modeled as ``effective'' flux quanta, and experience a reduced ``effective'' magnetic field~\cite{Jain89}. (The adjective ``effective'' reminds us that no real magnetic field is bound to the CFs -- an external magnetometer will not detect any additional magnetic field. The effective magnetic field is fully internal to the system of CFs and CFs themselves must be used to detect it. In the following, we will often omit the adjective effective for ease of language; it should be clear from the context which flux is real and which effective.) The vorticity of CFs is a topological property, from which many other topological properties of the FQH states follow, such as the fractionally quantized Hall resistance, and the fractional charge and braid statistics of the QPs.  The CFs form Landau-like levels called $\Lambda$ levels ($\Lambda$Ls) within the lowest Landau level (LLL) of electrons. The power of the CF theory is that, while electrons are strongly correlated and the model of noninteracting electrons neither captures the qualitative physics nor is a good starting point, the model of noninteracting CFs correctly predicts a large body of nontrivial observations and provides a starting point for many other observations. Specifically, the CF theory correctly predicts that the most prominent fractions occur along the sequences $\nu=n/(2pn\pm 1)$~\cite{Jain89,Jain07,Jain20} and the non-FQH states at even-denominator filling factors are Fermi-liquid metals of CFs~\cite{Halperin93,Halperin20,Shayegan20}. When the weak interactions between CFs are taken into account, the theory predicts paired CF states at half-filling in higher Landau levels (LLs) of semiconductor quantum wells or graphene~\cite{Moore91,Read00,Sharma23,Domaretskiy26}, in wide quantum wells~\cite{Shabani09A,Shabani13,Peterson08,Sharma21,Sharma24}, or in the presence of large LL mixing~\cite{Wang22,Zhao23,Wang23}; CF crystals at low fillings~\cite{Jiang90,Narevich01,Archer13} or high LL mixing~\cite{Zhao18,Ma20}; and also CF nematic-stripe and bubble phases~\cite{Lee02,Shingla23,Wang26}. 

Of most direct relevance to the present work are: (i) the ground state at $\nu=n/(2pn\pm 1)$ is understood as the $n$-filled-$\Lambda$L integer quantum Hall (IQH) state of CFs carrying $2p$ flux quanta; (ii) a QP is an isolated CF in the otherwise empty $\Lambda$L; and (iii) a QH is a missing CF in an otherwise filled $\Lambda$L. Exact diagonalization (ED) studies have shown that the wave functions for the ground states, as well as for QPs and QHs, are remarkably accurate representations of the corresponding Coulomb eigenstates~\cite{Jain07,Gattu24A}, leading to numerous detailed quantitative predictions. Furthermore, as we move further away from $\nu=n/(2pn\pm 1)$ by adding more and more QPs or QHs, the CF theory continues to provide a faithful and accurate account of the low-energy spectrum~\cite{Balram13}.

{\bf Fractional charge}: If the QP is simply an isolated CF, how can it have a fractional charge? How can this fractional charge be different for different FQH states of a sequence $\nu=n/(2pn+1)$, even though they are formed of the same CFs?  The CF theory actually provides a natural way to understand the fractional charge. To see this, let us consider the $\nu=n/(2pn +1)$ FQH state, which has $n$ lowest $\Lambda$Ls fully occupied. Now we add a CF to it by first adiabatically inserting $2p$ flux quanta, followed by the addition of an electron. An adiabatic insertion of a single flux quantum produces~\cite{Laughlin83} a charge $\nu e$ relative to the ground state. Adding the charges of $2p$ flux quanta and an electron produces the QP charge of
\begin{equation} \label{eq:q*}
  e^{\star}=(2p\nu -1)e = - \frac{e}{2pn+1} .
\end{equation}
This value, of course, agrees with that determined by explicit calculation of the total charge excess associated with a QP from the wave function. We added a total charge $-e$; the remainder of the charge goes to the boundary.

{\bf Fractional vorticity}: When combined with the CF theory, the fractional charge implies another remarkable property which, in our opinion, has not received the attention it deserves. Given that each electron has $2p$ vortices, modeled as $2p$ effective flux quanta, bound to it, the QP also carries a fractionally quantized effective flux given by
\begin{equation}\label{eq:fracflux}
  \phi^{\star}_{\QPsym}= \frac{2p}{2pn+1}\phi_{0}.
\end{equation}
This is the effective flux associated with the QP in excess of the uniform effective flux associated with the ground state. The corresponding QH carries the opposite flux $ \phi^{\star}_{\rm QH}= -(2p\phi_{0})/(2pn+1)$.

We again stress that the fractional effective flux actually represents fractional vorticity. It cannot be measured by an external magnetometer. Just like the effective magnetic field, the fractional flux is fully internal to the CFs, and CFs themselves must be used to measure it.

While a localized fractional charge implies localized fractional flux in the CF theory, because of the coupling between charge and flux, fractional flux is not necessarily a property of the QPs in non-CF approaches. Take, for example, Laughlin's wave function for the QP of the $1/3$ state. Numerical calculations~\cite{Kjonsberg99} have found that while the Laughlin QPs have well-localized fractional charge, they do not have well-defined braid statistics.  This suggests that the Laughlin QP does not carry a localized flux. Nardin et al.~\cite{Nardin23} have argued that the Laughlin QP can be viewed as a non-local superposition of Jain QPs, so in some sense the flux associated with the Laughlin QP is spread over the entire FQH droplet.

{\bf General derivation of braid statistics}: Next, we calculate the braid statistics of the QPs. For ``ideal anyons,'' the Berry phase associated with a closed loop of an anyon going around another anyon is independent of the size or shape of the loop and gives us the braid statistics. In contrast, the Berry phase associated with a closed loop of an FQH QP does not directly give us the braid statistics because it also contains an AB contribution arising from the QP charge. The AB contribution must be subtracted to extract the braid statistics. What makes it challenging in practice is that this requires subtracting two quantities that scale with the area enclosed by the loop to obtain the braid statistics, which is an order-one quantity. As a result, the two Berry phases must be determined with sufficiently high precision. Alternatively, one can measure the change in the Berry phase of a closed loop upon adding another QP, assuming the newly added QP does not affect the already present one.

Consider a closed loop ${\cal L}$ of a QP. Given that the QP is nothing but a CF, the Berry phase associated with the loop is given by
\begin{equation}\label{eq:Theta*}
  \begin{split}
    \Theta^*&=-\frac{2\pi B {A_{\cal L}}}{\phi_{0}}+2\pi\, 2p\int_{A_{\cal L}} d^2\vec{r}\, \rho(\vec{r})\\
    &\equiv -\frac{2\pi}{\phi_{0}}\int_{A_{\cal L}} d^2\vec{r}\, B^{\star}(\vec{r}) ,
  \end{split}
\end{equation}
where
\begin{equation} \label{eq:B*}
  B^{\star}(\vec{r})=B-2p \rho(\vec{r})\phi_{0}
\end{equation}
is the effective magnetic field experienced by the CF; $A_{\cal L}$ is the area enclosed by the loop ${\cal L}$, and $\rho(\vec{r})$ is the local density. The first term on the right-hand side of Eq.~\eqref{eq:Theta*} arises from the electron going around the loop, and the second term is due to the $2p$ effective flux quanta going around the loop, with a flux quantum going around an electron producing a phase $2\pi$.

Now, adding a QP in the interior changes the effective flux by the amount given in Eq.~\eqref{eq:fracflux} and causes the Berry phase to jump by
\begin{equation}\label{eq:braid_stat}
  \Delta \Theta^*=  2\pi \frac{2p}{2pn+1}
\end{equation}
This is the braid statistics of the QPs. The braid statistics of QHs is also the same, while that of a QP going around a QH differs by a sign; these results may be derived from the fact that a QP-QH pair behaves as a boson.

Note that we have not worried about the wave function of the QP in evaluating the Berry phase in Eq.~\eqref{eq:Theta*}. The finite size of the QP makes the exact definition of the loop or the area enclosed ambiguous.  In fact, the resulting uncertainty in the Berry phase of a complete loop of a QP is much larger than the contribution from the braid statistics. However, {\it provided that the wave function and the trajectory of the QP do not change when a new QP is inserted, the} change {\it in the Berry phase given in Eq.~\eqref{eq:braid_stat} remains sharply quantized.}

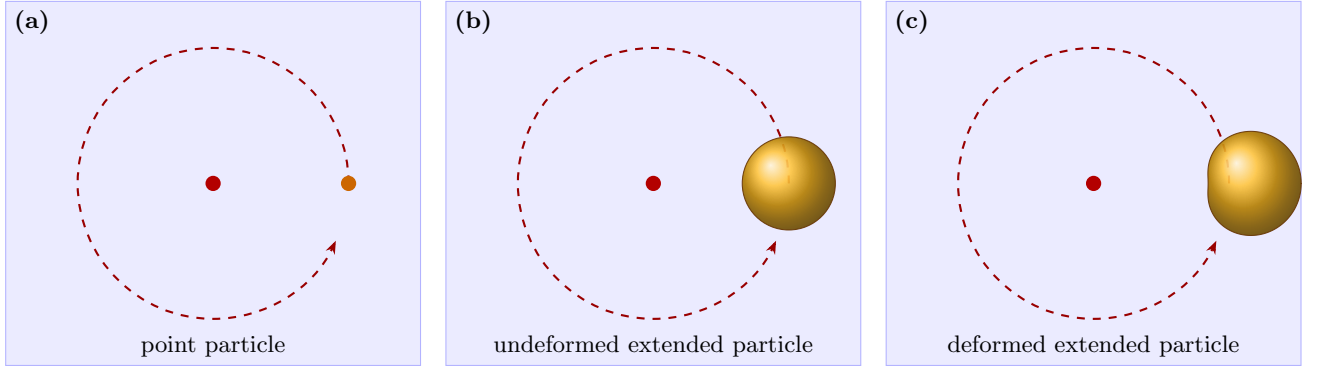
\begin{figure*}
  \centering
  \begin{tikzpicture}[scale=1.12]
    \foreach \xs/\plab/\panel in {0/point particle/a, 5.2/undeformed extended particle/b, 10.4/deformed extended particle/c} {
      \begin{scope}[xshift=\xs cm]
        \fill[blue!8] (-2.45,-2.15) rectangle (2.45,2.15);
        \draw[blue!30] (-2.45,-2.15) rectangle (2.45,2.15);
        \fill[red!70!black] (0,0) circle (0.09);
        \draw[dashed, red!60!black, thick, -{Stealth[length=5pt]}] (1.6,0) arc (0:335:1.6);
        \node[anchor=north west, font=\small\bfseries, inner sep=3pt] at (-2.45,2.15) {(\panel)};
        \node[anchor=south, font=\small, align=center, text width=4.6cm, inner sep=3pt] at (0,-2.15) {\plab};
      \end{scope}
    }
    \fill[orange!80!black] (0:1.6) circle (0.09);
    \begin{scope}[xshift=5.2cm]
      \shadedraw[ball color=orange!60!yellow, draw=orange!40!black, opacity=0.9] (0:1.6) circle (0.55);
    \end{scope}
    \begin{scope}[xshift=10.4cm]
      \shadedraw[ball color=orange!60!yellow, draw=orange!40!black, opacity=0.9]
      plot[domain=0:360, samples=120, smooth cycle] ({1.6+0.55*cos(\x)*(1+0.55*cos(\x))},{0.55*sin(\x)*(1+0.55*cos(\x))});
    \end{scope}
  \end{tikzpicture}
  \caption{The braid statistics is given by the difference between the Berry phases acquired by a quasiparticle (QP) along a circular loop (dashed) with and without another QP (red dot) localized at its center. The Berry phase is set by the number of CFs enclosed by the loop,
  because the orbiting QP, which is also a CF, sees a braid phase of $4p\pi$ per enclosed CF. (a)~If the orbiting QP were a point particle, the counting of CFs would be unambiguous. (b)~Even when the orbiting QP is in fact an extended, cloud-like object, so long as its shape is unaffected by the presence of the QP inside the loop, the Berry phase difference is still unambiguous. (c)~When the shape of the orbiting QP is deformed by the presence of the other QP, however, the Berry phase difference is no longer unambiguous.}
  \label{fig:point-vs-cloud}
\end{figure*}

How can a CF, which is a {\it fermion}, also behave, sometimes, as a fractionally charged anyon? We first note that the fermionic nature of CFs is fully established. It manifests through the appearance of the Jain sequences, which are the integer quantum Hall effect (IQHE) of CFs, and also through their Fermi liquid metals: both the IQHE and Fermi liquids are fundamentally properties of fermions. Furthermore, many even-denominator FQH states are viewed in terms of pairing of composite fermions, with CF pairs behaving as bosons. How can a CF in a partially filled $\Lambda$L be a fractionally-charged anyon? If we are describing the system in terms of {\it all} particles, relative to a ``vacuum'' state with no particles, then we have charge $-e$ CFs, with the property that each CF sees $2p$ vortices bound on all other CFs, which produces additional Berry phases as CFs move about. However, if we calculate the {\it difference} between charge densities of two systems with and without an excited CF, we obtain a charge excess that is a fraction of $-e$. Similarly, when we consider the {\it difference} between the Berry phases associated with a trajectory with and without another QP inside it, we obtain a fractional phase difference attributed to the braid statistics. There is no contradiction.

{\bf Numerical evaluation of braid statistics}: Confirming the above result with explicit wave functions has not been straightforward. Arovas {\it et al.}~\cite{Arovas84} evaluated the braid statistics of the Laughlin QHs~\cite{Laughlin83} analytically. Analytical calculations have not been possible for other QPs/QHs, but the braid statistics can be evaluated numerically. Kjonsberg and Leinaas reproduced the braid statistics of the Laughlin QH numerically~\cite{Kjonsberg97}: they found that when the QHs are not very close to one another or to the edge, the above procedure delivers a value that does not depend on the loop of the orbiting QH and agrees with the expectation. A subsequent study found that the Laughlin QPs~\cite{Laughlin83} do not satisfy a well-defined braid statistics; the calculated braid statistics fails to approach a constant value as the distance between two Laughlin QPs is increased~\cite{Kjonsberg99}.

Ref.~\cite{Kjonsberg99A} considered the Jain QP of the $\nu=1/3$ state (which is different from the Laughlin QP; explicit construction given below) and found that while the braid statistics does approach a constant value as the distance between the QPs is increased, it is not the expected value~\cite{Kjonsberg99A}. Later, it was realized that the discrepancy arose because the addition of a new QP shifts the location of the previously present QP ever so slightly due to the CF correlations present in the state~\cite{Jeon03,Jeon04}. The expected statistics was recovered after correcting for this shift. A similar correction is required to obtain the correct braid statistics of the 2/5 QPs~\cite{Jeon03,Jeon04}. Remarkably, the shift vanishes for large loops (when the two QPs are far from one another), but the correction to braid statistics remains finite. One may wonder why this was not an issue for the Laughlin QHs. It turns out that for the QHs in ``coherent states,'' of which the Laughlin QH is an example, there is no shift~\cite{Gattu24}. For all other cases, one must correct for the shift.

{\bf Lack of topological protection at the edge}: We note that when we add a CF at the edge, it no longer has a fractionally quantized charge. This is expected from the absence of a gap at the edge and is also seen routinely in numerical evaluations of the charge of a vortex near the edge~\cite{Kjonsberg97}. An intuitive understanding may be gained by noting that when a flux quantum is inserted adiabatically at the edge of an FQH state, there is much less charge available to push away in the outward direction, indicating a smaller overall charge accumulation than that when the CF is added in the interior. (Adiabaticity is ill-defined due to the absence of a gap, but as far as the wave function is concerned, we can locate the QP at the edge.)

\section{QP shape and braid statistics: First look}
\label{sec:first-look}

We will now consider QPs confined in the interior, so their topological properties are protected by a gap. Unlike its charge, the shape of a QP is not a topological property. In the absence of a potential, a QP is infinitely degenerate, as it may be moved without any energy cost. The shape of the QP is hence very susceptible to the potential it feels due to nearby localized QPs or disorder, which are likely to be present in any real experiment. As a result, in any experimental adiabatic loop, the shape of a QP will vary along its trajectory. We investigate the shape dependence of the braid statistics, which has so far largely been overlooked. (Refs.~\cite{Iyer24,Thamm24} examined how the finite temporal width of the pulses used to generate the tunneling current in time-domain braiding measurements affects the measured braiding signatures, and Ref.~\cite{Iyer24} also compared different pulse profiles. In our work, we focus on QPs localized entirely within the bulk.) We find, surprisingly, that the {\it Berry phase} of a closed loop depends significantly on the shape of the QP.

In this section, we illustrate the essential physics with a simple example, leaving the general case for later. Let us consider the QH of the $1/3$ state, which is the simplest such state. Ref.~\cite{Arovas84} considered the QH wave function $\prod_{l} z_l \prod_{i<j} (z_i-z_j)^3$ with $z_j=x_j-iy_j$ (corresponding to a magnetic field in the $+z$ direction), all lengths are in units of the magnetic length, and we have suppressed a Gaussian factor for simplicity. In the CF theory, this wave function is expressed as:
\begin{equation}
  \Phi_{m=0} =
  \begin{vmatrix}
    z_1 & \cdots & z_N\\
    z_1^2 & \cdots & z_N^2\\
    \vdots & \ddots & \vdots\\
    z_1^N & \cdots & z_N^N\\
  \end{vmatrix} \prod_{i<j} (z_i-z_j)^2,
\end{equation}
where the Slater determinant represents a state at $\nu=1$ with a hole in the angular momentum $m=0$ orbital. This corresponds in CF theory to a missing CF from the $m=0$ orbital in an otherwise full lowest $\Lambda$ level (CF Landau level). An $m=0$ hole at an arbitrary location $\eta$ is given by $\prod_{l} (z_l -\eta) \prod_{i<j} (z_i-z_j)^3$. Ref.~\cite{Arovas84} calculated the Berry phase associated with the closed circular loop of radius $R$ of an $m=0$ QH, and found that inserting another $m=0$ QH at the origin changes the Berry phase by $\Delta\Theta/(2\pi)=2/3$, thus yielding fractional braid statistics.

Now consider an $m=k$ QH at the origin, given by
\begin{equation}
  \Phi_{m=k} =
  \begin{vmatrix}
    1 & \cdots & 1\\
    z_1 & \cdots & z_N\\
    \vdots & \ddots & \vdots\\
    z_1^{k-1} & \cdots & z_N^{k-1}\\
    z_1^{k+1} & \cdots & z_N^{k+1}\\
    \vdots & \ddots & \vdots\\
    z_1^N & \cdots & z_N^N\\
  \end{vmatrix} \prod_{i<j} (z_i-z_j)^2,
\end{equation}
where the Slater determinant now represents the $\nu=1$ state with a hole in the $m=k$ orbital. QHs for all $k$ have the same charge. Now the QH located at $\eta$ cannot be obtained simply by replacing $z_j$ with $z_j-\eta$; we show later how to construct the wave function for a QH at an arbitrary location. An explicit calculation (details later) demonstrates that the relative braid statistics of an $m=k$ QH orbiting around an $m=k'$ (localized) QH is given by
\begin{equation}\label{eq:shapeshifter}
  \frac{\Delta \Theta}{2\pi} = \frac{2}{3}(1+2k)
\end{equation}
Thus, the braid statistics appears to depend explicitly on $k$, or the shape of the QH.

One might wonder why a sharply quantized, $k$-dependent value for the braid statistics is problematic. In fact, one could argue that the braid statistics will then have information about the internal structure of the QPs. To see the problem, however, suppose the QH is not rotationally invariant around its center, as would generically be the case if the potential localizing it is not rotationally invariant or there are other localized QPs in the vicinity. The actual QH wave function then will be a linear superposition of QH basis functions with different $k$, and the QH braid statistics will vary continuously with the QH shape and hence cease to be a meaningful quantum number. Analogous behavior follows for all of the QPs and QHs of the Jain FQH states.

{\bf Shape distortion:} A closer inspection shows that in addition to the overall shift in the location discussed above, the {\it shape} of the orbiting QP is, in general, also distorted when another QP is inserted at the origin. One may wonder whether the $ k$-dependence of the braid statistics is due to shape distortion.
The difficulty is illustrated in Fig.~\ref{fig:point-vs-cloud}.

To address this question, it would be necessary to be able to add a new QP in such a fashion as not to affect either the locations or the shapes of the already existing QPs. One way a new QP affects the previous ones is due to its Coulomb potential. Let us assume that this potential can be screened by placing metallic planes nearby, as has been done in the interference experiments. Now we ask: {\it Do the internal, long-range correlations in the FQH state due to the formation of CFs introduce shape deformations which cannot be eliminated while staying within the CF space? If that were the case, that would raise the question whether the braid statistics may be defined even in principle.}

We introduce two classes of modified wave functions to show that, for the Jain FQH states, it is possible to add a new QP without affecting either the location or the shape of an already present QP. Furthermore, we find that the braid statistics calculated with these wave functions is shape independent. These are the primary results of our paper.

We label the two classes I and II in the order in which they are developed in Secs.~\ref{sec:flux-dressed-cf-wavefunctions} and~\ref{sec:hf-construction}. Class I is somewhat technical and is deferred to Sec.~\ref{sec:flux-dressed-cf-wavefunctions}; Class II, whose essential idea can be seen in a toy model, is outlined here and developed more fully in Sec.~\ref{sec:hf-construction}.

\subsection{A toy model: Two anyons}

\begin{figure}
  \includegraphics[width=\columnwidth]{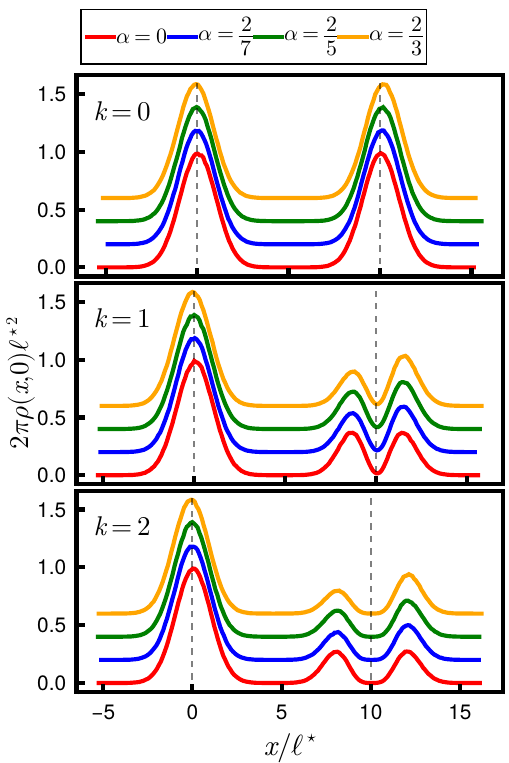}
  \caption{Density profile $2\pi\rho(x,0)\ellstar{}^{2}$ along the $x$ axis for the two-anyon wave function $\Psi^{k_{1},k_{2}}_{\omega_{1},\omega_{2}}$ of Eq.~\eqref{eq:intro-two-anyon-conventional} with $k_{1}=0$, $\omega_{1}=0$ and $k_{2}=k$, $\omega_{2}=\omega$. One anyon is localized at the origin, and the second is localized at $\omega=10\ellstar$ in the $k$ angular-momentum orbital about $\omega$. The top panel shows $k=0$, analogous to the situation considered by Jeon {\it et al.}~\cite{Jeon03,Jeon04}; here $\alpha$ is the statistical exponent. For $\alpha\neq0$, the Jastrow factor $(Z_{1}-Z_{2})^{\alpha}$ shifts the two density peaks away from each other, a displacement that can be accounted for by a position correction. The lower panels show $k=1$ and $k=2$. In these cases, vortex attachment produces not only a shift but also a deformation of the wavepacket localized near $\omega$, induced by the anyon at the origin. This deformation cannot be removed by a distance correction and is the origin of the shape-dependent Berry phase. We have shifted the plots for $\alpha=2/7, 2/5$ and $\alpha=2/3$ upwards by $0.20, 0.40$ and $0.60$ for ease of viewing. Here, $\ellstar$ is the magnetic length seen by the two anyons.
  }
  \label{fig:two-anyon-density}
\end{figure}

The basic idea underlying the modified wave functions can be motivated and illustrated by considering a toy problem of two anyons in the LLL. Modeling the anyons as electrons with fractional vortices bound to them (as appropriate within the LLL space), produces the wave function
\begin{equation}\label{eq:intro-two-anyon-conventional}
  \Psi^{k_{1},k_{2}}_{\omega_{1},\omega_{2}}
  =
  (Z_{1}-Z_{2})^{\alpha}
  \begin{vmatrix}
    \psi^{\omega_{1}}_{0,k_{1}}(Z_{1}) & \psi^{\omega_{1}}_{0,k_{1}}(Z_{2})\\
    \psi^{\omega_{2}}_{0,k_{2}}(Z_{1}) & \psi^{\omega_{2}}_{0,k_{2}}(Z_{2})
  \end{vmatrix},
\end{equation}
where
\begin{equation}\label{eq:intro-lll-packet}
  \psi^{\omega}_{0,k}(Z)
  \propto
  (Z-\omega)^{k}\,
  e^{-\frac{\abs{Z}^{2}}{4}+\frac{\conj{\omega}Z}{2}-\frac{\abs{\omega}^{2}}{4}}
\end{equation}
is the wave function of a single electron in the LLL with angular momentum $k$ about the point $\omega$, whose density $\propto\abs{Z-\omega}^{2k}e^{-\abs{Z-\omega}^{2}/2}$ is exponentially localized about $\omega$. We have set the magnetic length of the anyons, $\ell^*$, to unity. (Throughout the paper, capital $Z$ denotes the coordinate of an anyon of this toy model; lowercase $z$ is reserved for the electron coordinates of the FQH wave functions.) The generalization to anyons in higher LLs is given in Eq.~\eqref{eq:localized-orbital-expansion} and derived in Appendix~\ref{app:localized-electron-wavepackets}. In Appendix~\ref{app:berry-phase-two-anyon-model} we show that when the anyon localized at $\omega_{2}$ is taken along a circular loop about the origin, where the other anyon is localized ($\omega_{1}=0$), the Berry phase, in the limit $\abs{\omega_{2}}\to\infty$, is given by
\begin{equation}\label{eq:intro-two-anyon-berry}
  \frac{\Theta}{2\pi} = -\frac{\abs{\omega_{2}}^{2}}{2} - \alpha(1+2k_{2}).
\end{equation}
The first term is the AB contribution, $\Theta_{\rm AB}$. The remainder, $\Delta\Theta/2\pi=-\alpha(1+2k_{2})$, which one would expect to be the braid statistics $\Theta_{\rm braid}$, depends on the angular momentum $k_{2}$, i.e., it is shape dependent. (Note that we have not accounted for the density shift first noted by Jeon \emph{et al.}~\cite{Jeon03,Jeon04}; accounting for this shift does not by itself lead to a shape-independent statistics.) The origin of this shape dependence is a correlated shape deformation due to the vortex factor $(Z_{1}-Z_{2})^{\alpha}$: the presence of one localized anyon forces a deformation in the shape of the other, as seen in Fig.~\ref{fig:two-anyon-density}.

We now introduce a modified wave function
\begin{equation}\label{eq:intro-two-anyon-conventional-2}
  \tilde{\Psi}^{k_{1},k_{2}}_{\omega_{1},\omega_{2}}
  =
  (Z_{1}-Z_{2})^{\alpha}
  \begin{vmatrix}
    \tilde{\psi}^{\omega_{1}}_{0,k_{1}}(Z_{1}) & \tilde{\psi}^{\omega_{1}}_{0,k_{1}}(Z_{2})\\
    \tilde{\psi}^{\omega_{2}}_{0,k_{2}}(Z_{1}) & \tilde{\psi}^{\omega_{2}}_{0,k_{2}}(Z_{2})
  \end{vmatrix},
\end{equation}
where
\begin{equation}\label{eq:intro-lll-packet-2}
  \tilde{\psi}^{\omega_1}_{0,k_1}(Z)
  \propto
  \frac{(Z-\omega_1)^{k_1}}{(Z-\omega_2)^\alpha}\,
  \exp\left[-\frac{\abs{Z}^{2}}{4}+\frac{\conj{\omega}_1 Z}{2}-\frac{\abs{\omega_1}^{2}}{4}\right]
\end{equation}
We call $(Z-\omega_2)^{-\alpha}$ the inverse vortex factor: it is the reciprocal of the factor that would attach $\alpha$ vortices at $\omega_{2}$. Through this factor, each anyon has information about the locations of the other anyon. Similarly
\begin{equation}\label{eq:intro-lll-packet-3}
  \tilde{\psi}^{\omega_2}_{0,k_2}(Z)
  \propto
  \frac{(Z-\omega_2)^{k_2}}{(Z-\omega_1)^\alpha}\,
  \exp\left[-\frac{\abs{Z}^{2}}{4}+\frac{\conj{\omega}_2 Z}{2}-\frac{\abs{\omega_2}^{2}}{4}\right]
\end{equation}
The wave function in Eq.~\eqref{eq:intro-two-anyon-conventional-2} is then given by
\begin{eqnarray}
  &\tilde{\Psi}^{k_{1},k_{2}}_{\omega_{1},\omega_{2}} \propto \nonumber \\
  &(Z_{1}-Z_{2})^{\alpha}\left[ \frac{\psi^{\omega_{1}}_{0,k_{1}}(Z_{1}) \psi^{\omega_{2}}_{0,k_{2}}(Z_{2}) }{(Z_{1}-\omega_{2})^{\alpha}(Z_{2}-\omega_{1})^{\alpha}}-\{Z_1\leftrightarrow Z_2  \}\right].
\end{eqnarray}
Let us now write $Z_{i}=\omega_{i}+u_{i}$ and $\omega_{12}=\omega_{1}-\omega_{2}$, and treat $u_{i}/\omega_{12}$ as a small parameter: since the orbitals are exponentially localized about $\omega_{i}$, the wave function is appreciable only for $\abs{u_{i}}$ of the order of the magnetic length, which is small compared to $\abs{\omega_{12}}$ for well-separated anyons.
This yields
\begin{equation}\label{eq:intro-vortex-cancellation}
  \begin{split}
    &(Z_{1}-Z_{2})^{\alpha}\,
    (Z_{1}-\omega_{2})^{-\alpha}(Z_{2}-\omega_{1})^{-\alpha} \\
    &\quad=
    (-\omega_{12})^{-\alpha}
    \left[1+\alpha\frac{u_{1}u_{2}}{\omega_{12}^{2}}+\dots\right].
  \end{split}
\end{equation}
The key point is that all terms linear in $u_{i}/\omega_{12}$ cancel out. The factors $(Z_{1}-\omega_{2})^{-\alpha}(Z_{2}-\omega_{1})^{-\alpha}$ thus cancel the vortex factor $(Z_{1}-Z_{2})^{\alpha}$ to first order (Sec.~\ref{sec:hf-construction}), effectively eliminating the effect of the vortex factor on the densities of the two localized anyons while keeping their statistics intact. We also note, as discussed below, that this expression produces a wave function that has a well defined analytic structure of the LLL space.

\begin{figure}
  \includegraphics[width=\columnwidth]{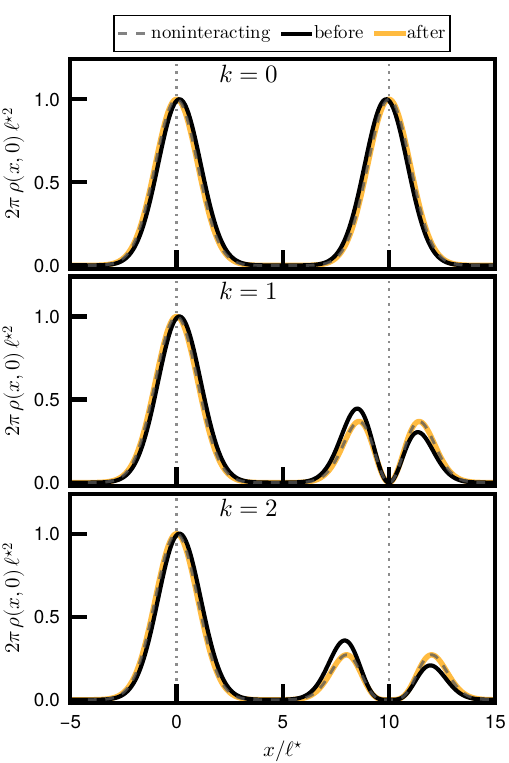}
  \caption{In Fig.~\ref{fig:two-anyon-density}, we saw how multiplication by the factor $(Z_1-Z_2)^\alpha$ distorts the density profile of two anyons. In this figure, the dashed  gray curve shows the density $\rho(x,0)$ of two independent particles: one localized at the origin in the $k_{1}=0$ angular-momentum state and the second at $\omega=10\ellstar$ in the $k$ state. The dotted vertical lines mark their positions. The orange curve shows the density profile of $\tilde{\Psi}^{k_{1},k_{2}}_{\omega_{1},\omega_{2}}$ in Eq.~\eqref{eq:intro-two-anyon-conventional-2} ; this is essentially identical to the gray dashed curve. The solid black curve is the density profile of $(Z_1-Z_2)^{-\alpha}\tilde{\Psi}^{k_{1},k_{2}}_{\omega_{1},\omega_{2}}$, i.e. of the Slater determinant on the right hand side of Eq.~\eqref{eq:intro-two-anyon-conventional-2}; this density is distorted in precisely such a fashion that the further distortion due to multiplication by the factor $(Z_1-Z_2)^\alpha$ produces an undistorted density profile. We have chosen $\alpha=2/3$ in this figure.}
  \label{fig:hf-det-before-after}
\end{figure}

Figure~\ref{fig:hf-det-before-after} shows the densities of $\Psi^{k_{1},k_{2}}_{\omega_{1},\omega_{2}}$ and the modified wave function $\tilde{\Psi}^{k_{1},k_{2}}_{\omega_{1},\omega_{2}}$ for $\alpha=2/3$. In the former, the wavepackets clearly show a distortion. The distortion is essentially eliminated in the latter.
Finally, the braid statistics extracted from the modified state is shape independent, $\Delta\Theta/2\pi=\alpha$ (Appendix~\ref{app:hf-braid}).

\subsection{Generalization to FQH QPs}

We next generalize this construction to FQH QPs. The wave function for two QPs is constructed as
\begin{equation}
  {\Psi}^{k_{1},k_{2}}_{\omega_{1},\omega_{2}}=\LLL\,\Phi^{k_{1},k_{2}}_{\omega_{1},\omega_{2}}\,\Phi_{1}^{2p}
\end{equation}
where $\Phi^{k_{1},k_{2}}_{\omega_{1},\omega_{2}}$ is the single Slater-determinant wave function of electrons at effective magnetic field $B^*$ with $n$ LLs fully occupied and two localized electrons in the $(n+1)^{st}$ LL at $\omega_1,\omega_2$ in angular momentum states $k_1,k_2$. FQH QPs are obtained by composite-fermionization, which amounts to multiplication by $\Phi_1^{2p}$ followed by LLL projection, where $\Phi_1=\prod_{i<j}(z_i-z_j) \exp[-\sum_j |z_j|^2/4l_1^2]$ where $l_1$ is the magnetic length corresponding to the magnetic field at filling factor unity.  The localized electrons in the $(n+1)^{\rm st}$ LL are placed in single particle states $\psi^{\omega_j}_{n,k_j}(z_j)$. The Berry phase calculated from this wave function produces a $k_j$ dependent braid statistics.

Taking a cue from the two-anyon problem, we consider a modified wave function
\begin{equation}
\tilde{\Psi}^{k_{1},k_{2}}_{\omega_{1},\omega_{2}}=\LLL\,\Phi^{k_{1},k_{2}}_{\omega_{1},\omega_{2}}\,\Phi_{1}^{2p}
\end{equation}
in which we make the replacement
\begin{equation}
  \psi^{\omega_1}_{n,k_1}(z)\rightarrow (z-\omega_2)^{-\alpha}\psi^{\omega_1}_{n,k_1}(z)
\end{equation}
\begin{equation}
  \psi^{\omega_2}_{n,k_2}(z)\rightarrow (z-\omega_1)^{-\alpha} \psi^{\omega_2}_{n,k_2}(z)
\end{equation}
(For $n\geq1$ the orbital $\psi^{\omega}_{n,k}$ lies outside the LLL, where a shift of angular momentum is not a multiplication by $z-\omega$; the operator form that replaces $(z-\omega_{b})^{-\alpha}$ in that case is given in Eq.~\eqref{eq:hf-jain-qp-orbital}.) One can treat $\alpha$ as a parameter and fix it by demanding that the QP shape after composite-fermionization is undistorted. While we do not have a proof, our guess is $\alpha=2p/(2pn+1)$, as each QP has $2p/(2pn+1)$ flux quanta bound to it, and the factor $(z-\omega_1)^{-2p/(2pn+1)}$ essentially eliminates the effect of that flux.
We find from explicit evaluation that $\alpha=2p/(2pn+1)$ indeed accomplishes that goal. Furthermore, we find, again from explicit calculation, that this wave function produces a $k$-independent braid statistics.

The modified orbitals of Eqs.~\eqref{eq:intro-lll-packet-2} and \eqref{eq:intro-lll-packet-3}, and likewise their FQH counterparts, do not lie within the LLL: the factor $(Z-\omega_{2})^{-\alpha}$ has a branch point at $\omega_{2}$. However, since the orbital is exponentially localized about $\omega_{1}$, we can expand $(Z-\omega_{2})^{-\alpha}$ as a series in $Z-\omega_{1}$, which is equivalent to a superposition of states in which the particle occupies various angular-momentum orbitals about $\omega_{1}$. This series is not an ordinary Taylor series but an asymptotic one: its error relative to the original function does not decrease indefinitely as more terms are included, but instead grows beyond a certain order. We keep the first $\approx d^{2}/2$ terms, where $d$ is the distance between the two anyons in units of the magnetic length; as far as we can tell, none of our results depends on this choice (Appendix~\ref{app:dressing-asymptotics}).

To summarize: The Berry phase of a QP braiding around another has contributions from (a) the AB phase and (b) the braid statistics; the latter is extracted from the difference between the Berry phases with and without the other QP inside the loop. We show that while the notion of braid statistics remains valid in principle, extracting it is highly challenging, because even small changes in the shape and the trajectory of the orbiting QP between the two loops alter this difference by amounts comparable to the order-one contribution from the expected, universal braid statistics.

\section{Localized QH and QP states at Jain fillings}
\label{sec:localized-qh-qp-states}

In this section, we construct a QH or a QP state localized at an arbitrary position $\omega$, with arbitrary angular momentum $k$ around this position. 

The central principle of CF theory is that \textit{the physics of strongly interacting electrons at fractional fillings $\nu=n/(2pn+1)$ maps onto the physics of noninteracting CFs at a reduced magnetic field $B^{\star}=B(1-2p\nu) \Leftrightarrow \nu^{\star}=n$} (here, $\nu^{\star}$ is the filling of CFs). This correspondence also holds at the wave function level: electron wave functions at fractional fillings are obtained by composite-fermionizing the corresponding IQH wave functions~\cite{Jain89,Jain07}.

Of most direct relevance to the present work is the fact that the ground state at $\nu=n/(2pn\pm 1)$ is understood as the $n$-filled-$\Lambda$L IQH state of CFs carrying $2p$ flux quanta; a QP is an isolated CF in the otherwise empty $\Lambda$L; and a QH is a missing CF in an otherwise filled $\Lambda$L. Exact diagonalization (ED) studies have shown that the Jain wave functions for the ground states as well as for QPs and QHs are remarkably accurate representations of the corresponding Coulomb eigenstates~\cite{Jain07,Gattu24A}, which has led to numerous detailed quantitative predictions. Furthermore, as we move further away from $\nu=n/(2pn\pm 1)$ by adding more and more QPs or QHs, the CF theory continues to provide a faithful and accurate account of the low-energy spectrum~\cite{Balram13}. This gives us confidence that the CF theory ought to provide a framework within which one can study the feasibility of topological concepts such as fractional statistics.

\begin{figure}
  \includegraphics[width=\columnwidth]{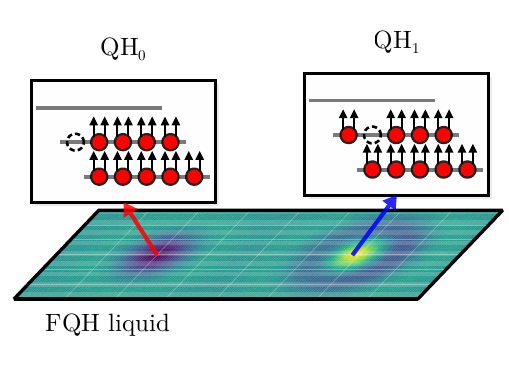}
  \caption{Schematic illustration of localized QHs in an FQH liquid at Jain fillings $\nu=n/(2pn+1)$. The color on the plane depicts the density of electrons in the bulk of an FQH liquid with two localized QHs, and the callouts show the corresponding CF occupation of the $\Lambda$Ls, with orbitals defined relative to the center of each QH. CFs are depicted as electrons with two attached flux tubes, while a QH, which is a missing CF, is indicated by a dashed circle.}
  \label{fig:schematic-qh-shape}
\end{figure}

In this section, we first review composite-fermionization by applying it to construct the CF wave function for the FQH ground state at $\nu=n/(2pn+1)$. We then construct localized QH and QP wave functions. In the IQH analog, a QH is a hole in the $(n-1)^{\rm th}$ LL and a QP is an additional electron in the $n^{\rm th}$ LL; the shape of the QP or the QH is determined by the single-particle problem within the relevant LL subject to the localizing potential. We express the eigenstates in terms of angular-momentum orbitals centered at the localization point, discuss the states selected by representative potentials, and then composite-fermionize the resulting IQH states to obtain localized QH and QP wave functions at fractional fillings; this construction is depicted schematically in Fig.~\ref{fig:schematic-qh-shape}. Finally, we generalize the construction to two-QH and two-QP states.

\subsection{Composite-fermionization}
Let $\Phi$ be a single Slater-determinant state of $N$ electrons occupying certain LL orbitals at field $B^{\star}$, corresponding to a single Slater-determinant state of $N$ CFs filling certain $\Lambda$L orbitals. Composite-fermionization of $\Phi$ consists of two steps: vortex attachment followed by projection into the LLL. Let us illustrate it using the example of the ground state at an integer filling.

We denote by $\Phi_n(\{z_i,\conj{z}_i\};B^{\star})$ the Slater determinant state of $N$ noninteracting electrons fully occupying the lowest $n$ LLs at the reduced field $B^{\star}$, i.e., the IQH ground state at filling $\nu=n$ ($\conj{z}_{i}$ is the complex conjugate of $z_i$):
\begin{equation}
  \label{eq:iqh-ground-state-disk}
  \begin{aligned}
    &\Phi_{n}(\{z_i,\conj{z}_i\};B^{\star})
    =
    \frac{1}{\sqrt{N!}}\\[-2pt]
    &\quad\times
    \begin{vmatrix}
      \eta_{0,0}(z_1) & \cdots & \eta_{0,0}(z_N)\\
      \eta_{0,1}(z_1) & \cdots & \eta_{0,1}(z_N)\\
      \vdots & \ddots & \vdots\\
      \eta_{0,M_{0}}(z_1) & \cdots & \eta_{0,M_{0}}(z_N)\\
      \eta_{1,-1}(z_1) & \cdots & \eta_{1,-1}(z_N)\\
      \vdots & \ddots & \vdots\\
      \eta_{1,M_{1}}(z_1) & \cdots & \eta_{1,M_{1}}(z_N)\\
      \vdots & \ddots & \vdots\\
      \eta_{n-1,-(n-1)}(z_1) & \cdots & \eta_{n-1,-(n-1)}(z_N)\\
      \vdots & \ddots & \vdots\\
      \eta_{n-1,M_{n-1}}(z_1) & \cdots & \eta_{n-1,M_{n-1}}(z_N)
    \end{vmatrix}.
  \end{aligned}
\end{equation}
Here $M_{\lambda}$ is the maximum angular momentum occupied in the $\lambda^{\rm th}$ LL; for $N$ electrons uniformly filling $n$ LLs in the disk geometry, $M_\lambda = N/n - 1 - \lambda$. In determinants we abbreviate $\eta_{\lambda,m}(z_i/\ellstar,\conj{z}_i/\ellstar)$ as $\eta_{\lambda,m}(z_i)$. The functions $\eta_{n,m}(z/\lB,\bar{z}/\lB)$ are the symmetric-gauge LL orbitals in the planar geometry:
\begin{equation}\label{eq:orbitals-definition-disk}
  \begin{split}
    \eta_{n, m}\left(\frac{z}{\lB}, \frac{\bar{z}}{\lB}\right) &= \frac{1}{\lB}\sqrt{\frac{n!}{2^{m+1}\pi (m+n)!}}\left(\frac{z}{\lB}\right)^{m}\\
    &\times L_{n}^{m}\left(\frac{\abs{z}^{2}}{2\lB^{2}}\right)\exp\left(-\frac{\abs{z}^{2}}{4\lB^{2}}\right).
  \end{split}
\end{equation}
Here, $n=0,1,2,\dots$ is the LL index, $m$ is the azimuthal angular momentum, and $\lB$ is the magnetic length. For the $n^{\mathrm{th}}$ LL, the allowed angular momenta satisfy $m\geq-n$.

For $\Phi_n$, the vortex-attachment step amounts to multiplying by the Jastrow factor
\begin{equation}
  \label{eq:vortex-attachment-factor}
  \Phi_{1}^{2p}
  =
  \left[\prod_{i<j=1}^{N}(z_i-z_j)
  \exp\left(-\sum_{i=1}^{N}\frac{\abs{z_i}^{2}}{4\ell_1^2}\right)\right]^{2p}.
\end{equation}
This factor attaches $2p$ vortices to each electron. Here, $\lB=\sqrt{\hbar c /eB}$ is the magnetic length and $\ell_{1}=\lB/\sqrt{\nu}$, and $\ellstar=\sqrt{\hbar c/eB^{\star}}=\sqrt{2pn+1}\,\lB$ is the effective magnetic length; this choice ensures that the Gaussians in $\Phi_1^{2p}$ and $\Phi_n$ combine to give the correct normalization at the physical field $B$, since $2p/(4\ell_1^2)+1/(4\ell^{\star 2}) = 1/(4\lB^2)$. Projecting the resulting state into the LLL gives
\begin{equation}
  \label{eq:jain-cf-ground-state-disk}
  \begin{split}
    \Psi_{\frac{n}{2pn+1}}
    &=
    \LLL
    \Phi_n(\{z_i,\conj{z}_i\};B^{\star})\Phi_1^{2p},
  \end{split}
\end{equation}
where $\LLL$ is the projection operator into the LLL (implemented in practice via the Jain--Kamilla method~\cite{Jain97}). This gives the CF wave function for the FQH ground state at $\nu=n/(2pn+1)$. The same prescription applies to arbitrary $\Lambda$L configurations and is used below to construct localized QH and QP wave functions.

\subsection{Localized hole / electron wavepackets at integer fillings $\nu=n$}

In this subsection we first show that, for the QH or QP of an IQH state, realistic localizing potentials can stabilize different angular-momentum states (angular momentum being defined relative to the center of the QH or QP), giving localized wavepackets of different shapes. We then show how such wavepackets can be constructed at an arbitrary position. The corresponding FQH QH and QP wave functions needed to evaluate the braid statistics are then obtained by composite-fermionization.

We assume that both the inter-electron interactions and the localizing potential are weak compared with the cyclotron energy (or, for CFs, the CF cyclotron energy), so that different LLs ($\Lambda$Ls) do not mix. At integer filling $\nu=n$, a QH is a positively charged hole in the $(n-1)^{\rm th}$ LL, while a QP is an added electron in the $n^{\rm th}$ LL; under this assumption, the many-body problem reduces to a single-particle problem within the relevant LL.

For simplicity, we focus on rotationally symmetric localizing potentials of the form $V(|r-\omega|)$, where $\omega$ is the center of the potential. When the potential is centered at the origin, angular momentum about the origin is a good quantum number, and the eigenstates are the symmetric-gauge LL orbitals $\eta_{n, m}(z/\lB,\bar{z}/\lB)$; different potentials select different angular-momentum orbitals, and hence wavepackets of different shapes. For a potential centered at a generic point $\omega$, the eigenstates are instead magnetic translations of these origin-centered orbitals. These single-particle orbitals are the building blocks of the many-body IQH states that we composite-fermionize below to obtain localized QH and QP wave functions at fractional fillings.

\subsubsection{Orbitals selected by a rotationally symmetric trap}

The objective of this subsection is to show that the lowest energy state of an electron restricted to an LL can have different angular momenta for a realistic (Gaussian or Coulomb type) localizing potential. This provides a motivation for considering QPs and QHs with different angular momenta. Another motivation comes from localizing potentials which are not rotationally invariant -- they will automatically produce QPs and QHs which are linear superpositions of states with different angular momenta.

We first consider a particle restricted to the $n^{\mathrm{th}}$ LL and subject to a rotationally symmetric localizing potential $V(r)$ centered at the origin. In the symmetric gauge, the orbitals $\eta_{n,m}(z/\lB,\bar{z}/\lB)$ have definite angular momentum about the origin, with $L_z=-\imath\hbar\partial_\theta$. Since the projected potential $\mathrm{P}_{n{\mathrm{LL}}}V(r)\mathrm{P}_{n{\mathrm{LL}}}$ preserves angular momentum, it is diagonal in this basis. Its matrix elements are
\begin{equation}
  \label{eq:projected-potential-matrix-element}
  V_{n, m} = \int d\tau V(r)\abs{\eta_{n, m}(r)}^{2}.
\end{equation}
Thus the localized eigenstates within this LL are the orbitals $\eta_{n,m}$, and the ground state is the orbital that minimizes $V_{n,m}$.

\begin{figure}
  \includegraphics{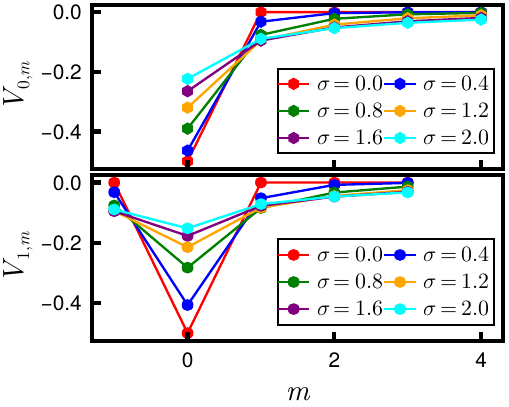}
  \caption{
    Eigenenergies $E_{n,m}=V_{n,m}$ for an electron in the Gaussian impurity potential $V(r)=-\exp(-r^2/2\sigma^{2})/(2\pi \sigma^{2})$, shown for orbitals in the $n=0$ and $n=1$ LLs (top and bottom panels, respectively) as a function of angular momentum $m$, for several values of $\sigma$ in units of $\lB$. For all $\sigma$, $E_{n,0}$ is the lowest, indicating that this potential traps an electron in the $m=0$ orbital regardless of which LL it occupies.
  }
  \label{fig:gaussian-impurity}
\end{figure}

As a first example, consider a Gaussian impurity potential $V(r)=-\exp(-r^2/2\sigma^{2})/(2\pi\sigma^{2})$. The eigenenergies $V_{n,m}$ are shown in Fig.~\ref{fig:gaussian-impurity} for $n=0$ and $n=1$. For this potential, the $m=0$ orbital has the largest density at the origin and is energetically favored.

\begin{figure}
  \includegraphics{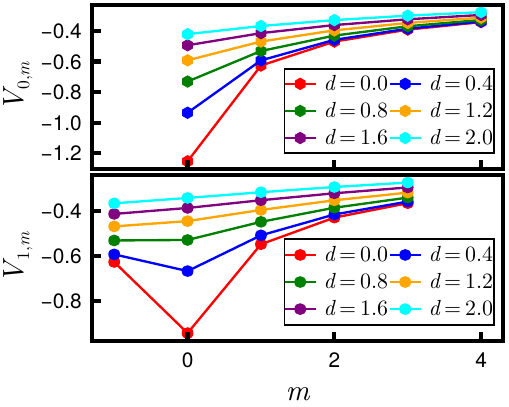}
  \caption{
    Eigenenergies $E_{n,m}=V_{n,m}$ for an electron in the Coulomb impurity potential $V(r)=-1/\sqrt{r^2+d^2}$, produced by a charge located a distance $d$ from the two-dimensional plane to which the electrons are restricted, shown for orbitals in the $n=0$ and $n=1$ LLs (top and bottom panels, respectively) as a function of angular momentum $m$, for several values of $d$ in units of $\lB$. In the $n=0$ LL, the $m=0$ state is the ground state for all $d$. In the $n=1$ LL, for $d \gtrsim 0.8\lB$, the $m=-1$ state becomes the ground state.
  }
  \label{fig:coulomb-impurity}
\end{figure}

As a second example, consider a Coulomb impurity $V(r)=-1/\sqrt{r^2+d^2}$ positioned a distance $d$ away from the two-dimensional plane to which the electrons are restricted. The lowest-energy orbital can now change as $d$ is varied, as shown in Fig.~\ref{fig:coulomb-impurity}. When the impurity is close, i.e., $d\to 0$, the $m=0$ orbital is favored. For sufficiently large $d$, however, the favored orbital is the one most tightly concentrated near the origin, corresponding to $m=-n$.

These examples show that realistic localizing potentials can select wavepackets with different angular momenta, and hence different shapes.

The microscopic potential that stabilizes a localized FQH QH or QP wavepacket will generally differ from the single-particle IQH examples considered here, because the density profile of an FQH QH or QP is not simply that of a hole or electron at the effective magnetic length $\ellstar$. The examples above illustrate how local potentials can select different orbital shapes.

\subsubsection{Angular momentum orbitals about a generic point}

If the potential is centered at a point $\omega$, taking the form $V(|r-\omega|)$, it is convenient to work with angular momentum measured about $\omega$. The corresponding orbitals are magnetic translations of the orbitals at the origin. Note the sign: $\mathcal{T}_{-\omega}$ shifts the orbital center from the origin to $\omega$, while $\mathcal{T}_{\omega}$ would shift it to $-\omega$. With the convention
\begin{equation}
  \label{eq:magnetic-translation-operator}
  \mathcal{T}_{\omega}\psi(z,\conj{z}) = \exp\left(\frac{-\conj{\omega}z+\omega\conj{z}}{4\lB^{2}}\right)\psi(z+\omega,\conj{z}+\conj{\omega})
\end{equation}
the translated angular momentum orbital is (as derived in Appendix~\ref{app:localized-electron-wavepackets})
\begin{equation}
  \label{eq:localized-orbital-expansion}
  \begin{split}
    \psi_{n, m}^{\omega}(z, \bar{z}) &= \mathcal{T}_{-\omega}\eta_{n, m}(\frac{z}{\lB},\frac{\conj{z}}{\lB}) \\ &= \exp\left(\frac{\conj{\omega}z-\omega\conj{z}}{4\lB^{2}}\right) \eta_{n,m} (\frac{z-\omega}{\lB},\frac{\conj{z}-\conj{\omega}}{\lB})\\
    & = \sqrt{2\pi}\sum_{M=-n}^{\infty}\conj{\eta}_{m+n,M-m}(\frac{\omega}{\lB},\frac{\conj{\omega}}{\lB})\\
    &\hspace{5.5em}\times\eta_{n, M}(\frac{z}{\lB},\frac{\conj{z}}{\lB})
  \end{split}
\end{equation}
These orbitals are the single-particle building blocks for localized QH and QP wavepackets. The expansion has two simple checks. First, when $\omega=0$, only the term with $M=m$ survives, so $\psi_{n,m}^{0}$ reduces to the orbital $\eta_{n,m}$ centered at the origin. Second, the coefficients in the expansion are themselves LL orbitals as functions of $\omega$, which ensures that these orbitals form a closed set under magnetic translations.

For rotationally asymmetric localizing potentials, the eigenstates will generally be linear combinations of the orbitals $\psi_{n,m}^{\omega}$. We do not consider this additional complication here (see Sec.~\ref{subsec:asymmetric-qp}); the main points below do not rely on this simplification.

\subsection{Localized quasihole / quasiparticle wave functions at fillings $\nu=n$ and $\nu=n/(2pn+1)$}
The single-particle solution now determines the corresponding IQH many-body wave functions. We use $m$ for the physical angular momentum of the hole or electron, and $k$ as a unified label for the shape of the localized wavepacket. For a hole in the $(n-1)^{\rm th}$ LL, $m=k-(n-1)$; for an added electron in the $n^{\rm th}$ LL, $m=k-n$. These many-body IQH wave functions then allow us to construct the corresponding QH and QP wave functions at $\nu=n/(2pn+1)$ via composite-fermionization.

For a QH at $\nu=n$ localized at $\omega$, we remove an electron from the angular-momentum orbital labeled by $k$, equivalently the orbital with $m=k-(n-1)$ centered about $\omega$ in the $(n-1)^{\rm th}$ LL. This hole is created by the action of a linear combination of annihilation operators $c_{n-1,m}$ which annihilate an electron in the $\eta_{n-1,m}$ orbital, with coefficients fixed by Eq.~\eqref{eq:localized-orbital-expansion}. The many-body wave function for the state $\mathrm{QH}_{k}$ can be written as (up to a normalization constant)
\begin{equation}
  \label{eq:iqh-localized-hole-determinant}
  \begin{aligned}
    &\Phi_{n}^{{\QHsym_k}, \omega}(\{z_i,\conj{z}_i\};B^{\star})
    =\\
    &\scalebox{0.94}{$\displaystyle{\setlength{\arraycolsep}{2.5pt}
        \begin{vmatrix}
          \eta_{0,0}(z_1) & \cdots & \eta_{0,0}(z_N) & 0\\
          \eta_{0,1}(z_1) & \cdots & \eta_{0,1}(z_N) & 0\\
          \vdots & \ddots & \vdots & \vdots\\
          \eta_{n-2,M_{n-2}}(z_1) & \cdots & \eta_{n-2,M_{n-2}}(z_N) & 0\\
          \eta_{n-1,-(n-1)}(z_1) & \cdots & \eta_{n-1,-(n-1)}(z_N) & \eta_{k, -k}(\omega)\\
          \vdots & \ddots & \vdots & \vdots\\
          \eta_{n-1,M_{n-1}}(z_1) & \cdots & \eta_{n-1,M_{n-1}}(z_N) & \eta_{k, M_{n-1}+n-1-k}(\omega)
    \end{vmatrix}}$}.
  \end{aligned}
\end{equation}

The last column in Eq.~\eqref{eq:iqh-localized-hole-determinant} contains the expansion coefficients of the localized orbital $\psi^\omega_{n-1,k-(n-1)}$ in the $\{\eta_{n-1,m'}\}$ basis [cf.\ Eq.~\eqref{eq:localized-orbital-expansion}]; expanding the determinant along this column yields the state with exactly the orbital $\psi^\omega_{n-1,k-(n-1)}$ removed.

Similarly, a QP localized at $\omega$ is obtained by adding an electron in the angular-momentum orbital labeled by $k$, equivalently the orbital with $m=k-n$ centered about $\omega$ in the $n^{\rm th}$ LL. The many-body wave function for the state $\mathrm{QP}_{k}$ can be written as
\begin{equation}
  \label{eq:iqh-localized-electron-determinant}
  \begin{aligned}
    &\Phi_{n}^{{\QPsym_k},\omega}(\{z_i,\conj{z}_i\};B^{\star}) = \\
    &
    \begin{vmatrix}
      \eta_{0,0}(z_1) & \cdots & \eta_{0,0}(z_N)\\
      \eta_{0,1}(z_1) & \cdots & \eta_{0,1}(z_N)\\
      \vdots & \ddots & \vdots\\
      \eta_{n-1,M_{n-1}}(z_1) & \cdots & \eta_{n-1,M_{n-1}}(z_N)\\
      \psi_{n,k-n}^{\omega}(z_1)
      & \cdots &
      \psi_{n,k-n}^{\omega}(z_N)
    \end{vmatrix}.
  \end{aligned}
\end{equation}
In the last row, $\psi_{n,k-n}^{\omega}$ is the angular-momentum orbital centered about $\omega$ defined in Eq.~\eqref{eq:localized-orbital-expansion}. Note that the orbital index $m = k-n$ is negative for $k < n$.

Composite-fermionizing these IQH states gives the corresponding localized QH and QP wave functions at Jain fillings $\nu = n/(2pn+1)$:
\begin{align}
  \label{eq:localized-qh-wavepacket}
  \Psi^{\mathrm{QH}_{k}, \omega}_{\frac{n}{2pn+1}} &= \LLL \Phi_{n}^{\mathrm{QH}_{k}, \omega}(\{z_i,\conj{z}_i\};B^{\star}) \Phi_{1}^{2p}\\
  \label{eq:localized-qp-wavepacket}
  \Psi^{\mathrm{QP}_{k}, \omega}_{\frac{n}{2pn+1}} &= \LLL \Phi_{n}^{\mathrm{QP}_{k}, \omega}(\{z_i,\conj{z}_i\};B^{\star}) \Phi_{1}^{2p}
\end{align}

To study braiding statistics we need states with two localized QHs or QPs. We can similarly create states of two QHs and two QPs localized at $\omega_{1}, \omega_{2}$ at Jain fillings $\nu = n/(2pn+1)$:
\begin{equation}
  \label{eq:localized-two-qh-wavepacket}
  \begin{split}
    &\Psi^{\mathrm{QH}_{k_{1}}, \omega_{1};\mathrm{QH}_{k_{2}}, \omega_{2}}_{\frac{n}{2pn+1}}= \\
    &\LLL
    \Phi_{n}^{\mathrm{QH}_{k_{1}}, \omega_{1};\mathrm{QH}_{k_{2}}, \omega_{2}}
    (\{z_i,\conj{z}_i\};B^{\star}) \Phi_{1}^{2p},
  \end{split}
\end{equation}
\begin{equation}
  \label{eq:localized-two-qp-wavepacket}
  \begin{split}
    &\Psi^{\mathrm{QP}_{k_{1}}, \omega_{1};\mathrm{QP}_{k_{2}}, \omega_{2}}_{\frac{n}{2pn+1}}= \\
    &\LLL
    \Phi_{n}^{\mathrm{QP}_{k_{1}}, \omega_{1};\mathrm{QP}_{k_{2}}, \omega_{2}}
    (\{z_i,\conj{z}_i\};B^{\star}) \Phi_{1}^{2p}.
  \end{split}
\end{equation}

The corresponding IQH Slater determinant for two QHs is
\begin{equation}
  \label{eq:iqh-localized-two-hole-determinant}
  \setlength{\arraycolsep}{3pt}
  \begin{aligned}
    &\Phi_{n}^{{\mathrm{QH}_{k_{1}}, \omega_{1};\mathrm{QH}_{k_{2}}, \omega_{2}}}
    (\{z_i,\conj{z}_i\};B^{\star})= \\
    &\scalebox{0.95}{$\displaystyle{\setlength{\arraycolsep}{3pt}
        \begin{vmatrix}
          \eta_{0,0}(z_1) & \cdots & \eta_{0,0}(z_N) & 0 & 0\\
          \eta_{0,1}(z_1) & \cdots & \eta_{0,1}(z_N) & 0 & 0\\
          \vdots & \ddots & \vdots & \vdots & \vdots\\
          \eta_{n-2,M_{n-2}}(z_1) & \cdots & \eta_{n-2,M_{n-2}}(z_N) & 0 & 0\\
          \eta_{n-1,-(n-1)}(z_1) & \cdots & \eta_{n-1,-(n-1)}(z_N) & h^{(1)}_{-(n-1)} & h^{(2)}_{-(n-1)}\\
          \vdots & \ddots & \vdots & \vdots & \vdots\\
          \eta_{n-1,M_{n-1}}(z_1) & \cdots & \eta_{n-1,M_{n-1}}(z_N) & h^{(1)}_{M_{n-1}} & h^{(2)}_{M_{n-1}}
    \end{vmatrix}}$},
  \end{aligned}
\end{equation}
where $h^{(a)}_{m} \equiv \eta_{k_a,\,m+(n-1)-k_a}(\omega_a)$ for $a=1,2$ are the expansion coefficients of the localized orbitals removed at $\omega_1$ and $\omega_2$ [cf.\ Eq.~\eqref{eq:iqh-localized-hole-determinant}].

For two QPs, the corresponding IQH Slater determinant is
\begin{equation}
  \label{eq:iqh-localized-two-electron-determinant}
  \begin{aligned}
    &\Phi_{n}^{{\mathrm{QP}_{k_{1}}, \omega_{1}; \mathrm{QP}_{k_{2}}, \omega_{2}}}
    (\{z_i,\conj{z}_i\};B^{\star}) = \\
    &
    \begin{vmatrix}
      \eta_{0,0}(z_1) & \cdots & \eta_{0,0}(z_N)\\
      \eta_{0,1}(z_1) & \cdots & \eta_{0,1}(z_N)\\
      \vdots & \ddots & \vdots\\
      \eta_{n-1,M_{n-1}}(z_1) & \cdots & \eta_{n-1,M_{n-1}}(z_N)\\
      \psi_{n,k_1-n}^{\omega_1}(z_1) & \cdots & \psi_{n,k_1-n}^{\omega_1}(z_N)\\
      \psi_{n,k_2-n}^{\omega_2}(z_1) & \cdots & \psi_{n,k_2-n}^{\omega_2}(z_N)
    \end{vmatrix}.
  \end{aligned}
\end{equation}

In the next section, we use these wave functions to evaluate the Berry phases acquired during braiding and to determine how the QH or QP shapes affect those phases.

\begin{figure}
  \includegraphics[width=\columnwidth]{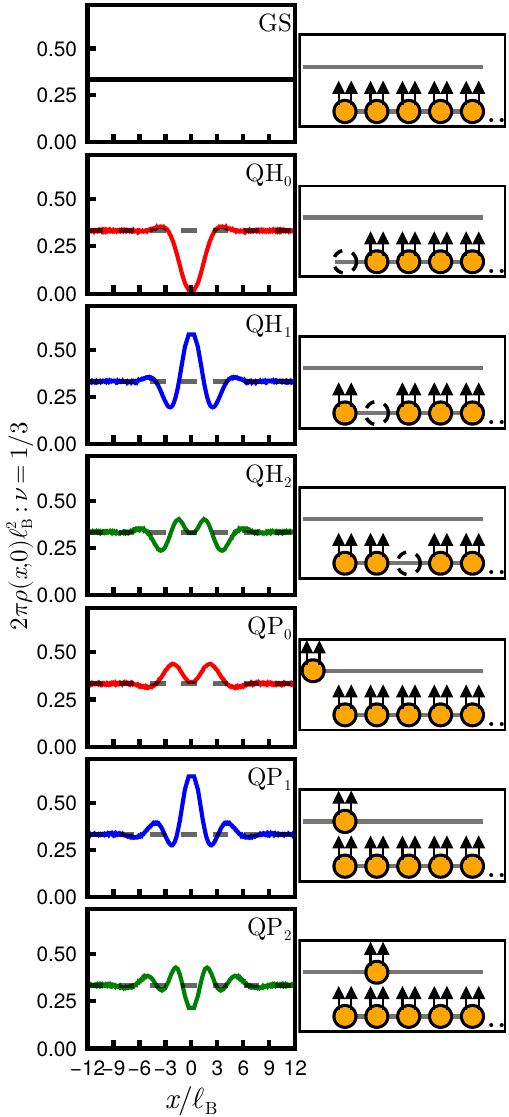}
  \caption{
    Electron density $\rho(x,0)$ along the $x$ axis for $\nu=1/3$ FQH liquid. The ground state (GS) corresponds to CFs filling the $n=0$ $\Lambda$L. The QH states $\mathrm{QH}_{0,1,2}$ are obtained by removing one CF from the $m=0,1,2$ orbitals, respectively, of this level. The QP states $\mathrm{QP}_{0,1,2}$ are obtained by adding one CF to the $m=-1,0,1$ orbitals, respectively, of the $n=1$ $\Lambda$L. The index $k$ in $\mathrm{QH}_{k} / \mathrm{QP}_{k}$ specifies the angular momentum and hence the shape of the QH or QP. The schematics on the right show the corresponding CF occupations. The dashed horizontal line marks the uniform background density $\nu/(2\pi\lB^{2})$, with $\lB = \sqrt{\hbar c/eB}$.
  }
  \label{fig:schematic-qh-cf-occupations}
\end{figure}

\begin{figure}
  \includegraphics[width=\columnwidth]{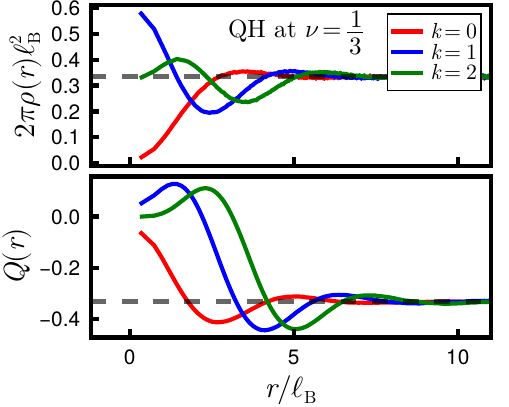}
  \includegraphics[width=\columnwidth]{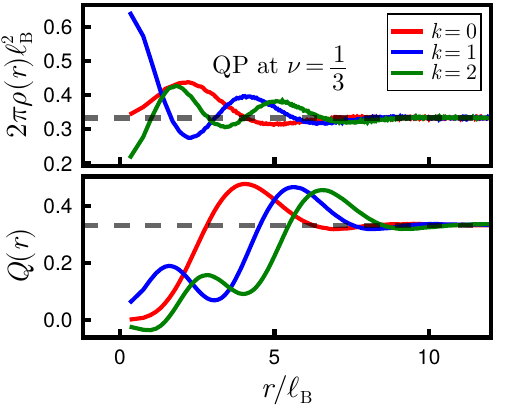}
  \caption{Density $2\pi \rho(r)\lB^{2}$ and integrated charge $Q(r)$ for localized QH and QP wavepackets at $\nu=1/3$ in the $k=0,1,2$ angular-momentum states about the origin. The QH states correspond to a CF missing from an otherwise full $n=0$ $\Lambda$L in the $m=0,1,2$ orbitals, while the QP states correspond to an additional CF in the $n=1$ $\Lambda$L in the $m=-1,0,1$ orbitals. Dashed lines indicate the background filling $\nu=1/3$ and the expected integrated charges $\mp 1/3$ in units of the electron charge $-e$.}
  \label{fig:qh-qp-1-3-shapes}
\end{figure}

\begin{figure}
  \includegraphics[width=\columnwidth]{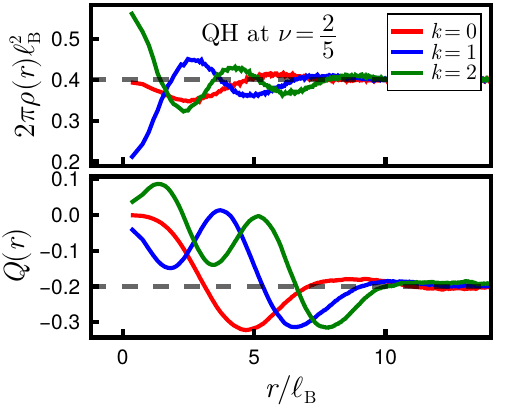}
  \includegraphics[width=\columnwidth]{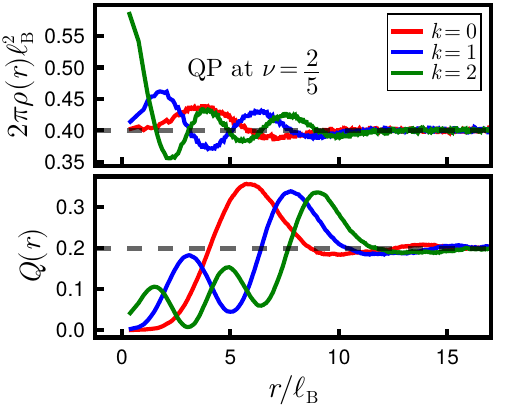}
  \caption{Density $2\pi \rho(r)\lB^{2}$ and integrated charge $Q(r)$ for localized QH and QP wavepackets at $\nu=2/5$ in different angular-momentum states about the origin. Dashed lines indicate the background filling $\nu=2/5$ and the expected integrated charges $\mp 1/5$ in units of the electron charge $-e$.}
  \label{fig:qh-qp-2-5-shapes}
\end{figure}

Figure~\ref{fig:schematic-qh-cf-occupations} shows the QH and QP density profiles at $\nu=1/3$ for $k=0,1,2$ wavepackets localized at the origin, together with the corresponding CF occupations. Figures~\ref{fig:qh-qp-1-3-shapes} and \ref{fig:qh-qp-2-5-shapes} compare the localized QH and QP density profiles and integrated charge distributions for $k=0,1,2$ at $\nu=1/3$ and $\nu=2/5$. In all cases, the integrated charge converges to $\pm e/(2pn+1)$ at large $r$, confirming the expected fractional charge.

\section{QP shape dependence of braid statistics: Conventional model}
\label{sec:shape-dependence-braid-statistics}
In this section, we show that the Berry phase associated with a closed QP loop is sensitive to the QP shape. 

\subsection{Setup}
We begin with a single localized QH or QP, specified by the angular-momentum label $k$, taken adiabatically around the origin along the circular path
\begin{equation}\label{eq:braid-loop-path}
  \omega(t)=\omega \exp(\imath \phi(t)),\qquad \phi(0)=0,\phi(T)=2\pi.
\end{equation}

We compute the associated Berry phase $\Theta(\omega)$  from the wave functions in Eqs.~\eqref{eq:localized-qh-wavepacket} and \eqref{eq:localized-qp-wavepacket}. As shown in Sec.~\ref{subsec:single-qh-qp-berry-phase}, this phase is independent of $k$ and is given by:
\begin{equation}
  \frac{\Theta(\omega)}{2\pi}
    =
    \pm \frac{\abs{\omega}^{2}}{2\ellstar{}^{2}}.
\end{equation}
This is equal to the AB phase acquired by a charge $-e$ CF moving in an effective field $B^*=B/(2pn+1)$ or the phase acquired by a point particle of charge $e^{\star}=\pm e/(2pn+1)$ in the external field $B$.

We then introduce a second localized QH or QP inside the loop. More generally, we consider the Berry phase $\Theta(\omega_{1},\omega_{2})$  acquired when the two-QH / two-QP configuration is rotated about the origin,
\begin{equation}
  \begin{aligned}
    \omega_{1}(t)&=\omega_{1}\exp(\imath \phi(t)),
    &
    \omega_{2}(t)&=\omega_{2}\exp(\imath \phi(t)),\\
    \abs{\omega_{1}}&<\abs{\omega_{2}}.
  \end{aligned}
\end{equation}
The standard geometry, in which one QH or QP is fixed at the origin while the other is taken around it, is recovered by setting $\omega_{1}=0$.

After subtracting the Berry phases of the two QHs, we extract the  remainder in the limit of large separation,

\begin{equation}\label{eq:braid-statistics-large-separations}
    \Delta\Theta = \lim_{\abs{\omega_{1}-\omega_{2}}\to\infty}\left[\Theta(\omega_{1},\omega_{2})-\Theta(\omega_{1})-\Theta(\omega_{2})\right].
\end{equation}
We shall see whether this yields the expected braid statistics.

Let us first keep one QH / QP fixed at $\omega_{1}=0$ while the second QH / QP at $\omega_{2}$ orbits around it. In this standard geometry, we find from explicit calculation that: 
\begin{equation}\label{eq:theta-standard-geometry}
  \frac{\Delta\Theta^{\QHsym}}{2\pi}=\alpha(1+2k_{2}),
  \qquad
  \frac{\Delta\Theta^{\QPsym}}{2\pi}=-\alpha(1+2k_{2}),
\end{equation}
where $\alpha=2p/(2pn+1)$. Surprisingly, the braid statistics thus appears to depend upon the shape of the orbiting QH / QP. This expression is obtained before applying the density-shift correction of Refs.~\cite{Jeon03,Jeon04}. But as we see later, applying the density shift correction does not cure the shape dependence. Thus the braid statistics extracted from these wave functions depends on the shape of the outer orbiting QH or QP.

One might wonder why we consider the general geometry, in which both QHs / QPs rotate, rather than the standard geometry in which one QH / QP is held fixed at the origin while the other executes a closed loop around it.  The reason is that the general geometry sheds light on a feature of Eq.~\eqref{eq:theta-standard-geometry}, where the statistics depends on the shape of the orbiting QH / QP but not on that of the fixed one. This is puzzling because in a frame attached to the moving QH / QP, it is the other QH / QP that executes the loop. Therefore, the two shapes specified by $k_{1}$ and $k_{2}$ ought to enter on an equal footing. The general geometry shows that they indeed do: as seen in Eqs.~\eqref{eq:theta-two-qhs-k} and \eqref{eq:theta-two-qps-k}, the statistics is symmetric in $k_{1}$ and $k_{2}$. For two QHs / QPs placed at $\mp\omega/2$ and rotated in this locked configuration it depends only on the sum $k_{1}+k_{2}$. The standard geometry is a special configuration for which the shape of the central QH / QP drops out.

It is rather surprising that the braid statistics depends on $k$ of the QPs. It is possible to extend the argument given in Ref.~\cite{Nardin23A} to show, under mild assumptions, that the braid statistics ought to be independent of $k$. What could be going wrong? Going back to the discussion around Eq.~\eqref{eq:braid_stat}, a crucial assumption is that the insertion of a new QP must not change the wave function or the location of the orbiting QP. 

We are therefore led to examine the density profiles of the one-QH/QP wavepackets and the two-QH/two-QP wave functions. We find that, except when both QHs or QPs are localized in $k=0$ states, these two-QH/two-QP wave functions acquire a correlated shape deformation. This deformation becomes locally invisible when the two QHs / QPs are far-separated (the deformation in the density decays as $\sim 1/d$, where $d$ is the distance between the two QPs / QHs) and therefore cannot be detected directly by an ideal local density probe in that limit. Nevertheless, it makes a finite contribution to the Berry phase.

\subsection{Berry phase for a single QH or QP}
\label{subsec:single-qh-qp-berry-phase}
We now evaluate the Berry phase acquired by a single QH or QP when it is taken around a circular loop about the origin, using the wave functions in Eqs.~\eqref{eq:localized-qh-wavepacket} and \eqref{eq:localized-qp-wavepacket}.

The key simplification, first noted in Ref.~\cite{Pu22}, is that in a thermodynamic limit the overlap between two different wave functions corresponding to a single QH or QP localized at two different positions at filling $\nu=n/(2pn+1)$ reduces to the overlap between the two corresponding wave functions of a single IQH hole or particle localized at the same two positions at filling $\nu^{\star}=n$ in the effective field $B^{\star}=B/(2pn+1)$. To see this, consider the QH case. Using Eq.~\eqref{eq:localized-orbital-expansion}, the localized QH state can be expanded as

  \begin{equation}
    \Psi^{\QH{k}, \omega}_{\frac{n}{2pn+1}}
    =
    \sum_{m'=-(n-1)}^{\infty}
    \eta_{k, m'+(n-1)-k}(\omega)\,\Psi^{\QHsym, m'}_{\frac{n}{2pn+1}}.
  \end{equation}

Here, $\Psi^{\QHsym, m'}_{\frac{n}{2pn+1}}$ denotes the CF state corresponding to a missing CF from the $m'$ angular-momentum orbital in the $(n-1)^{\rm th}$ $\Lambda$L,

  \begin{equation}
    \Psi^{{\QHsym}, m'}_{\frac{n}{2pn+1}}=\LLL \Phi_{n}^{\QHsym, m'}\Phi_{1}^{2p},
  \end{equation}

where $\Phi_{n}^{\QHsym, m'}$ is the normalized IQH wave function with an electron missing from the $m'$ angular-momentum orbital in the $(n-1)^{\rm th}$ LL. The corresponding CF wave function is unnormalized.

The overlap between the two unnormalized localized QH wavepackets centered at $\omega$ and $\omega'$ is therefore

  \begin{equation}
    \begin{split}
      &\overlap{\Psi^{\QH{k}, \omega'}_{\frac{n}{2pn+1}}}{\Psi^{\QH{k}, \omega}_{\frac{n}{2pn+1}}} \\
      &=
      \sum_{m'=-(n-1)}^{\infty}
      \conj{\eta}_{k, m'+(n-1)-k}(\omega')
      \eta_{k, m'+(n-1)-k}(\omega) \\
      & \times \overlap{\Psi^{\QHsym, m'}_{\frac{n}{2pn+1}}}{\Psi^{\QHsym, m'}_{\frac{n}{2pn+1}}}.
    \end{split}
  \end{equation}

Ref.~\cite{Pu22} showed that in the thermodynamic limit $\overlap{\Psi^{\QHsym, m'}_{\frac{n}{2pn+1}}}{\Psi^{\QHsym, m'}_{\frac{n}{2pn+1}}}$ (the normalization constant of $\Psi^{{\QHsym}, m'}_{\frac{n}{2pn+1}}$) is independent of $m'$. Physically, this is a consequence of the translational symmetry of the bulk FQH droplet; see Appendix~\ref{app:jain-qh-qp-wavefunctions} for further details.

This implies the remarkable identity (after some algebra, see Ref.~\cite{Pu22} for more details):
  \begin{equation}\label{eq:psi-phi-overlap-identity}
    \begin{split}
      &\tfrac{\overlap{\Psi^{\QH{k}, \omega'}_{\frac{n}{2pn+1}}}{\Psi^{\QH{k}, \omega}_{\frac{n}{2pn+1}}}}
      {\sqrt{\overlap{\Psi^{\QH{k}, \omega'}_{\frac{n}{2pn+1}}}{\Psi^{\QH{k}, \omega'}_{\frac{n}{2pn+1}}}
      \overlap{\Psi^{\QH{k}, \omega}_{\frac{n}{2pn+1}}}{\Psi^{\QH{k}, \omega}_{\frac{n}{2pn+1}}}}} =
      \overlap{\Phi_{n}^{\QH{k}, \omega'}}{\Phi_{n}^{\QH{k}, \omega}},
    \end{split}
  \end{equation}
the left-hand side involves the (strongly) correlated many-body wave functions, with the vortices attached by $\Phi_{1}^{2p}$ and the LLL projection, yet it equals the overlap of two Slater determinants at $B^{\star}$. The right-hand side, moreover, is effectively a single-particle problem: the two Slater determinants differ only in the orbital of the hole in the $(n-1)^{\rm th}$ LL (for a QP, the orbital of the additional electron in the $n^{\rm th}$ LL), so the many-body overlap reduces to the overlap of two localized single-particle wavepackets in the $m=k-(n-1)$ angular-momentum orbital centered at $\omega$ and $\omega'$ at magnetic field $B^{\star}$ (as first noted in Ref.~\cite{Pu22}). Evaluating this single-particle overlap gives

  \begin{equation}
    \begin{split}
      &\frac{\overlap{\Psi^{\QH{k}, \omega'}_{\frac{n}{2pn+1}}}{\Psi^{\QH{k}, \omega}_{\frac{n}{2pn+1}}}}{\sqrt{\overlap{\Psi^{\QH{k}, \omega'}_{\frac{n}{2pn+1}}}{\Psi^{\QH{k}, \omega'}_{\frac{n}{2pn+1}}}
      \overlap{\Psi^{\QH{k}, \omega}_{\frac{n}{2pn+1}}}{\Psi^{\QH{k}, \omega}_{\frac{n}{2pn+1}}}}} \\
      &=
      L_{k}\left(\frac{\abs{\omega-\omega'}^{2}}{2\ellstar{}^{2}}\right)
      \exp\left(
        \frac{\conj{\omega'}\omega}{2\ellstar{}^{2}}
        -\frac{\abs{\omega}^{2}}{4\ellstar{}^{2}}
        -\frac{\abs{\omega'}^{2}}{4\ellstar{}^{2}}
      \right).
    \end{split}
  \end{equation}

The single-particle identity used to evaluate this overlap is
\begin{equation}\label{eq:ll1-ll2-overlap}
  \begin{split}
    &\sum_{m=0}^{\infty}
    \eta_{k_1,m-k_1}(\omega_1)
    \overline{\eta_{k_2,m-k_2}(\omega_2)}
    \\
    & =
    \frac{1}{2\pi \ellstar{}^{2}}
    \exp\left[
      \frac{\omega_1\overline{\omega}_2}{2\ellstar{}^{2}}
      -\frac{|\omega_1|^2}{4\ellstar{}^{2}}
      -\frac{|\omega_2|^2}{4\ellstar{}^{2}}
    \right]\sqrt{\frac{\min(k_{1}, k_{2})!}{\max(k_{1}, k_{2})!}}\\
    & \times
    \begin{cases}
      \displaystyle
      \left(
        \frac{\overline{\omega}_2-\overline{\omega}_1}{\sqrt{2}\ellstar}
      \right)^{k_1-k_2}
      L_{k_2}^{k_1-k_2}
      \left(
        \frac{|\omega_1-\omega_2|^2}{2\ellstar{}^{2}}
      \right),\;k_1\ge k_2,
      \\[1.3em]
      \displaystyle
      \left(
        \frac{\omega_1-\omega_2}{\sqrt{2}\ellstar}
      \right)^{k_2-k_1}
      L_{k_1}^{k_2-k_1}
      \left(
        \frac{|\omega_1-\omega_2|^2}{2\ellstar{}^{2}}
      \right),
      \;k_2\ge k_1.
    \end{cases}
  \end{split}
\end{equation}
For an infinitesimal step along the circular path, $\omega'=\omega\exp(\imath d\phi)$ with $d\phi\to0$, this overlap becomes

  \begin{equation}
    \begin{split}
      &\frac{\overlap{\Psi^{\QH{k}, \omega'}_{\frac{n}{2pn+1}}}{\Psi^{\QH{k}, \omega}_{\frac{n}{2pn+1}}}}{\sqrt{\overlap{\Psi^{\QH{k}, \omega'}_{\frac{n}{2pn+1}}}{\Psi^{\QH{k}, \omega'}_{\frac{n}{2pn+1}}}
      \overlap{\Psi^{\QH{k}, \omega}_{\frac{n}{2pn+1}}}{\Psi^{\QH{k}, \omega}_{\frac{n}{2pn+1}}}}} \\
      &\to
      \exp\left(-\imath \frac{\abs{\omega}^{2}d\phi}{2\ellstar{}^{2}}\right).
    \end{split}
  \end{equation}

Therefore, the Berry phase for a QH is

\begin{equation}\label{eq:single-qh-berry-phase}
    \frac{\Theta}{2\pi} = \frac{\abs{\omega}^{2}}{2\ellstar{}^{2}} = \frac{B\pi \abs{\omega}^{2}}{(2pn+1)\phi_{0}}.
\end{equation}
For a QP, the analogous IQH state contains an added electron in the translated orbital $\psi_{n,k-n}^{\omega}$. The same overlap argument gives the opposite sign:

\begin{equation}\label{eq:single-qp-berry-phase}
    \frac{\Theta}{2\pi} = - \frac{\abs{\omega}^{2}}{2\ellstar{}^{2}} = -\frac{B\pi \abs{\omega}^{2}}{(2pn+1)\phi_{0}}.
\end{equation}
Thus the Berry phase acquired by a QH or QP localized in a definite angular-momentum state specified by $k$ is independent of $k$, and equals the Aharonov--Bohm phase of a point particle of charge $e^{\star}$. This need not remain true when the localized QH or QP is a superposition of angular momentum states about the point of localization, as can occur for a non-centrosymmetric trapping potential or in the presence of a long-range potential produced by either an impurity or another QH / QP in the FQH liquid. We return to this in Sec.~\ref{subsec:asymmetric-qp}.

We have shown that a single QH or QP in a definite angular-momentum state about its localization center acquires the same Berry phase as a point particle of the corresponding charge. We now ask how this result is modified when a second QH or QP lies inside the loop.

\subsection{Berry phase for simultaneous rotation of two QHs or two QPs}

We now use the two-QH and two-QP wave functions in Eqs.~\eqref{eq:localized-two-qh-wavepacket} and \eqref{eq:localized-two-qp-wavepacket} to compute the Berry phase $\Theta$  acquired when the localized QHs or QPs at $\omega_{1}$ and $\omega_{2}$ are {\it simultaneously} rotated about the origin along circular paths. We need $\Theta$, the Berry phase acquired by the microscopic wave function for two localized QHs / QPs on top of an FQH state in the large-separation limit, which requires the asymptotic overlaps of the corresponding microscopic two-QH and two-QP wave functions. In Appendix~\ref{app:jain-qh-qp-wavefunctions}, we show the remarkable result that the overlaps between two many-body wave functions corresponding to two QHs / QPs localized at two different pairs of positions are equal to the corresponding overlaps in an effective two-particle problem, where the two particles are anyons which see a fractional number of vortices at each other's location; this effective model is valid only for computing the overlaps needed to extract the braid statistics $\Delta\Theta$ defined in Eq.~\eqref{eq:braid-statistics-large-separations}, and need not reproduce other observables, such as the density. We construct this effective model below, as we believe it is of independent interest for understanding two-QH and two-QP wave functions more broadly; readers interested only in the result may skip ahead to Eqs.~\eqref{eq:theta-two-qhs-k} and \eqref{eq:theta-two-qps-k}. We then examine the density profiles of the two-QH and two-QP wave functions directly, to understand physically why the braid statistics depends on shape.
\subsubsection*{Effective model for two-QH and two-QP state overlaps}

Let $\Psi_{\omega_{1},\omega_{2}}$ denote the microscopic two-QH or two-QP wave function with localization centers $\omega_{1}$ and $\omega_{2}$ [Eqs.~\eqref{eq:localized-two-qh-wavepacket} and \eqref{eq:localized-two-qp-wavepacket}]. Along the braid loop $\omega_{a}(t)=\omega_{a}e^{\imath\phi(t)}$, the Berry phase is
  \begin{equation}\label{eq:berry-phase-microscopic-definition}
    \Theta
    =
    -\imath\oint dt\,
    \frac{
      \bra{\Psi_{\omega_{1}(t),\omega_{2}(t)}}
      \partial_{t}
      \ket{\Psi_{\omega_{1}(t),\omega_{2}(t)}}
    }{
      \overlap{\Psi_{\omega_{1}(t),\omega_{2}(t)}}
      {\Psi_{\omega_{1}(t),\omega_{2}(t)}}
    }.
  \end{equation}
  Equivalently, denoting by $\mathcal{O}(d\phi)=\overlap{\Psi_{\omega_{1},\omega_{2}}}{\Psi_{\omega_{1}e^{\imath d\phi},\omega_{2}e^{\imath d\phi}}}$ the overlap between the microscopic wave functions describing two QHs / QPs localized at $(\omega_{1},\omega_{2})$ and at $(\omega_{1}e^{\imath d\phi},\omega_{2}e^{\imath d\phi})$,
  \begin{equation}\label{eq:berry-phase-overlap-definition}
    \frac{\Theta}{2\pi}
    =
    \lim_{d\phi\to0}\frac{1}{\imath d\phi}
    \left[\frac{\mathcal{O}(d\phi)}{\mathcal{O}(0)}-1\right].
  \end{equation}
Evaluating $\Theta$ therefore requires the overlaps between microscopic two-QH / two-QP wave functions localized at infinitesimally rotated pairs of positions. We now show that these overlaps can be obtained from an effective model of two anyons.

Consider first a general two-QH wave function at $\nu=n/(2pn+1)$, which we call $\Psi^{f}_{\frac{n}{2pn+1}}$. We expand it in the basis wave functions $\Psi_{\frac{n}{2pn+1}}^{\QHsym, m_{1}; \QHsym, m_{2}}$, where the two QHs correspond to missing CFs in the $m=m_{1}$ and $m=m_{2}$ orbitals of the $(n-1)^{\mathrm{th}}$ $\Lambda$L:

  \begin{equation}
    \Psi^{f}_{\frac{n}{2pn+1}}
    =
    \sum_{m_{1},m_{2}=-(n-1)}^{\infty}
    f_{m_{1},m_{2}}\Psi_{\frac{n}{2pn+1}}^{\QHsym, m_{1}; \QHsym, m_{2}}.
  \end{equation}

The unnormalized microscopic basis wave function is

  \begin{equation}\label{eq:cf-two-qhs-in-orbitals}
    \begin{split}
      &\Psi_{\frac{n}{2pn+1}}^{\QHsym, m_{1}; \QHsym, m_{2}} = \LLL \Phi_{n}^{\QHsym, m_{1}; \QHsym, m_{2}} \Phi_{1}^{2p}.
    \end{split}
\end{equation}
Here, $\Phi_{n}^{\QHsym, m_{1}; \QHsym, m_{2}}$ is the normalized IQH state with $n$ filled LLs and two holes in the $m=m_{1}$ and $m=m_{2}$ orbitals of the $(n-1)^{\mathrm{th}}$ LL. For the localized two-QH wave function in Eq.~\eqref{eq:localized-two-qh-wavepacket}, the coefficients are
\begin{equation}
  \begin{split}
    &f_{m_{1},m_{2}}=\\
    &\eta_{k_{1},m_{1}+(n-1)-k_{1}}\left(\frac{\omega_{1}}{\ellstar},\frac{\conj{\omega}_{1}}{\ellstar}\right)
    \eta_{k_{2},m_{2}+(n-1)-k_{2}}\left(\frac{\omega_{2}}{\ellstar},\frac{\conj{\omega}_{2}}{\ellstar}\right) \\
    &-
    \eta_{k_{1},m_{2}+(n-1)-k_{1}}\left(\frac{\omega_{1}}{\ellstar},\frac{\conj{\omega}_{1}}{\ellstar}\right)
    \eta_{k_{2},m_{1}+(n-1)-k_{2}}\left(\frac{\omega_{2}}{\ellstar},\frac{\conj{\omega}_{2}}{\ellstar}\right).
  \end{split}
\end{equation}
We assume that $f$ has non-negligible support only for $\abs{m_{1}-m_{2}}\gg 1$, corresponding to two well-separated QHs. In order to calculate the braid statistics, we move the two localized QHs infinitesimally along circular loops about the origin and calculate the Berry phase acquired via Eq.~\eqref{eq:berry-phase-overlap-definition}, in the limit of large separation.

In this limit, the overlap of two microscopic two-QH states $\Psi^{f}_{\frac{n}{2pn+1}}$ and $\Psi^{g}_{\frac{n}{2pn+1}}$, specified by coefficients $f_{m_{1},m_{2}}$ and $g_{m_{1},m_{2}}$ as above, is reproduced by the overlap of the corresponding effective two-anyon states $\Psi^{f}_{\eff}$ and $\Psi^{g}_{\eff}$. The effective state is built from the same coefficients that specify the microscopic state:

  \begin{equation}\label{eq:two-qh-effective-state}
    \begin{split}
      &\Psi^{f}_{\eff}(Z_{1},\conj{Z}_{1};Z_{2},\conj{Z}_{2})= \\
      &(\conj{Z}_{1}-\conj{Z}_{2})^{\alpha} \sum_{m_{1}, m_{2}=0}^{\infty}
      f_{m_{1}-(n-1),m_{2}-(n-1)} \\
      &
      \begin{vmatrix}
        \conj{\eta}_{0, m_{1}}(\frac{Z_1}{\ellstar}, \frac{\conj{Z}_{1}}{\ellstar}) &  \conj{\eta}_{0, m_{1}}(\frac{Z_2}{\ellstar}, \frac{\conj{Z}_{2}}{\ellstar}) \\
        \conj{\eta}_{0, m_{2}}(\frac{Z_1}{\ellstar}, \frac{\conj{Z}_{1}}{\ellstar}) &  \conj{\eta}_{0, m_{2}}(\frac{Z_2}{\ellstar}, \frac{\conj{Z}_{2}}{\ellstar}) \\
      \end{vmatrix},
    \end{split}
  \end{equation}

and identically for $\Psi^{g}_{\eff}$ with $f\to g$. Here $Z_{1}, Z_{2}$ are the coordinates of the two effective anyons; they are unrelated to the electron coordinates $z_{i}$ of the microscopic wave functions.

We derive these results in Appendix~\ref{app:jain-qh-qp-wavefunctions} by working in the spherical geometry, where a two-QH / two-QP state can be expressed as a linear superposition of unnormalized total-angular-momentum eigenstates; knowledge of the normalization constants of these eigenstates allows for arbitrary overlap computations. Through comparison with explicit evaluations from the microscopic wave functions, performed using Monte Carlo methods, we show that these normalization constants are very well predicted by a model of two anyons which see an $\alpha$ vortex at each other's location (Figs.~\ref{fig:two-qh-norms} and \ref{fig:two-qp-norms}). Here and below, $\alpha=2p/(2pn+1)$ for two QH / QP states at filling $\nu=n/(2pn+1)$.
The two-QP case is analogous. The basis states $\Psi_{\frac{n}{2pn+1}}^{\QPsym, m_{1}; \QPsym, m_{2}}$ correspond to two additional CFs in the $m=m_{1}$ and $m=m_{2}$ orbitals of the $n^{\mathrm{th}}$ $\Lambda$L. The corresponding unnormalized microscopic basis wave function is

  \begin{equation}\label{eq:cf-two-qp-in-orbitals}
    \begin{split}
      &\Psi_{\frac{n}{2pn+1}}^{\QPsym, m_{1}; \QPsym, m_{2}} = \LLL \Phi_{n}^{\QPsym, m_{1}; \QPsym, m_{2}} \Phi_{1}^{2p}.
    \end{split}
\end{equation}
Here, $\Phi_{n}^{\QPsym, m_{1}; \QPsym, m_{2}}$ is the normalized IQH state with $n$ filled LLs and two additional particles in the $m=m_{1}$ and $m=m_{2}$ orbitals of the $n^{\mathrm{th}}$ LL. A general two-QP state $\Psi^{f}_{\frac{n}{2pn+1}}$ is expanded in coefficients $f_{m_{1},m_{2}}$ with $m_{1},m_{2}\geq -n$. Its unnormalized effective state is

  \begin{equation}\label{eq:two-qp-effective-state}
    \begin{split}
      &\Psi^{f}_{\eff}(Z_{1},\conj{Z}_{1};Z_{2},\conj{Z}_{2})= \\
      &({Z}_{1}-{Z}_{2})^{\alpha} \sum_{m_{1}, m_{2}=0}^{\infty}
      f_{m_{1}-n,m_{2}-n} \\
      &
      \begin{vmatrix}
        {\eta}_{0, m_{1}}(\frac{Z_1}{\ellstar}, \frac{\conj{Z}_{1}}{\ellstar}) &  {\eta}_{0, m_{1}}(\frac{Z_2}{\ellstar}, \frac{\conj{Z}_{2}}{\ellstar}) \\
        {\eta}_{0, m_{2}}(\frac{Z_1}{\ellstar}, \frac{\conj{Z}_{1}}{\ellstar}) &  {\eta}_{0, m_{2}}(\frac{Z_2}{\ellstar}, \frac{\conj{Z}_{2}}{\ellstar}) \\
      \end{vmatrix}.
    \end{split}
\end{equation}

For the microscopic wave function in Eq.~\eqref{eq:localized-two-qh-wavepacket}, the corresponding QH effective state is

  \begin{equation}
    \begin{split}
      &\Psi^{\QHsym_{k_{1}},\omega_{1};\QHsym_{k_{2}},\omega_{2}}_{\frac{n}{2pn+1};\eff}\\
      &=\exp\left[
        -\frac{\abs{Z_{1}}^{2}+\abs{Z_{2}}^{2}+\abs{\omega_{1}}^{2}+\abs{\omega_{2}}^{2}}{4\ellstar{}^{2}}
      \right] \\
      &\times
      \begin{vmatrix}
        \left(\frac{\conj{Z}_{1}-\conj{\omega}_{1}}{\ellstar}\right)^{k_{1}}
        \exp\left(\frac{\omega_{1}\conj{Z}_{1}}{2\ellstar{}^{2}}\right)
        &
        \left(\frac{\conj{Z}_{2}-\conj{\omega}_{1}}{\ellstar}\right)^{k_{1}}
        \exp\left(\frac{\omega_{1}\conj{Z}_{2}}{2\ellstar{}^{2}}\right)
        \\
        \left(\frac{\conj{Z}_{1}-\conj{\omega}_{2}}{\ellstar}\right)^{k_{2}}
        \exp\left(\frac{\omega_{2}\conj{Z}_{1}}{2\ellstar{}^{2}}\right)
        &
        \left(\frac{\conj{Z}_{2}-\conj{\omega}_{2}}{\ellstar}\right)^{k_{2}}
        \exp\left(\frac{\omega_{2}\conj{Z}_{2}}{2\ellstar{}^{2}}\right)
      \end{vmatrix} \\
      & \times (\conj{Z}_{1}-\conj{Z}_{2})^{\alpha}.
    \end{split}
\end{equation}
For the two-QP microscopic wave function in Eq.~\eqref{eq:localized-two-qp-wavepacket}, the corresponding effective state is

  \begin{equation}
    \begin{split}
      &\Psi^{\QPsym_{k_{1}},\omega_{1};\QPsym_{k_{2}},\omega_{2}}_{\frac{n}{2pn+1};\eff}\\
      &=
      \exp\left[
        -\frac{\abs{Z_{1}}^{2}+\abs{Z_{2}}^{2}+\abs{\omega_{1}}^{2}+\abs{\omega_{2}}^{2}}{4\ellstar{}^{2}}
      \right]\\
      &\times
      \begin{vmatrix}
        \left(\frac{Z_{1}-\omega_{1}}{\ellstar}\right)^{k_{1}}
        \exp\left(\frac{\conj{\omega}_{1}Z_{1}}{2\ellstar{}^{2}}\right)
        &
        \left(\frac{Z_{2}-\omega_{1}}{\ellstar}\right)^{k_{1}}
        \exp\left(\frac{\conj{\omega}_{1}Z_{2}}{2\ellstar{}^{2}}\right)
        \\
        \left(\frac{Z_{1}-\omega_{2}}{\ellstar}\right)^{k_{2}}
        \exp\left(\frac{\conj{\omega}_{2}Z_{1}}{2\ellstar{}^{2}}\right)
        &
        \left(\frac{Z_{2}-\omega_{2}}{\ellstar}\right)^{k_{2}}
        \exp\left(\frac{\conj{\omega}_{2}Z_{2}}{2\ellstar{}^{2}}\right)
      \end{vmatrix} \\
      & \times (Z_{1}-Z_{2})^{\alpha}.
    \end{split}
\end{equation}
Here, we have dropped a $k_{1}, k_{2}$-dependent normalization constant.

In these effective models, replacing the Jastrow factors by their absolute values or by their complex conjugates would give the same overlaps. We choose the anti-holomorphic Jastrow factor for QHs and the holomorphic one for QPs for convenience.

To evaluate $\Theta$, we use the effective overlap model above. The details of the analytic calculation for different QH or QP shapes specified by $k_{1}$ and $k_{2}$ are given in Appendix~\ref{app:berry-phase-two-anyon-model}. We summarize the result below. Define
\begin{equation}\label{eq:gamma-one-geometry-factor}
  \chi=\operatorname{Re}\left(\frac{\omega_{\cm}}{\omega_{\rel}}\right),
  \qquad
  \omega_{\cm}=\frac{\omega_{1}+\omega_{2}}{\sqrt{2}},
  \quad
  \omega_{\rel}=\frac{\omega_{1}-\omega_{2}}{\sqrt{2}}.
\end{equation}
For two-QH states, in the limit $|\omega_{\rel}| \to \infty$, we find

\begin{equation}\label{eq:gamma-1-two-qhs-k}
  \begin{split}
    \frac{\Theta}{2\pi}
      =
      \frac{\abs{\omega_{1}}^{2}}{2\ellstar{}^{2}}
      +
      \frac{\abs{\omega_{2}}^{2}}{2\ellstar{}^{2}}
      +
      \alpha\left[
        1+k_{1}+k_{2}
        +(k_{1}-k_{2})\chi
      \right].
  \end{split}
\end{equation}
Similarly, for two QPs,

\begin{equation}\label{eq:gamma-1-two-qps-k}
  \begin{split}
    \frac{\Theta}{2\pi}
      =
      -\frac{\abs{\omega_{1}}^{2}}{2\ellstar{}^{2}}
      -
      \frac{\abs{\omega_{2}}^{2}}{2\ellstar{}^{2}}
      -
      \alpha\left[
        1+k_{1}+k_{2}
        +(k_{1}-k_{2})\chi
      \right].
  \end{split}
\end{equation}
The first two terms are the two single-particle Berry phases, $\Theta_{\rm AB}=\Theta(\omega_{1})+\Theta(\omega_{2})$; the remainder is $\Delta\Theta$ of Eq.~\eqref{eq:braid-statistics-large-separations}.
Thus, the braid statistics extracted from these two-QH wave functions is
\begin{equation}\label{eq:theta-two-qhs-k}
  \begin{split}
    \frac{\Delta\Theta^{\QHsym}}{2\pi}
    =
    \alpha\left[
      1+k_{1}+k_{2}
      +(k_{1}-k_{2})\chi
    \right],
  \end{split}
\end{equation}
and for two QPs it is
\begin{equation}\label{eq:theta-two-qps-k}
  \begin{split}
    \frac{\Delta\Theta^{\QPsym}}{2\pi}
    =
    -\alpha\left[
      1+k_{1}+k_{2}
      +(k_{1}-k_{2})\chi
    \right].
  \end{split}
\end{equation}

The connection to the more traditional setup is obtained by setting $\omega_{1}=0$, so that the QH or QP at $\omega_{2}$ is transported around the one fixed at the origin. In this case the braid statistics is
\begin{equation}\label{eq:theta-natural-setup}
  \Delta\Theta^{\QHsym}=2\pi\alpha(1+2k_{2}),\qquad
  \Delta\Theta^{\QPsym}=-2\pi\alpha(1+2k_{2}).
\end{equation}
Thus the braid statistics depends on the shape of the orbiting QH or QP.

The expected, universal value of the braid statistics is $2\pi\alpha$ for both QHs and QPs [Eq.~\eqref{eq:braid_stat}]. For QPs, part of the deviation of $\Delta\Theta$ from $2\pi\alpha$ is present even for $k_{1}=k_{2}=0$: it equals $-4\pi\alpha$ and is the contribution of the shift in the position of the orbiting QP identified in Refs.~\cite{Jeon03,Jeon04}. The $k$-dependent remainder is the subject of this work.

 Taken literally, this conclusion would imply that braiding one localized QH or QP around another in the bulk can give non-universal answers that depend on the shape selected by the local environment; we return to this point in Sec.~\ref{sec:discussion-outlook}.

\subsubsection*{How can Berry phase depend on the QP shape?}

As discussed at the beginning of this section, from the perspective of CF theory a dependence of the braid statistics on the shape of the orbiting QP is possible only if there is a \emph{correlated} shape deformation or shift in position: the insertion of a new QP must either deform the shape of the orbiting QP or displace it from its original location.

The shape dependence can also be viewed in the light of the relation, derived by Nardin \emph{et al.}~\cite{Nardin23A} and reviewed in Appendix~\ref{app:berry-density}, between the Berry phase and the density. Under the assumption that the many-body wave function is an eigenstate of the combined rotation of the electron coordinates $z_{1},\dots,z_{N}$ and the localization coordinates $\omega_{1},\dots,\omega_{k}$, the Berry phase acquired under the braid loop equals, up to an integer multiple of $2\pi$, the Aharonov--Bohm phase acquired by a classical fluid with the same density; the Berry phase is thus specified entirely by the density distribution on top of the FQH liquid. This concretizes the intuitive picture suggested by the CF theory, namely that the Berry phase acquired by a QH / QP is tied to the density profile; we expect this picture to hold more generally, for example when the wave function is not strictly confined to the LLL, or when the QHs / QPs are not localized in angular-momentum eigenstates. The spin-statistics relation of Ref.~\cite{Nardin23A} rests on one further assumption: that the density of two QHs / QPs is the sum of their individual densities, i.e., that the two are uncorrelated. Under this assumption the expected statistics is always recovered. From both points of view, it follows that the shape-dependent braid statistics found here must be the consequence of a correlated shift or a hitherto unknown shape deformation. We identify this correlated deformation below.

\begin{figure}
  \begin{overpic}[width=0.9\columnwidth,percent]{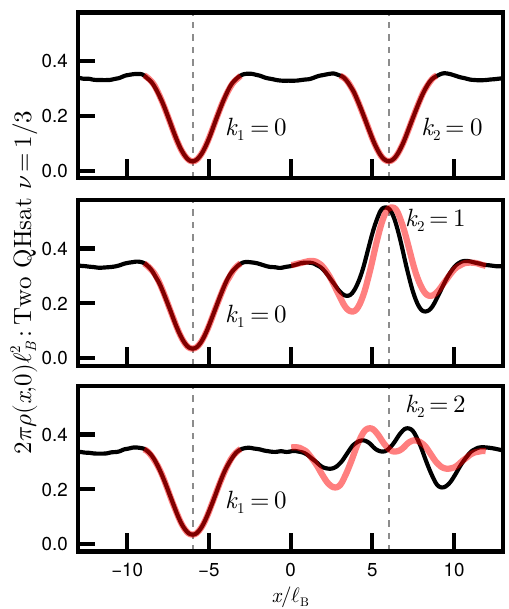}
    \put(-1,20){\color{white}\rule{7.8\unitlength}{68\unitlength}}
    \put(2.9,54){\rotatebox{90}{\makebox(0,0){\large $2\pi\rho(x,0)\lB^{2}$\,:\,Two QHs at $\nu=1/3$}}}
  \end{overpic}
  \caption{Density profiles $2\pi \rho(x,0)\lB^{2}$ for two QHs at filling factor $\nu=1/3$, obtained from the wave function in Eq.~\eqref{eq:localized-two-qh-wavepacket}. The profiles are shown (solid black curves) along the axis connecting the intended localization centers, marked by the vertical dashed lines at $\omega_1=-6\lB$ and $\omega_2=6\lB$. The label $k$ specifies the angular momentum of each localized QH. The semi-transparent red curve is a reflection of the density within the localization window, included as a guide to shape distortions. As can be seen, when the localized QH occupies a $k\neq 0$ angular momentum orbital, in the presence of another QH, it acquires a correlated shape deformation.}
  \label{fig:two-qh-1-3-shapes}
\end{figure}

\begin{figure}
  \includegraphics[width=0.9\columnwidth]{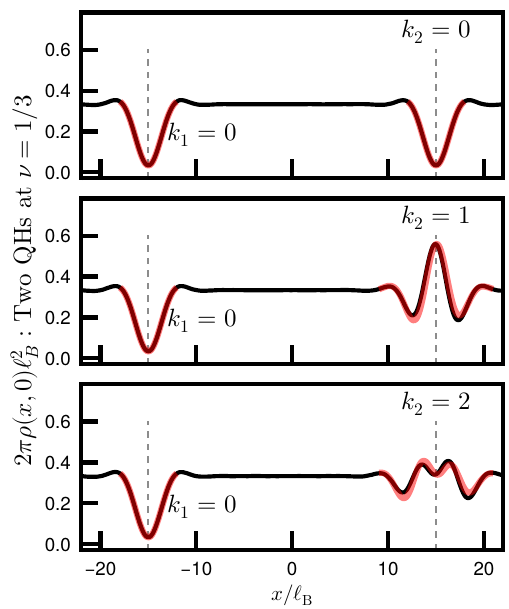}
  \caption{Same as Fig.~\ref{fig:two-qh-1-3-shapes}, but at $2.5$ times the separation: the intended localization centers are $\omega_{1}=-15\lB$ and $\omega_{2}=15\lB$, with $N=144$ electrons. The $k_{2}\neq0$ profiles remain distorted, though much less so than in Fig.~\ref{fig:two-qh-1-3-shapes}. Repeating the calculation at separations $d=12\lB$, $24\lB$, and $30\lB$, we find that the distortion decays as $1/d$ (see text).}
  \label{fig:two-qh-1-3-far}
\end{figure}

\begin{figure}
  \begin{overpic}[width=0.9\columnwidth,percent]{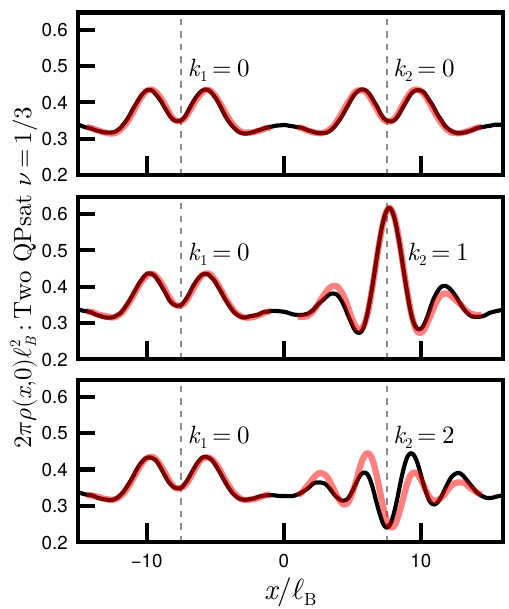}
    \put(-1,20){\color{white}\rule{7.8\unitlength}{68\unitlength}}
    \put(2.9,54){\rotatebox{90}{\makebox(0,0){\large $2\pi\rho(x,0)\lB^{2}$\,:\,Two QPs at $\nu=1/3$}}}
  \end{overpic}
  \caption{Density profiles $2\pi \rho(x,0)\lB^{2}$ for two QPs at filling factor $\nu=1/3$, obtained from the wave function in Eq.~\eqref{eq:localized-two-qp-wavepacket}. The profiles are shown along the axis connecting the intended localization centers, marked by the vertical dashed lines at $\omega_1=-7.5\lB$ and $\omega_2=7.5\lB$. The label $k$ specifies the angular momentum of each localized QP. The semi-transparent red curve is a reflection of the density within the localization window, included as a guide to density shifts and shape distortions.}
  \label{fig:two-qp-1-3-shapes}
\end{figure}

\begin{figure}
  \begin{overpic}[width=0.9\columnwidth,percent]{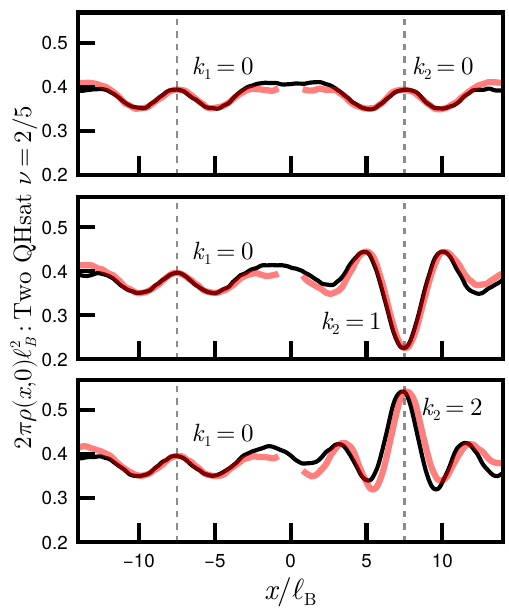}
    \put(-1,20){\color{white}\rule{7.8\unitlength}{68\unitlength}}
    \put(2.9,54){\rotatebox{90}{\makebox(0,0){\large $2\pi\rho(x,0)\lB^{2}$\,:\,Two QHs at $\nu=2/5$}}}
  \end{overpic}
  \caption{Density profiles $2\pi \rho(x,0)\lB^{2}$ for two QHs at filling factor $\nu=2/5$, obtained from the wave function in Eq.~\eqref{eq:localized-two-qh-wavepacket}. The profiles are shown along the axis connecting the intended localization centers, marked by the vertical dashed lines at $\omega_1=-7.5\lB$ and $\omega_2=7.5\lB$. The label $k$ specifies the angular momentum of each localized QH. The semi-transparent red curve is a reflection of the density within the localization window, included as a guide to shape distortions.}
  \label{fig:two-qh-2-5-shapes}
\end{figure}

\begin{figure}
  \begin{overpic}[width=0.9\columnwidth,percent]{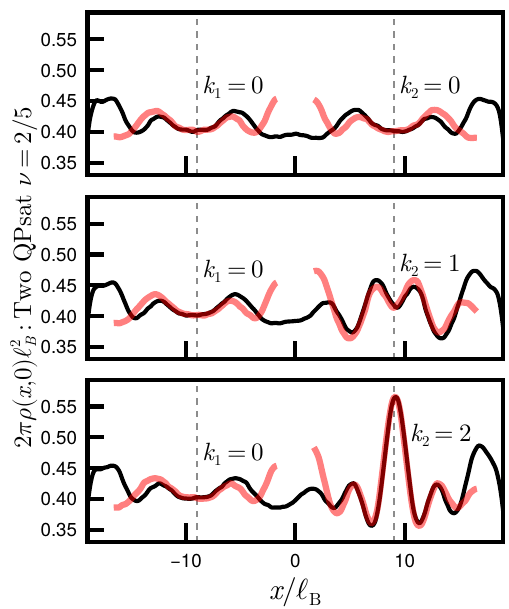}
    \put(-1,20){\color{white}\rule{7.8\unitlength}{68\unitlength}}
    \put(2.9,54){\rotatebox{90}{\makebox(0,0){\large $2\pi\rho(x,0)\lB^{2}$\,:\,Two QPs at $\nu=2/5$}}}
  \end{overpic}
  \caption{Density profiles $2\pi \rho(x,0)\lB^{2}$ for two QPs at filling factor $\nu=2/5$, obtained from the wave function in Eq.~\eqref{eq:localized-two-qp-wavepacket}. The profiles are shown along the axis connecting the intended localization centers, marked by the vertical dashed lines at $\omega_1=-9\lB$ and $\omega_2=9\lB$. The label $k$ specifies the angular momentum of each localized QP. The semi-transparent red curve is a reflection of the density within the localization window, included as a guide to density shifts and shape distortions.}
  \label{fig:two-qp-2-5-shapes}
\end{figure}

The dependence of the two-QH and two-QP density profiles on $k_{1},k_{2}$ is shown in Figs.~\ref{fig:two-qh-1-3-shapes}--\ref{fig:two-qp-2-5-shapes}. The localization positions $\omega_{1}$ and $\omega_{2}$ were chosen to make any displacement or shape distortion as visible as possible while keeping the localized QHs or QPs from overlapping. Even with this choice, the distortions are visually small except for the QH profiles at $\nu=1/3$. In the limit in which the two QHs and two QPs are far from each other, in all cases, this distortion would become negligible and undetectable by local density probes. The approach to this limit is, however, algebraic. Figure~\ref{fig:two-qh-1-3-far} shows the two-QH density profiles at $\nu=1/3$ at separation $d=30\lB$, computed with $N=144$ electrons: the $k_{2}\neq0$ distortions remain visible although they are clearly less prominent than at $d=12\lB$ (Fig.~\ref{fig:two-qh-1-3-shapes}).

Two effects are visible. First, two localized QPs are pushed away from their intended centers, irrespective of $k_{1}$ and $k_{2}$, as first shown in Refs.~\cite{Jeon03,Jeon04}; no analogous displacement is observed for two localized QHs. Second, when either localized QH or QP occupies an angular-momentum orbital other than the $k=0$ orbital about its intended localization position, its shape is distorted by the presence of the other QH or QP. The reflected density profiles in Figs.~\ref{fig:two-qh-1-3-shapes}--\ref{fig:two-qp-2-5-shapes} are meant as a visual guide to this second effect. Comparing with the single-QH/QP density profiles in Figs.~\ref{fig:qh-qp-1-3-shapes} and \ref{fig:qh-qp-2-5-shapes} makes clear that the two-QH/two-QP density is not simply the sum of two independently localized wavepacket densities.

The shape dependence of the braid statistics in Eqs.~\eqref{eq:theta-two-qhs-k} and \eqref{eq:theta-two-qps-k} is therefore a consequence of these correlated density deformations. This deformation cannot be cured by the distance correction of Refs.~\cite{Jeon03,Jeon04}. Shifting $\omega_{1}$ and $\omega_{2}$ can correct a displacement of the wavepacket center, but it cannot turn a distorted density profile into the sum of two independent wavepacket densities.

Even though the correlated shape deformation becomes locally invisible as the separation between the two localized QHs or QPs becomes large, it makes a finite contribution to the Berry phase and produces the shape-dependent braid statistics above.

This is not in conflict with the density--Berry-phase relation reviewed in Appendix~\ref{app:berry-density}, which gives the phase not as a \emph{local} functional of the density but as its second moment about the center of the braid loop, $\Theta/2\pi=-\int d^{2}r\,\rho(\bm{r})\left(r^{2}/2\lB^{2}-1\right)$, up to an integer. Two states with the same number of electrons and the same localized charge have a density difference that integrates to zero, so the second moment reduces to the dipole moment of the deformation weighted by the separation, $[\Theta^{(1)}-\Theta^{(2)}]/2\pi\simeq-(d/\lB^{2})\int d^{2}r\,\delta\rho\,u_{\parallel}$, with $u_{\parallel}$ measured along the line joining the two QHs / QPs. A deformation of amplitude $\propto1/d$ has a dipole moment $\propto1/d$, and therefore makes a finite, $d$-independent contribution to the phase even as it becomes locally invisible.

To summarize, a single localized QH or QP in a definite angular-momentum orbital about its localization center has the Berry phase of a point particle with charge $e^{\star}=\pm e/(2pn+1)$ at $\nu=n/(2pn+1)$. By contrast, the two-QH and two-QP wave functions of Eqs.~\eqref{eq:localized-two-qh-wavepacket} and \eqref{eq:localized-two-qp-wavepacket}, obtained by composite-fermionizing the corresponding noninteracting IQH two-hole or two-electron states, do not give the sum of the two individual QH or QP Berry phases plus a universal statistical contribution: the statistical contribution depends on the shapes of the QHs or QPs, owing to correlated deformations of the two localized density profiles.

\subsection{Non-centrosymmetric traps and unscreened charges}
\label{subsec:asymmetric-qp}
Until this point in the paper we have focused primarily on the special cases in which a localized QH / QP occupies a single angular-momentum state $k$ about its localization center (recall that different angular-momentum orbitals give wavepackets of different shapes). In an experimental system this need not be the case. To understand the effect of the QH / QP being localized in a superposition of angular-momentum states, we consider two distinct sources of such a superposition. The first is a trap that is not centrosymmetric; for concreteness, we take an STM tip carrying an in-plane dipole moment. The second is the long-range potential of a charged impurity, or of another QH / QP in the FQH liquid. As above, it suffices to treat the corresponding single-particle problem of one additional electron at the effective field $B^{\star}$ and filling $\nu^{\star}=n$ to understand the qualitative aspects of the physics at hand.

Consider first the STM tip, and place it at the origin. We assume that the STM tip is sufficiently far so as to stabilize the $k=0$ wavepacket (see Fig.~\ref{fig:coulomb-impurity}) when the dipole moment is $0$; we treat the dipole moment perturbatively and assume that it lies in the plane and points out along $\bm{\hat n}=(\cos\vartheta,\sin\vartheta)$. In this scenario, the dipole contribution reduces to:
  \begin{equation}\label{eq:tip-dipole-potential}
      V_{\bm{p}}(\bm{r})
      =
      \frac{1}{4\pi\epsilon}
      \frac{\bm{p}\cdot\bm{r}}{\left(r^{2}+h^{2}\right)^{3/2}}
      \simeq
      g\,\bm{\hat n}\cdot\bm{r},
      \qquad
      g=\frac{p}{4\pi\epsilon h^{3}} ,
\end{equation}

We can express the electron coordinate operator $\bm{r}$ in terms of the inter-LL and intra-LL ladder operators~\cite{Girvin84,Jain07}; the action of the former is trivial when the electron is restricted to the $n^{\rm th}$ LL. The intra-LL operators $b,b^{\dagger}$ obey $[b,b^{\dagger}]=1$ and lower and raise the angular momentum about the trap center, $b^{\dagger}\ket{\psi_{k}}=\sqrt{k+1}\ket{\psi_{k+1}}$. The dipole term can be written as
  \begin{equation}\label{eq:dipole-ladder}
      g\,\bm{\hat n}\cdot\bm{r}
      \;\to\;
      \frac{g\,\ellstar}{\sqrt{2}}
      \left(e^{\imath\vartheta}\,b^{\dagger}+e^{-\imath\vartheta}\,b\right),
  \end{equation}
  which connects only orbitals whose angular momenta differ by one, with $\bra{\psi_{1}}V_{\bm{p}}\ket{\psi_{0}}=g\,\ellstar\,e^{\imath\vartheta}/\sqrt{2}$. Thus, to first order in $g/\Delta$, where $\Delta=V_{n,-n+1}-V_{n, -n}$ is the difference in energies between $k=1$ and $k=0$ angular momentum states with respect to the centrosymmetric part of the potential, the state becomes
  \begin{equation}\label{eq:dipole-perturbed-state}
      \frac{1}{\sqrt{1+s^{2}}}
      \left[\psi_{0}+s\,e^{\imath\vartheta}\,\psi_{1}\right],
      \qquad
      s=-\frac{g\,\ellstar}{\sqrt{2}\,\Delta} .
  \end{equation}
The dipole thus admixes the $k=1$ channel with amplitude $s$ and phase $\vartheta$ (set by the direction of the dipole), displacing the wavepacket by $\sim s\,\ellstar$ along the dipole axis.

This is the solution with the tip placed at the origin. If the tip is now moved to $\omega$, then by the construction of Sec.~\ref{sec:localized-qh-qp-states} (see also Appendix~\ref{app:localized-electron-wavepackets}), the perturbed ground state becomes
  \begin{equation}\label{eq:tip-state-at-omega}
      \Psi_{\omega}
      =
      \frac{1}{\sqrt{1+s^{2}}}
      \left[\psi^{\omega}_{0}+s\,e^{\imath\vartheta}\,\psi^{\omega}_{1}\right],
  \end{equation}
with the same amplitude $s$ and the same phase $e^{\imath\vartheta}$, so long as the dipole keeps pointing along $\bm{\hat n}$. The point to note is that a localizing potential which is not centrosymmetric mixes the angular-momentum states of the QH / QP: the wavepacket is no longer a definite-$k$ state, but a superposition whose internal orientation is set by the tip. We now ask what this does to the Berry phase.

The Berry phase is the phase accumulated by the state as the tip is carried around the loop, that is, the continuum limit of the product of the overlaps between the states at successive points,
  \begin{equation}\label{eq:berry-accumulation}
      \Theta
      =
      -\imath\oint\overlap{\Psi_{\omega}}{d\Psi_{\omega}} .
  \end{equation}
  Evaluating $\overlap{\Psi_{\omega}}{\Psi_{\omega+d\omega}}$ with the identity Eq.~\eqref{eq:ll1-ll2-overlap} and keeping the terms of first order in $d\omega$,
  \begin{equation}\label{eq:tip-berry-connection}
      \begin{split}
        d\Theta
        =\;
        &-\frac{1}{2\ellstar{}^{2}}\,\mathrm{Im}\left[\conj{\omega}\,d\omega\right]
        -\frac{\sqrt{2}\,s}{1+s^{2}}\,\frac{\mathrm{Im}\left[e^{\imath\vartheta}\,d\omega\right]}{\ellstar}\\
        &+\frac{s^{2}}{1+s^{2}}\,d\vartheta .
      \end{split}
  \end{equation}
  The first term is the Aharonov--Bohm phase of the enclosed area. The second is the interference of the $k=0$ and $k=1$ wavepackets $\sim \mathcal{O}(\omega^{1})$. The third is present only if the orientation of the admixture changes along the path; for the tip it is fixed, $d\vartheta=0$, and this term drops out. On the circular path $\omega=|\omega|e^{\imath\phi}$, for which $d\omega=\imath\,\omega\,d\phi$, Eq.~\eqref{eq:tip-berry-connection} gives
  \begin{equation}\label{eq:tip-berry-density}
      \frac{d\Theta}{d\phi}
      =
      -\frac{|\omega|^{2}}{2\ellstar{}^{2}}
      -\frac{\sqrt{2}\,s}{1+s^{2}}\,\frac{|\omega|}{\ellstar}\,\cos(\vartheta+\phi) .
  \end{equation}
  The second term integrates to zero around the loop, leaving
  \begin{equation}\label{eq:tip-fixed-orientation-berry}
      \frac{\Theta}{2\pi}=-\frac{|\omega|^{2}}{2\ellstar{}^{2}} ,
  \end{equation}
unchanged from Eq.~\eqref{eq:single-qp-berry-phase}. The tip therefore modifies the Berry-phase density along the loop, Eq.~\eqref{eq:tip-berry-density}, but not its integral.

Now let the trap be centrosymmetric, but place a charge inside the loop, such as an unscreened impurity. Expanded about $\omega$, its potential is again a uniform in-plane field of the form of Eq.~\eqref{eq:tip-dipole-potential}, now pointing along the line joining $\omega$ to the charge, so the wavepacket is polarized exactly as before, Eq.~\eqref{eq:dipole-perturbed-state}. The difference is that the orientation of the admixture is no longer fixed in the laboratory frame: the charge is seen from $\omega$ along the radial direction, so $\vartheta$ follows the particle around the loop, its orientation at the start of the loop being $\vartheta_{0}$,
  \begin{equation}\label{eq:corotating-orientation}
      \vartheta(\phi)=\vartheta_{0}-\phi .
  \end{equation}
  Two things then change in Eq.~\eqref{eq:tip-berry-connection}. The interference term is no longer oscillatory, since $\cos(\vartheta+\phi)=\cos\vartheta_{0}$ is now constant, and the last term no longer vanishes, since $d\vartheta=-d\phi$. Both survive the loop integral,
  \begin{equation}\label{eq:k01-corotating-berry}
      \frac{\Theta}{2\pi}
      =
      -\frac{|\omega|^{2}}{2\ellstar{}^{2}}
      -\frac{\sqrt{2}\,s}{1+s^{2}}\,\frac{|\omega|}{\ellstar}\,\cos\vartheta_{0}
      -\frac{s^{2}}{1+s^{2}} .
  \end{equation}
Unlike the case considered earlier (where the asymmetry arises from the asymmetry in the localizing trap), this correction grows with the size of the loop. Screening long-range interactions is therefore essential for extracting both the QP/QH charge and the braid statistics, which poses a further experimental challenge.

\section{Modified two-QH / two-QP CF wave functions: Class I}
\label{sec:flux-dressed-cf-wavefunctions}

In the conventional formulation of Secs.~\ref{sec:localized-qh-qp-states} and~\ref{sec:shape-dependence-braid-statistics}, we begin with the noninteracting parent problem at effective field $B^{\star}$: $n$ filled Landau levels plus two electrons (or holes) in localized wavepackets $\psi^{\omega_{1}}_{k_{1}}$ and $\psi^{\omega_{2}}_{k_{2}}$ [Eq.~\eqref{eq:intro-lll-packet}]. Composite-fermionization through vortex attachment on uncorrelated IQH QPs produces correlated FQH QPs, shifting their density peaks and, for $k\neq0$, deforming their profiles (Figs.~\ref{fig:two-anyon-density} and~\ref{fig:two-qh-1-3-shapes}). 

A natural question is whether it is possible to construct uncorrelated FQH QPs within the CF theory. 
We answer in the affirmative, and construct two classes of one-parameter family of two-QP and two-QH wave functions. In one limit, these wave functions reduce to the conventional wave functions in Eqs.~\eqref{eq:localized-two-qh-wavepacket} and \eqref{eq:localized-two-qp-wavepacket}. For a special value of the parameter, these represent uncorrelated FQH QHs or QPs, i.e. the two-QH / two-QP density profiles have no correlated deformation or shifts at large separations. 
The braid statistics extracted from these states is then shape-independent and takes the expected value.

We describe the essential idea before plunging into details. Motivated by the two-anyon framework introduced in Sec.~\ref{sec:first-look}, we introduce a fractional inverse vortex in the QP or QH coordinates which, just as in the two-anyon framework, eliminates the correlated shift and deformation. There are two ways to do this. 

In the Class I or the ``flux-dressed wave functions,'' we introduce the fractional inverse vortex {\it after} composite-fermionization. To accomplish this, we first express the wave function of the two QPs or the two QHs localized at $\omega_{1}$ and $\omega_{2}$ in terms of the center-of-mass and relative localization coordinates, $\omega_{\cm}$ and $\omega_{\rel}$. The expansion coefficients contain the LLL wave functions $\conj{\eta}_{0,m}(\omega_{\rel})$, where $m$ is the relative angular momentum index. We then make the replacement $\conj{\eta}_{0,m}(\omega_{\rel})\rightarrow \conj{\eta}_{0,m+\alpha}(\omega_{\rel})$ remembering that fractional angular momentum in the relative coordinate is equivalent to fractional statistics.  An appropriate choice of $\alpha$ cancels the vortex deformation. The resulting microscopic state is a superposition of a large number of CF basis states.

In the Class~II construction, we introduce a fractional inverse vortex prior to composite-fermionization, which amounts to building complex non-local correlations between the QPs or QHs of the IQH state. This construction contains a single Slater determinant and generalizes to many QPs or QHs. 

In both constructions, the strength of the inverse vortex, namely $\alpha$ may be replaced by a variational parameter $\alpha'$.  Setting $\alpha'=0$ recovers the conventional wave function, whereas the special value $\alpha'=2p/(2pn+1)$ cancels the shift and deformation, leaving uncorrelated densities at large separation and eliminating the shape dependence of $\Delta\Theta$. Table~\ref{tab:constructions} in Sec.~\ref{sec:discussion-outlook} summarizes the three constructions.

\subsection{Two-QP wave functions}
\label{subsec:flux-dressed-two-qp}

We first develop the construction for QPs. In Sec.~\ref{sec:shape-dependence-braid-statistics}, we showed that the many-particle wave function consisting of two QPs can be modeled as an effective two-particle wave function insofar as the overlaps between states with different QP locations are concerned. This effective wave function consists of $\alpha=2p/(2pn+1)$ vortices attached in the relative coordinate of the two particles. 
The density shifts and deformations seen in the full many-body wave function of electrons consisting of two QPs are closely mirrored by those of the corresponding effective two-QP wave function, as shown in Fig.~\ref{fig:two-anyon-density}. We therefore begin with the effective model and show how it may be modified to eliminate the density shift and deformation. We will then come back to the full many-body wave function. 

The simplest case is two localized QPs with $k_{1}=k_{2}=0$. The corresponding effective state factorizes into center-of-mass and relative-coordinate components:

  \begin{equation}\label{eq:two-anyon-conventional-k00}
    \begin{split}
      &\Psi^{\QP{0}, \omega_{1}; \QP{0}, \omega_{2}}_{\frac{n}{2pn+1}, \eff}(Z_{1}, \bar{Z}_{1}, Z_{2}, \bar{Z}_{2})\\
      &=
      \left[\sum_{M=0}^{\infty}
        \conj{\eta}_{0,M}(\omega_{\cm})
      \eta_{0,M}(Z_{\cm}) \right]\\
      &\times
      \left[\sum_{m=1,3,\dots}
        \conj{\eta}_{0,m}(\omega_{\rel})
      \eta_{0,m}(Z_{\rel})Z_{\rel}^{\alpha}\right].
    \end{split}
  \end{equation}

Here
\begin{equation*}
  \omega_{\cm}=\frac{\omega_{1}+\omega_{2}}{\sqrt{2}},
  \qquad
  \omega_{\rel}=\frac{\omega_{1}-\omega_{2}}{\sqrt{2}},
\end{equation*}
and similarly for $Z_{\cm}$ and $Z_{\rel}$. Because the vortex factor depends on $Z_{\rel}$ alone, it acts on the relative coordinate only. The factor $Z_{\rel}^{\alpha}$ attaches $\alpha$ vortices. It is responsible for the statistical phase, but also produces the density shift, and for $k_{1}^{2}+k_{2}^{2}\neq 0$ it produces a shape deformation as well.

We ask whether the two-anyon state can be written so that, apart from the statistical phase, it has the same density structure as two noninteracting electrons in the LLL. If the interaction between the two anyons is screened, then under the projected localizing potential $\LLL[V_{1}(|r-\omega_{1}|)+V_{2}(|r-\omega_{2}|)]\LLL$, the density should match that of two noninteracting electrons localized at $\omega_{1}$ and $\omega_{2}$.

For $k_{1}=k_{2}=0$, the corresponding noninteracting state has the form:

  \begin{equation}\label{eq:two-electron-cm-rel-k00}
    \begin{split}
      &\Psi^{\QP{0}, \omega_{1}; \QP{0}, \omega_{2}}_{\frac{n}{2pn+1}, {\rm eff}}(Z_{1}, \bar{Z}_{1}, Z_{2}, \bar{Z}_{2})\\
      &=
      \left[\sum_{M=0}^{\infty}
        \conj{\eta}_{0,M}(\omega_{\cm})
      \eta_{0,M}(Z_{\cm})\right]\\
      &\times \left[\sum_{m=1,3,\dots}
        \conj{\eta}_{0,m}(\omega_{\rel})
      \eta_{0,m}(Z_{\rel})\right].
    \end{split}
  \end{equation}

If the two-anyon state can be rewritten to match the above wave function (at least in structure), then the density shift should be absent at the level of the effective model.

This can be achieved by noting that
\begin{equation}\label{eq:relative-index-shift}
  \eta_{0,m}(Z_{\rel})Z_{\rel}^{\alpha}
  =
  N_{\alpha,m}\eta_{0,m+\alpha}(Z_{\rel})
\end{equation}
with
\begin{equation}\label{eq:anyon-constant}
  N_{\alpha,m}=\sqrt{2^{\alpha}\frac{\Gamma(m+\alpha+1)}{\Gamma(m+1)}}.
\end{equation}

Therefore, we replace the coefficient $\conj{\eta}_{0,m}(\omega_{\rel})$ in Eq.~\eqref{eq:two-anyon-conventional-k00} by
\begin{equation}\label{eq:qp-relative-coefficient-replacement}
  \conj{\eta}_{0,m}(\omega_{\rel})
  \to
  e^{\imath\alpha\arg[\omega_{\rel}]}
  \frac{\conj{\eta}_{0,m+\alpha}(\omega_{\rel})}{N_{\alpha,m}}.
\end{equation}
With this replacement, the effective two-anyon state becomes

  \begin{equation}
    \begin{split}
      &\Psi^{\QP{0}, \omega_{1}; \QP{0}, \omega_{2}}_{\frac{n}{2pn+1}, \eff}(Z_{1}, \bar{Z}_{1}, Z_{2}, \bar{Z}_{2})\\
      &=
      e^{\imath\alpha\arg[\omega_{\rel}]}
      \left[\sum_{M=0}^{\infty}
        \conj{\eta}_{0,M}(\omega_{\cm})
      \eta_{0,M}(Z_{\cm})\right] \\
      &\left[\sum_{m=1,3,\dots}
        \conj{\eta}_{0,m+\alpha}(\omega_{\rel})
      \eta_{0,m+\alpha}(Z_{\rel})\right].
    \end{split}
  \end{equation}

\begin{figure}
  \includegraphics[width=\columnwidth]{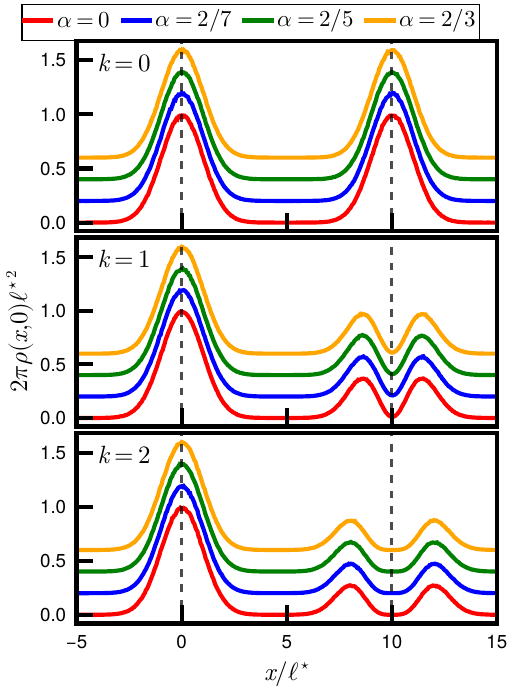}
  \caption{
    Density profile $2\pi\rho(x,0)\ellstar{}^{2}$ along the $x$ axis for the effective two-anyon state corresponding to the two-QP wave function
 $\Psi^{\QP{0}, \omega_{1}; \QP{k}, \omega_{2}}_{\frac{n}{2pn+1}, \fd}$
  in Eq.~\eqref{eq:two-qp-corrected}. Other details are the same as in Fig.~\ref{fig:two-anyon-density}. The curves for $\alpha=2/7, 2/5$ and $\alpha=2/3$ are respectively shifted up by $0.20, 0.40$ and $0.60$. For every $\alpha$ shown, including $\alpha=2/7,2/5,$ and $2/3$, corresponding respectively to $\nu=3/7,2/5,$ and $1/3$, the flux-dressed states show no density shift or deformation, irrespective of $k$. Correspondingly, each state gives a shape-independent statistical contribution to the Berry phase equal to $+2\pi \alpha$.}
  \label{fig:two-anyon-corrected-density}
\end{figure}

This state has the same density profile as two noninteracting electrons in the large-$\omega_{\rel}$ limit (see Fig.~\ref{fig:two-anyon-corrected-density}), but gives braid statistics $\Delta\Theta/(2\pi)=\alpha$ because of the phase $\exp(\imath\alpha\arg[\omega_{\rel}])$. Here $\Delta\Theta$ denotes the statistical contribution after subtracting the single-particle Berry phases. Correspondingly, in units where lengths are measured in $\ellstar$, we find

\begin{equation}\label{eq:gamma-one-flux-dressed-qp-k00}
    \frac{\Theta}{2\pi}
    =
    -\frac{|\omega_{\cm}|^{2}}{2}
    -\frac{|\omega_{\rel}|^{2}}{2}
    +\alpha
\end{equation}
in comparison to Eq.~\eqref{eq:gamma-1-two-qps-k}.

We now translate this effective-state prescription into a prescription for the full many-body CF wave function consisting of two QPs. In the conventional construction, the two-QP state with $k_{1}=k_{2}=0$ is written as
  \begin{equation}\label{eq:two-qp-conventional-k00-expansion}
    \begin{split}
      &\Psi^{\QP{0}, \omega_{1}; \QP{0}, \omega_{2}}_{\frac{n}{2pn+1}} = \\
      &\sum_{m_{1}, m_{2}=0}^{\infty}\conj{\eta}_{0, m_{1}}(\omega_{1})\conj{\eta}_{0, m_{2}}(\omega_{2})\Psi^{\QPsym, m_{1}-n;\, \QPsym, m_{2}-n}_{\frac{n}{2pn+1}}.
    \end{split}
  \end{equation}
To connect this expansion with the center-of-mass and relative-coordinate form of Eq.~\eqref{eq:two-anyon-conventional-k00}, we use the identity
\begin{equation}\label{eq:cm-rel-product-identity}
  \begin{split}
  &\conj{\eta}_{0,m_{1}}(\omega_{1})\,\conj{\eta}_{0,m_{2}}(\omega_{2}) \\
  &=
  \sum_{m=0}^{L} A_{m}(m_{1},m_{2})\,
  \conj{\eta}_{0,L-m}(\omega_{\cm})\,\conj{\eta}_{0,m}(\omega_{\rel}),
  \end{split}
\end{equation}
which follows from expanding $\conj{\omega}_{1}^{m_{1}}\conj{\omega}_{2}^{m_{2}}$ in powers of $\conj{\omega}_{\cm}$ and $\conj{\omega}_{\rel}$; here $L=m_{1}+m_{2}$ and
\begin{equation}\label{eq:cm-relative-binomial-coefficient}
  A_{m}(m_{1},m_{2})
  =
  \left[x^{m}\right]
  (1+x)^{m_{1}}(1-x)^{m_{2}}
  \sqrt{
    \frac{
      m!(L-m)!
    }{
      m_{1}!m_{2}!2^{L}
  }},
\end{equation}
where $[x^m]p(x)$ denotes the coefficient of $x^m$ in the polynomial $p(x)$. Substituting Eq.~\eqref{eq:cm-rel-product-identity} into Eq.~\eqref{eq:two-qp-conventional-k00-expansion}, the conventional wave function becomes
  \begin{equation}\label{eq:two-qp-conventional-cm-rel}
    \begin{split}
      &\Psi^{\QP{0}, \omega_{1}; \QP{0}, \omega_{2}}_{\frac{n}{2pn+1}} = \\
      &\sum_{m_{1}, m_{2}=0}^{\infty}\sum_{m=0}^{L} A_{m}(m_{1},m_{2})\,
      \conj{\eta}_{0,L-m}(\omega_{\cm})\,\conj{\eta}_{0,m}(\omega_{\rel}) \\
      &\times
      \Psi^{\QPsym, m_{1}-n;\, \QPsym, m_{2}-n}_{\frac{n}{2pn+1}}
    \end{split}
  \end{equation}
We now make, term by term, the replacement of Eq.~\eqref{eq:qp-relative-coefficient-replacement} in the relative-coordinate orbital,
\begin{equation}\label{eq:two-qp-fd-replacement-microscopic}
  \conj{\eta}_{0,m}(\omega_{\rel})
  \to
  e^{\imath\alpha\arg[\omega_{\rel}]}\,
  \frac{\conj{\eta}_{0,m+\alpha}(\omega_{\rel})}{N_{\alpha,m}}.
\end{equation}
The result is our proposal for the two-QP wave function at $\nu=n/(2pn+1)$,
  \begin{equation}\label{eq:two-qp-corrected-k1-0-k2-0}
    \begin{split}
      &\Psi^{\QP{0}, \omega_{1};
      \QP{0}, \omega_{2}}_{\frac{n}{2pn+1}, \fd} \\
      &=
      \sum_{m_{1},m_{2}=0}^{\infty}
      e^{\imath\alpha\arg[\omega_{\rel}]} \\
      &\times
      \conj{\mathcal{A}}_{m_{1},m_{2}}(\omega_{1},\omega_{2})
      \Psi^{\QPsym, m_{1}-n;\, \QPsym, m_{2}-n}_{\frac{n}{2pn+1}},
    \end{split}
  \end{equation}
with the coefficients
\begin{equation}\label{eq:two-qp-fd-coefficients-k00}
  \begin{split}
    &\mathcal{A}_{m_{1},m_{2}}(\omega_{1}, \omega_{2}) = \\
    & \sum_{m=0}^{L}\frac{A_{m}(m_{1},m_{2})}{N_{\alpha,m}}{\eta}_{0, L-m}(\omega_{\cm}){\eta}_{0, m+\alpha}(\omega_{\rel}).
  \end{split}
\end{equation}
The subscript ``fd'' stands for flux-dressed: the relative angular momentum of the two QPs has been shifted by $\alpha$, the fractional flux bound to each of them. Unlike the coefficients $\conj{\eta}_{0,m_{1}}(\omega_{1})\conj{\eta}_{0,m_{2}}(\omega_{2})$ of Eq.~\eqref{eq:two-qp-conventional-k00-expansion}, the coefficients $\mathcal{A}_{m_{1},m_{2}}$ are not separable in $m_{1}$ and $m_{2}$: the shift of the relative angular momentum by $\alpha$ ties the two indices together through $m$, and $\mathcal{A}_{m_{1},m_{2}}$ does not factorize into a function of $\omega_{1}$ times a function of $\omega_{2}$; the product form is recovered only for $\alpha=0$. The flux-dressed state is therefore a correlated superposition of the CF Slater determinants $\Psi^{\QPsym, m_{1}-n;\, \QPsym, m_{2}-n}_{\frac{n}{2pn+1}}$.

This form shows that the microscopic wave function belongs to a one-parameter family. We could replace the physical exponent $\alpha$ in the relative-coordinate coefficients and in the overall phase by a free parameter $\alpha'$:
\begin{equation}
  \begin{split}
    N_{\alpha,m}
    &\to N_{\alpha',m}, \\
    \eta_{0,m+\alpha}(\omega_{\rel})
    &\to\eta_{0,m+\alpha'}(\omega_{\rel}), \\
    e^{\imath\alpha\arg[\omega_{\rel}]}
    &\to e^{\imath\alpha'\arg[\omega_{\rel}]}.
  \end{split}
\end{equation}
The choice $\alpha'=0$ recovers the conventional CF wave function, while $\alpha'=\alpha$ gives Eq.~\eqref{eq:two-qp-corrected-k1-0-k2-0}.

For the physical choice $\alpha'=\alpha=2p/(2pn+1)$, the corresponding effective two-anyon wave function has no density shift, as shown in Fig.~\ref{fig:two-anyon-corrected-density}: the density peaks remain at $\omega_{1}$ and $\omega_{2}$. The same choice also gives the expected braid statistics.

The corresponding microscopic density profiles are compared with those obtained from the conventional wave functions in Fig.~\ref{fig:corrected-two-qp-density}. As can be seen, the analogy carries over: curing the density shift in the effective two-anyon wave function also cures it in the microscopic two-QP wave function. In this sense, the proposed construction removes the density shift noted in Refs.~\cite{Jeon03,Jeon04}.

We now extend the same idea to $k_{1}^{2}+k_{2}^{2}\neq0$, where the construction is designed to remove not only the density shift but also the shape deformation. Before doing so, we clarify the phase $\exp(\imath \alpha \arg[\omega_{\rel}])$ in the proposed wave function.

\subsubsection*{Phase convention}
At this point, the phase in Eqs.~\eqref{eq:qp-relative-coefficient-replacement} and \eqref{eq:two-qp-corrected-k1-0-k2-0} may appear to have been put in by hand. For $\alpha'=\alpha$, this phase is precisely what gives the expected braid statistics in the Berry phase. It is therefore natural to ask whether this phase is physical, or merely a convention: can we choose a different phase and obtain a different result?

Suppose that the state from Eq.~\eqref{eq:two-qp-corrected-k1-0-k2-0}, denoted by $\ket{\omega_{1},\omega_{2}}$, is an exact eigenstate of a Hamiltonian $H[\omega_{1},\omega_{2}]$, and that $H[\omega_{1},\omega_{2}]$ is invariant under a simultaneous global rotation of all electron coordinates and of $\omega_{1},\omega_{2}$. If the localizing potentials are moved adiabatically so that the Hamiltonian follows $H[\omega_{1}(t),\omega_{2}(t)]$, with $\omega_{i}(t)=\omega_{i}\exp(\imath\phi(t))$ as in the braid loops used to calculate $\Theta$, the time-evolved state equals $\ket{\omega_{1}(t),\omega_{2}(t)}$, up to the ordinary dynamical phase:
\begin{equation}
  \begin{split}
    \ket{\Phi(t)}
    &=
    \mathcal{T}
    \exp\left(-\frac{\imath}{\hbar}\int_{0}^{t}H(\tau)d\tau\right)
    \ket{\omega_{1},\omega_{2}}\\
    &=
    \exp\left[-\frac{\imath}{\hbar}\int_{0}^{t}E(\tau)d\tau\right]
    \ket{\omega_{1}(t),\omega_{2}(t)}.
  \end{split}
\end{equation}
This can be seen directly from the real-space form of the wave function. With the above phase convention, and suppressing Gaussians, the microscopic two-QP wave function contains terms of the form
\begin{equation}
  \dots + f_{m_1,m_{2}, \dots, m_{N}}(|\omega_{\cm}|,|\omega_{\rel}|)
  z_{1}^{m_{1}}z_{2}^{m_{2}}\cdots z_{N}^{m_{N}}+\dots .
\end{equation}
Consequently, the unitary evolution under the Hamiltonian $H[\omega_{1}(t), \omega_{2}(t)]$ amounts to an inverse global rotation of the electron coordinates, $z_{i}\to z_{i}\exp(-\imath\phi(t))$, together with the ordinary dynamical phase. Hence, the Berry phase along the braid loop can be computed directly from the chosen family of states as
\begin{equation}
  \Theta =
  -\imath\oint dt\,
  \frac{
    \bra{\omega_{1}(t),\omega_{2}(t)}
    \partial_{t}
    \ket{\omega_{1}(t),\omega_{2}(t)}
  }{
    \overlap{\omega_{1}(t),\omega_{2}(t)}
    {\omega_{1}(t),\omega_{2}(t)}
  }.
\end{equation}
If instead we had not included the phase $\exp(\imath\alpha\arg[\omega_{\rel}])$, then the real-space wave function would contain an extra factor $\exp(-\imath\alpha\arg[\omega_{\rel}])$. This factor would not evolve with time, and the Berry phase computed from that gauge would have to be corrected as
\begin{equation}
  \Theta =
  -\imath\oint dt\,
  \frac{
    \bra{\omega_{1}(t),\omega_{2}(t)}
    \partial_{t}
    \ket{\omega_{1}(t),\omega_{2}(t)}
  }{
    \overlap{\omega_{1}(t),\omega_{2}(t)}
    {\omega_{1}(t),\omega_{2}(t)}
  }
  +2\pi\alpha.
\end{equation}
Thus the statistical contribution cannot be gauged away.

We now turn to the case $k_{1}^{2}+k_{2}^{2}\neq0$. The solution follows from the same principle. We match the two-anyon effective state, up to the phase term, with the solution for two noninteracting electrons trapped by localizing potentials so as to occupy the $k_{1}$ angular-momentum orbital about $\omega_{1}$ and the $k_{2}$ angular-momentum orbital about $\omega_{2}$. \textit{Equivalently, we solve the problem of two otherwise noninteracting fermions with $\alpha$ vortices attached and then composite-fermionize the result.} This gives the microscopic two-QP wave function

  \begin{equation}\label{eq:two-qp-corrected}
    \begin{split}
      &\Psi^{\QP{k_{1}}, \omega_{1};
      \QP{k_{2}}, \omega_{2}}_{\frac{n}{2pn+1}, \fd} \\
      &=
      \sum_{m_{1},m_{2}=0}^{\infty}
      e^{\imath\alpha\arg[\omega_{\rel}]} \\
      &\times
      \conj{\mathcal{A}}^{k_{1}, k_{2}}_{m_{1},m_{2}}(\omega_{1},\omega_{2})
      \Psi^{\QPsym, m_{1}-n;\, \QPsym, m_{2}-n}_{\frac{n}{2pn+1}}.
    \end{split}
  \end{equation}

\begin{figure}
  \textbf{(a)}
  \begin{overpic}[width=0.9\columnwidth,percent]{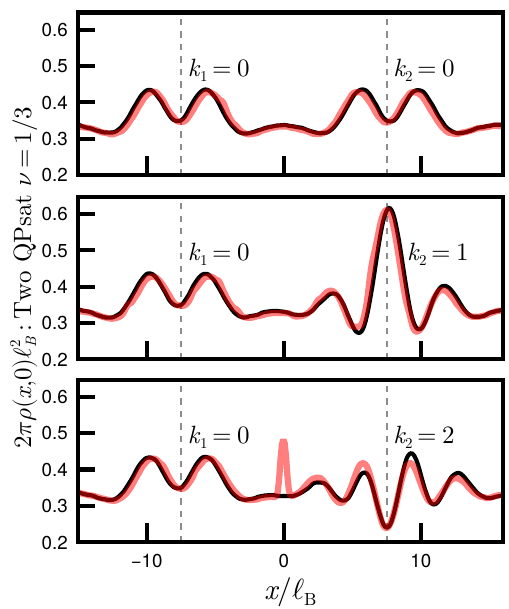}
    \put(-1,20){\color{white}\rule{7.8\unitlength}{68\unitlength}}
    \put(2.9,54){\rotatebox{90}{\makebox(0,0){\large $2\pi\rho(x,0)\lB^{2}$\,:\,Two QPs at $\nu=1/3$}}}
  \end{overpic}\\
  \textbf{(b)}
  \begin{overpic}[width=0.9\columnwidth,percent]{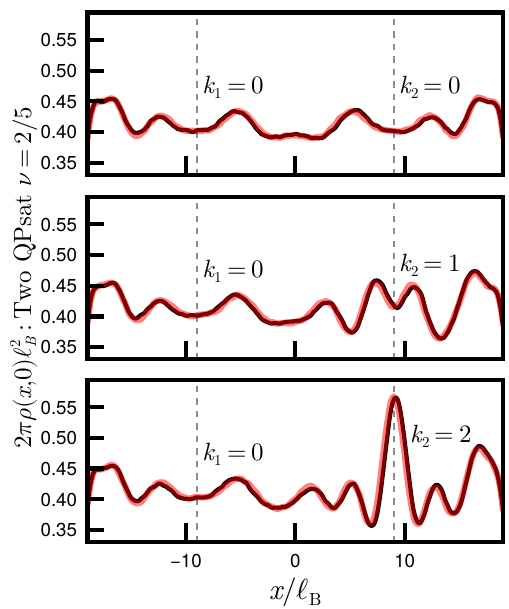}
    \put(-1,20){\color{white}\rule{7.8\unitlength}{68\unitlength}}
    \put(2.9,54){\rotatebox{90}{\makebox(0,0){\large $2\pi\rho(x,0)\lB^{2}$\,:\,Two QPs at $\nu=2/5$}}}
  \end{overpic}
  \caption{Density profiles $2\pi \rho(x,0)\lB^{2}$ for two QPs at (a) $\nu=1/3$ and (b) $\nu=2/5$. Black curves correspond to the conventional two-QP wave functions of Eq.~\eqref{eq:localized-two-qp-wavepacket}; semi-transparent red curves correspond to the flux-dressed two-QP wave functions of Eq.~\eqref{eq:two-qp-corrected}. The shape specified by $k$ and the intended localization positions are shown in the panels, with the latter marked by vertical dashed lines. For the flux-dressed QP wave functions, the displacement and shape distortion are removed: at these separations, the two-QP density is well described as the sum of the corresponding single-QP density profiles.}
  \label{fig:corrected-two-qp-density}
\end{figure}

The coefficient matrix $\mathcal{A}^{k_{1},k_{2}}_{m_{1},m_{2}}$, with $L=m_{1}+m_{2}$ and $K=k_{1}+k_{2}$, is
\begin{equation}\label{eq:two-qp-fd-coefficients}
  \begin{split}
    &\mathcal{A}^{k_{1}, k_{2}}_{m_{1}, m_{2}} =
    \sum_{s=0}^{K}\sum_{m=0}^{L}
    A_{s}(k_{1}, k_{2})A_{m}(m_{1}, m_{2}) \\
    &\times
    \eta_{K-s, L-m+s-K}
    \left(\frac{\omega_{\cm}}{\ellstar}, \frac{\overline{\omega}_{\cm}}{\ellstar}\right) \\
    &\times
    \frac{1}{N_{\alpha, m}}
    \eta_{s, m+\alpha-s}
    \left(\frac{\omega_{\rel}}{\ellstar}, \frac{\overline{\omega}_{\rel}}{\ellstar}\right).
  \end{split}
\end{equation}
The density profile of the corresponding two-anyon effective state is shown in Fig.~\ref{fig:two-anyon-corrected-density}. The microscopic two-QP densities for the flux-dressed construction are compared with the conventional ones in Fig.~\ref{fig:corrected-two-qp-density}, for $\nu=1/3$ and $\nu=2/5$ and for the angular-momentum states shown there. In each case, the proposed wave function removes both the density shift and the shape distortion: the two-QP density is well described as the sum of the corresponding single-QP density profiles.

In the effective model, the two anyons behave as uncorrelated localized wavepackets when the separation between them is large. Correspondingly, the Berry phase takes the shape-independent value

\begin{equation}\label{eq:gamma-one-flux-dressed-qp}
    \frac{\Theta}{2\pi}
    =
    -\frac{|\omega_{\cm}|^2}{2\ellstar{}^{2}}
    -\frac{|\omega_{\rel}|^2}{2\ellstar{}^{2}}
    +\alpha,
\end{equation}
and the extracted braid statistics satisfies $\Delta\Theta_{\fd}^{\QPsym}/(2\pi)=\alpha$, the expected universal value (see Appendix~\ref{app:berry-phase-two-anyon-model} for further details).

We have thus constructed a one-parameter family of wave functions for two-QP states. This family can be viewed as composite-fermionizing the solution to a system of electrons filling the first $n$ LLs, together with two additional electrons localized by the two potentials and carrying $\alpha$ attached vortices. When $\alpha=0$, we recover the conventional two-QP wave functions, in which the two QPs suffer from a correlated density deformation that leads to shape-dependent braid statistics. For two-QP states at filling $\nu=n/(2pn+1)$, the physical value $\alpha=2p/(2pn+1)$ instead gives a state in which the two-QP density becomes uncorrelated at large separation, i.e., it is well described by the sum of the individual QP density profiles. Consequently, the construction recovers shape-independent braid statistics and cures both the density deformation and the density shift simultaneously.

\subsection{Two-QH wave functions}
\label{subsec:flux-dressed-two-qh}
The QH construction follows the same logic. We write the conventional two-QH wave function from Eq.~\eqref{eq:localized-two-qh-wavepacket} in terms of center-of-mass and relative coordinates:

  \begin{equation}
    \begin{split}
      &\Psi^{\QHsym_{k_{1}};\QHsym_{k_{2}}}_{\frac{n}{2pn+1}} = \sum_{K_{\rm cm}, M_{\rm cm}; K_{\rm rel}, M_{\rm rel}}C_{K_{\rm cm}, M_{\rm cm}; K_{\rm rel}, M_{\rm rel}} \\
      & \eta_{K_{\rm cm}, M_{\rm cm}}(\omega_{\cm})
      \eta_{K_{\rm rel}, M_{\rm rel}}(\omega_{\rel})\Psi^{K_{\rm cm}, M_{\rm cm}; K_{\rm rel}, M_{\rm rel}}_{\frac{n}{2pn+1}}.
    \end{split}
    \label{eq:two-qh-cm-rel-expansion}
\end{equation}
Here, $C$ is a coefficient tensor whose precise form is not important for the present argument, and $\Psi^{K_{\rm cm}, M_{\rm cm}; K_{\rm rel}, M_{\rm rel}}_{\frac{n}{2pn+1}}$ is the electronic state accompanying $\eta_{K_{\rm cm}, M_{\rm cm}}(\omega_{\cm})\eta_{K_{\rm rel}, M_{\rm rel}}(\omega_{\rel})$. We ask whether the relative angular momentum in $\omega_{\rel}$ can be shifted so that, at large separation, the two-QH density reduces to the sum of the individual QH densities. This is achieved by the replacement of $\eta_{K_{\rm rel}, M_{\rm rel}} \to \eta_{K_{\rm rel}, M_{\rm rel}-\alpha}N_{-\alpha, M_{\rm rel} + K_{\rm rel}}\exp(\imath \alpha \arg[\omega_{\rm rel}])$ in Eq.~\eqref{eq:two-qh-cm-rel-expansion}:

  \begin{equation}
    \begin{split}
      &\Psi^{\QHsym_{k_{1}};\QHsym_{k_{2}}}_{\frac{n}{2pn+1},\fd} = e^{\imath \alpha \arg[\omega_{\rel}]}\\
      & \times \sum_{K_{\rm cm}, M_{\rm cm}; K_{\rm rel}, M_{\rm rel}}C_{K_{\rm cm}, M_{\rm cm}; K_{\rm rel}, M_{\rm rel}}N_{-\alpha, M_{\rm rel}+K_{\rm rel}} \\
      & \eta_{K_{\rm cm}, M_{\rm cm}}(\omega_{\cm}){\eta_{K_{\rm rel}, M_{\rm rel}-\alpha}}(\omega_{\rel})\Psi^{K_{\rm cm}, M_{\rm cm}; K_{\rm rel}, M_{\rm rel}}_{\frac{n}{2pn+1}},
    \end{split}
\end{equation}
which lowers the relative angular momentum in $\omega_{\rel}$ by $\alpha$ units, analytically continued away from integer $\alpha$. \textit{Equivalently, we solve the problem of two noninteracting holes with flux tubes of strength $-\alpha\phi_{0}$ bound to them, and then composite-fermionize the corresponding hole state.}

Explicitly, our proposal for the two-QH wave function at $\nu=n/(2pn+1)$ is

  \begin{equation}\label{eq:two-qh-corrected}
    \begin{split}
      &\Psi^{\QH{k_{1}}, \omega_{1};
      \QH{k_{2}}, \omega_{2}}_{\frac{n}{2pn+1}, \fd} \\
      &=
      \sum_{m_{1},m_{2}=0}^{\infty}
      e^{\imath\alpha\arg[\omega_{\rel}]} \\
      &\times
      \mathcal{B}^{k_{1}, k_{2}}_{m_{1},m_{2}}(\omega_{1},\omega_{2})
      \Psi^{\QHsym, m_{1}-n+1;\, \QHsym, m_{2}-n+1}_{\frac{n}{2pn+1}}.
    \end{split}
  \end{equation}

Here, $\Psi^{\QHsym, m_{1}; \QHsym, m_{2}}_{\frac{n}{2pn+1}}$ are the states from Eq.~\eqref{eq:cf-two-qhs-in-orbitals} corresponding to missing CFs from the $m_{1}, m_{2}$ angular momentum orbitals in the $(n-1)^{\mathrm{th}}$ $\Lambda$L. The coefficient matrix $\mathcal{B}^{k_{1},k_{2}}_{m_{1},m_{2}}$, with $L=m_{1}+m_{2}$ and $K=k_{1}+k_{2}$, is
\begin{equation}
  \begin{split}
    &\mathcal{B}^{k_{1}, k_{2}}_{m_{1}, m_{2}} =
    \sum_{s=0}^{K}\sum_{m=0}^{L}
    A_{s}(k_{1}, k_{2})A_{m}(m_{1}, m_{2}) \\
    &\times
    \eta_{K-s, L-m+s-K}
    \left(\frac{\omega_{\cm}}{\ellstar}, \frac{\overline{\omega}_{\cm}}{\ellstar}\right) \\
    &\times
    N_{-\alpha, m}\eta_{s, m-\alpha-s}
    \left(\frac{\omega_{\rel}}{\ellstar}, \frac{\overline{\omega}_{\rel}}{\ellstar}\right).
  \end{split}
\end{equation}
The distinction between QPs and QHs appears in the relative-coordinate factor: QP coefficients contain $N_{\alpha,m}^{-1}\eta_{s,m+\alpha-s}$, whereas QH coefficients contain $N_{-\alpha,m}\eta_{s,m-\alpha-s}$. At the wave function level, this is the difference between adding CFs to form QPs and removing CFs to form QHs.

\begin{figure}
  \textbf{(a)}
  \begin{overpic}[width=0.9\columnwidth,percent]{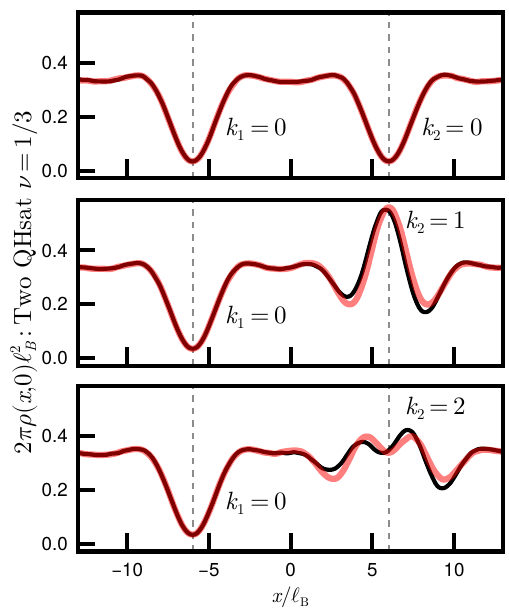}
    \put(-1,20){\color{white}\rule{7.8\unitlength}{68\unitlength}}
    \put(2.9,54){\rotatebox{90}{\makebox(0,0){\large $2\pi\rho(x,0)\lB^{2}$\,:\,Two QHs at $\nu=1/3$}}}
  \end{overpic}\\
  \textbf{(b)}
  \begin{overpic}[width=0.9\columnwidth,percent]{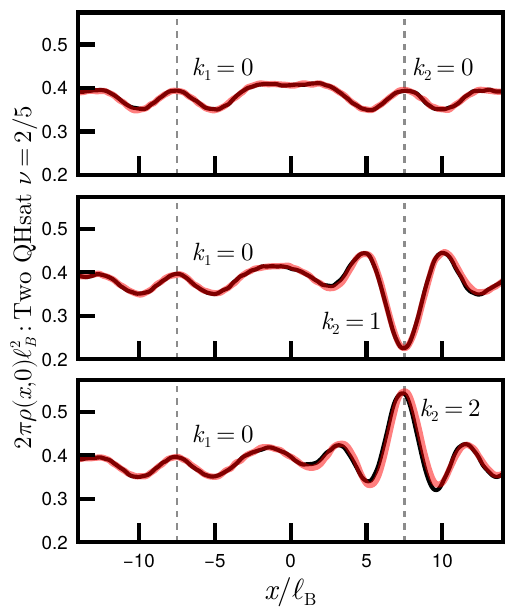}
    \put(-1,20){\color{white}\rule{7.8\unitlength}{68\unitlength}}
    \put(2.9,54){\rotatebox{90}{\makebox(0,0){\large $2\pi\rho(x,0)\lB^{2}$\,:\,Two QHs at $\nu=2/5$}}}
  \end{overpic}
  \caption{Density profiles $2\pi \rho(x,0)\lB^{2}$ for two QHs at (a) $\nu=1/3$ and (b) $\nu=2/5$. Black curves show the conventional two-QH wave functions of Eq.~\eqref{eq:localized-two-qh-wavepacket}; semi-transparent red curves show the flux-dressed two-QH wave functions of Eq.~\eqref{eq:two-qh-corrected}. The shape specified by $k$ and the intended localization positions are indicated in the panels, with the latter marked by vertical dashed lines. For the flux-dressed QH wave functions, the shape distortion of the conventional profiles is removed: at these separations, the two-QH density is well described as the sum of the corresponding single-QH density profiles.}
  \label{fig:corrected-two-qh-density}
\end{figure}

The microscopic two-QH densities for the flux-dressed construction are compared with the conventional ones in Fig.~\ref{fig:corrected-two-qh-density}, for $\nu=1/3$ and $\nu=2/5$ and for the angular-momentum states shown there. In each case, the flux-dressed wave function removes the shape distortion: the two-QH density is well described as the sum of the corresponding single-QH density profiles.

In the effective model, the two anyons behave as uncorrelated localized wavepackets when the separation between them is large. Correspondingly, the braid statistics takes the shape-independent value $\Delta\Theta_{\fd}^{\QHsym}/(2\pi)=\alpha$ (see Appendix~\ref{app:berry-phase-two-anyon-model} for further details).

It is useful to consider a special case to better understand how the flux-dressed wave functions differ from the conventional ones. For $k_{1}=0$ and $k_{2}=k$, Eq.~\eqref{eq:two-qh-corrected} reduces to a nonlocal superposition of conventional two-QH wave functions, up to real normalization constants that depend on $\abs{\omega_{\rel}}$:

  \begin{equation}\label{eq:two-qh-dressed-k}
    \begin{split}
      &\Psi^{\QH{0}, \omega_{1};
      \QH{k}, \omega_{2}}_{\frac{n}{2pn+1}, \fd}
      =
      \sum_{r=0}^{k}
      \sqrt{\frac{k!}{r!}}
      \binom{-\alpha}{k-r}
      \left(\frac{\ellstar}{\omega_{\rel}}\right)^{k-r} \\
      &\hspace{4.0em}\times
      \Psi^{\QH{0}, \omega_{1};
      \QH{r}, \omega_{2}}_{\frac{n}{2pn+1}, \mathrm{conv}}.
    \end{split}
  \end{equation}

Thus, in the QH case, the difference between the flux-dressed wave function and the conventional wave functions is a simple linear transformation. The unusual feature is that this transformation depends on the relative coordinate $\omega_{\rel}$, and is therefore nonlocal in the positions of the two localized QHs. Equation~\eqref{eq:two-qh-dressed-k} also makes clear why $k=0$ is special: for $k=0$, the flux-dressed construction is identical to the conventional one. This is not true for QPs.

Therefore, just as for two-QP states, the flux-dressed construction defines a one-parameter family parameterized by $\alpha$, with $\alpha=0$ recovering the conventional two-QH wave functions and the physical value $\alpha=2p/(2pn+1)$ giving states whose density is uncorrelated at large separation. Although the form of the coefficients differs from the QP case, reflecting the addition of holes rather than the addition of particles, the outcome is the same: the physical value of $\alpha$ removes the shape distortion and recovers shape-independent braid statistics. Unlike the QP case, there is no density shift to cure.

\section{Modified Two-QP / Two-QH wave functions: Class II}
\label{sec:hf-construction}

The flux-dressed wave functions of the preceding section remove the correlated shape deformation and recover shape-independent braid statistics. They are, however, extremely complicated: the coefficients $\conj{\mathcal{A}}^{k_{1},k_{2}}_{m_{1},m_{2}}$ in Eq.~\eqref{eq:two-qp-corrected} are nonseparable in $m_{1}$ and $m_{2}$, so the flux-dressed states are superpositions of a thermodynamically large number of CF Slater determinants. In this section, we ask: what is the minimal ingredient within these wave functions that cures the correlated deformation of the conventional construction? We show that a \emph{single} Slater determinant of suitably modified orbitals suffices, and that this simplification can be generalized to an arbitrary number of QPs or QHs.

These modified orbitals carry an inverse vortex factor: the orbital of the anyon localized at $\omega_{a}$ in the angular-momentum-$k_{a}$ state is $\varphi_{a}(Z)\propto(Z-\omega_{b})^{-\alpha}\,\psi^{\omega_{a}}_{0,k_{a}}(Z)$, where $\psi^{\omega_{a}}_{0,k_{a}}$ is the localized wavepacket of Eq.~\eqref{eq:intro-lll-packet}, $\omega_{b}$ is the localization center of the other anyon, and $\alpha$ is the statistical exponent. The factor $(Z-\omega_{b})^{-\alpha}$ is the reciprocal of the one that attaches $\alpha$ vortices at $\omega_{b}$. Once these orbitals are antisymmetrized and the vortices are attached, the inverse vortex factors cancel, at leading order, the effect of the vortex attachment on the density profile. The braid statistics is not canceled along with it: it is a property of the vortex attachment factor of Eq.~\eqref{eq:vortex-attachment-factor}, which remains in place. What the construction undoes is the correlated shift and deformation of the density, not the statistics itself.

\subsection{The deformation before vortex attachment}

\begin{figure}
  \includegraphics[width=\columnwidth]{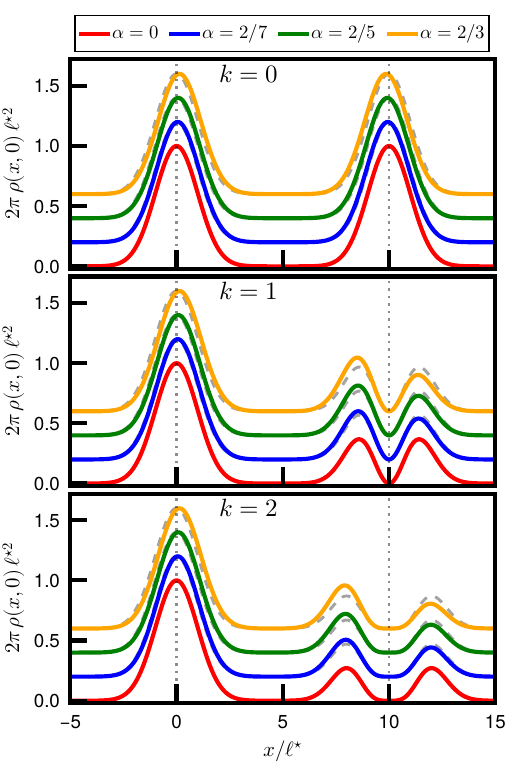}
  \caption{Density profile $2\pi\rho(x,0)\ellstar{}^{2}$ along the $x$ axis for the flux-dressed two-QP effective state of Eq.~\eqref{eq:two-qp-corrected} \emph{before} vortex attachment, i.e., for $\Psi_{\fd}/(Z_{1}-Z_{2})^{\alpha}$. One anyon is localized at the origin in the $k_{1}=0$ angular-momentum state and the second at $\omega=10\ellstar$ in the $k$ state. Curves for $\alpha=2/7, 2/5$ and $\alpha=2/3$ are shifted up by $0.20, 0.40$ and $0.60$; the gray dashed curves show the noninteracting ($\alpha=0$) reference at the same shift. The profiles carry the mirror image of the shift and deformation of Fig.~\ref{fig:two-anyon-density}: the wavepackets are displaced toward each other and, for $k\neq0$, deformed with the opposite sense.}
  \label{fig:two-anyon-prevortex-qp}
\end{figure}

\begin{figure}
  \includegraphics[width=\columnwidth]{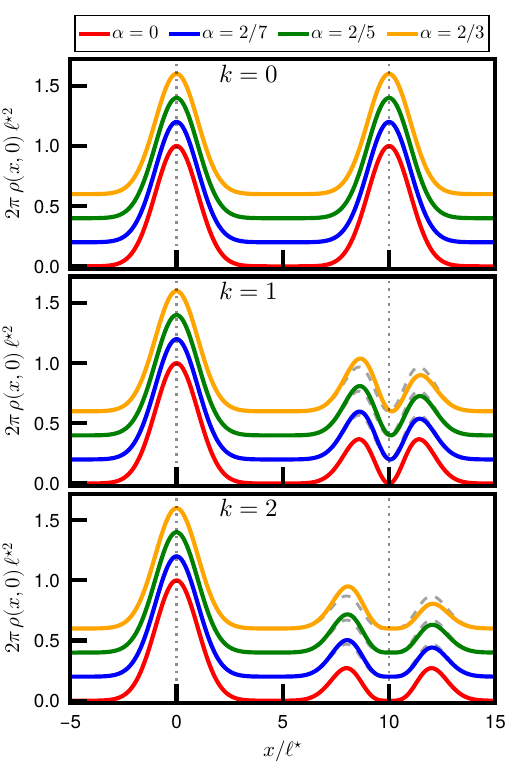}
  \caption{Same as Fig.~\ref{fig:two-anyon-prevortex-qp} for the flux-dressed two-QH effective state of Eq.~\eqref{eq:two-qh-corrected}. For $k=0$ the profile coincides with the noninteracting reference exactly, since the flux-dressed and conventional constructions are identical there; the mirror-image distortion appears only for $k\neq0$ and is weaker than in the QP case.}
  \label{fig:two-anyon-prevortex-qh}
\end{figure}

To identify the minimal ingredient, we return to the effective two-anyon model and examine the flux-dressed states \emph{before} vortex attachment. Consider the effective wave function corresponding to our proposal for two QPs localized at $\omega_{1}$ and $\omega_{2}$ in the $k_{1}$ and $k_{2}$ angular-momentum states, Eq.~\eqref{eq:two-qp-corrected}, and strip off the Jastrow factor, i.e., consider $\Psi_{\fd}/(Z_{1}-Z_{2})^{\alpha}$; for two QHs we analogously divide Eq.~\eqref{eq:two-qh-corrected} by $(\conj{Z}_{1}-\conj{Z}_{2})^{\alpha}$. This is an ordinary two-fermion state. Figures~\ref{fig:two-anyon-prevortex-qp} and \ref{fig:two-anyon-prevortex-qh} show the resulting density profiles. Comparing with Fig.~\ref{fig:two-anyon-density}: before vortex attachment, the flux-dressed states carry the \emph{mirror image} of the shift and shape deformation that vortex attachment produces in the conventional states. The two wavepackets are pulled toward each other, and for $k\neq0$ the lobe facing the other anyon is enhanced; these are precisely the distortions that the factor $(Z_{1}-Z_{2})^{\alpha}$ subsequently undoes.

This observation suggests that the essential content of the flux-dressed construction is not the detailed correlated structure of the coefficients $\conj{\mathcal{A}}^{k_{1},k_{2}}_{m_{1},m_{2}}$ of Eq.~\eqref{eq:two-qp-fd-coefficients} [or $\mathcal{B}^{k_{1},k_{2}}_{m_{1},m_{2}}$ of Eq.~\eqref{eq:two-qh-corrected} for QHs], but rather that, prior to composite-fermionization, the densities of the localized QPs / QHs carry a shift and a deformation which precisely undo those produced by the vortex attachment factor of Eq.~\eqref{eq:vortex-attachment-factor}.

\subsection{Two QPs}
\label{subsec:hf-two-qp}

As in the preceding section, we work in the effective two-anyon model, whose density profiles closely parallel those of the two QPs of the FQH liquid. We first ask whether the two-anyon problem admits a single-Slater-determinant state, albeit with modified orbitals, that after vortex attachment gives a wave function with uncorrelated density profiles. This is indeed the case. The modified orbitals are
  \begin{equation}\label{eq:hf-two-qp-orbital}
    \begin{split}
      \varphi_{a}(Z)
      &\propto
      \left(\frac{Z-\omega_{b}}{\ellstar}\right)^{-\alpha}\psi^{\omega_{a}}_{0,k_{a}}(Z)
      \\
      &\propto
      \sum_{j\geq0}\binom{-\alpha}{j}
      \sqrt{\frac{2^{j}\,(k_{a}+j)!}{k_{a}!}} \\
      &\quad\times
      \left(\frac{\ellstar}{\omega_{a}-\omega_{b}}\right)^{j}
      \psi^{\omega_{a}}_{0,k_{a}+j}(Z),
    \end{split}
  \end{equation}
where $b$ labels the localization center of the \textit{other anyon}. The second line is obtained by expanding $(Z-\omega_{b})^{-\alpha}$ about $\omega_{a}$ with the binomial series and noting that $u_{a}^{j}\,\psi^{\omega_{a}}_{0,k_{a}}\propto\psi^{\omega_{a}}_{0,k_{a}+j}$, with $u_{a}=Z-\omega_{a}$. Thus the inverse vortex factor mixes in higher angular-momentum wavepackets about $\omega_{a}$, with amplitudes that fall off as powers of $\ellstar/d$, where $d=\abs{\omega_{1}-\omega_{2}}$ is the separation; the leading admixture is $-\alpha\sqrt{2(k_{a}+1)}\,\ellstar/(\omega_{a}-\omega_{b})$. Each modified orbital remains a wavepacket localized at $\omega_{a}$ in the $k_{a}$ angular-momentum state. Strictly, $(Z-\omega_{b})^{-\alpha}$ has a branch point at $\omega_{b}$, so the modified orbital does not lie within the LLL, and the series in Eq.~\eqref{eq:hf-two-qp-orbital} is asymptotic rather than convergent, as discussed in Sec.~\ref{sec:first-look}. Following the remark made there, we truncate it at the first $j^{\star}\approx d^{2}/(2\ellstar{}^{2})$ terms (Appendix~\ref{app:dressing-asymptotics}); all numerical results below use this truncation, and we find no difference attributable to it.

We place the two electrons in these orbitals and attach the vortices to obtain (the superscript HF indicates a single Slater determinant of modified orbitals, in analogy with a Hartree--Fock state)
  \begin{equation}\label{eq:hf-two-qp-state}
    \Psi^{\mathrm{HF}}
    =
    e^{\imath\alpha\arg[\omega_{\rel}]}\,
    (Z_{1}-Z_{2})^{\alpha}
    \begin{vmatrix}
      \varphi_{1}(Z_{1}) & \varphi_{1}(Z_{2})\\
      \varphi_{2}(Z_{1}) & \varphi_{2}(Z_{2})
    \end{vmatrix}.
  \end{equation}
  We now show that this state has, asymptotically, the same density as that of two noninteracting electrons localized at $\omega_{1}$ and $\omega_{2}$ in the angular-momentum states $k_{1}$ and $k_{2}$. Consider the determinant term by term. The first term is
  \begin{equation}\label{eq:hf-direct-term}
    \begin{split}
      &(Z_{1}-Z_{2})^{\alpha}\,\varphi_{1}(Z_{1})\varphi_{2}(Z_{2}) \\
      &\quad\propto
      \frac{(Z_{1}-Z_{2})^{\alpha}\;\psi^{\omega_{1}}_{0,k_{1}}(Z_{1})\,\psi^{\omega_{2}}_{0,k_{2}}(Z_{2})}{(Z_{1}-\omega_{2})^{\alpha}\,(Z_{2}-\omega_{1})^{\alpha}}.
    \end{split}
  \end{equation}
  Each orbital in Eq.~\eqref{eq:hf-two-qp-orbital} is a superposition of wavepackets localized about its own center, so this term is appreciable only for $Z_{1}$ near $\omega_{1}$ and $Z_{2}$ near $\omega_{2}$. We may therefore expand in $u_{i}=Z_{i}-\omega_{i}$. To first order in $u_{i}/\omega_{12}$, with $\omega_{12}=\omega_{1}-\omega_{2}$,
  \begin{equation}\label{eq:hf-factor-expansions}
    \begin{split}
      (Z_{1}-Z_{2})^{\alpha}
      &=
      \omega_{12}^{\alpha}\left[1+\alpha\frac{u_{1}-u_{2}}{\omega_{12}}+\dots\right],
      \\
      (Z_{1}-\omega_{2})^{-\alpha}
      &=
      \omega_{12}^{-\alpha}\left[1-\alpha\frac{u_{1}}{\omega_{12}}+\dots\right],
      \\
      (Z_{2}-\omega_{1})^{-\alpha}
      &=
      (-\omega_{12})^{-\alpha}\left[1+\alpha\frac{u_{2}}{\omega_{12}}+\dots\right].
    \end{split}
  \end{equation}
  The first-order terms cancel in the product. Carrying the expansion to second order gives
  \begin{equation}\label{eq:hf-hartree}
    \begin{split}
      &(Z_{1}-Z_{2})^{\alpha}\,
      (Z_{1}-\omega_{2})^{-\alpha}(Z_{2}-\omega_{1})^{-\alpha} \\
      &\quad=
      (-\omega_{12})^{-\alpha}
      \left[1+\alpha\frac{u_{1}u_{2}}{\omega_{12}^{2}}+\dots\right].
    \end{split}
  \end{equation}
The first term of the determinant therefore reduces to $\psi^{\omega_{1}}_{0,k_{1}}(Z_{1})\,\psi^{\omega_{2}}_{0,k_{2}}(Z_{2})$, up to a constant and the two-body correction $\alpha u_{1}u_{2}/\omega_{12}^{2}$. The second term contributes in the same manner as the first with the interchange of $Z_{1}$ and $Z_{2}$ (and an accompanying $-$ sign).

The two-body correction, which cannot be removed by any choice of single-particle orbitals, being a genuine two-body correlation, does not, however, affect the density at leading order. Figure~\ref{fig:hf-det-before-after} shows the density of this state before and after vortex attachment, for $\alpha=2/3$, $k_{1}=0$, and $k_{2}=0,1,2$. Before vortex attachment, the determinant carries the same mirror-image deformation as the pre-vortex flux-dressed states: each wavepacket is weighted by $\abs{Z-\omega_{b}}^{-2\alpha}$, the wavepackets are pulled toward each other, and for $k\neq0$ the lobe facing the other anyon is enhanced (compare Fig.~\ref{fig:two-anyon-prevortex-qp}). After vortex attachment, the density coincides with that of two noninteracting electrons. At $d=10\ellstar$, the density of $\Psi^{\mathrm{HF}}$ agrees with that of the flux-dressed state, and hence with the sum of the two individual wavepacket densities.

The correspondence with the FQH liquid carries over. Composite-fermionizing the determinant of the modified orbitals gives a microscopic two-QP wave function; the explicit form, valid for any number of QPs, is given in Sec.~\ref{sec:hf-many}. Figure~\ref{fig:hf-two-qp-density} shows its density at $\nu=1/3$ and $\nu=2/5$: the correlated displacement and shape deformation of the conventional two-QP wave functions are removed, and the two-QP density is well described by the sum of the corresponding single-QP density profiles.

In Appendix~\ref{app:hf-braid} we evaluate the braid statistics of $\Psi^{\mathrm{HF}}$. We find $\Delta\Theta/(2\pi)=\alpha$ in the large-separation limit, independent of $k_{1}$ and $k_{2}$. Thus, once the shape deformation is removed, the shape dependence of the braid statistics disappears as well and we recover wave functions which give a universal braid statistics contribution to the Berry phase.

\subsection{Two QHs}

For two QHs we ask the same question. The modified orbital now takes a different form:
  \begin{equation}\label{eq:hf-two-qh-orbital}
    \xi_{a}(Z)
    =
    \sum_{j=0}^{k_{a}}\binom{-\alpha}{j}
    \sqrt{\frac{k_{a}!}{(k_{a}-j)!}}
    \left(\frac{\sqrt{2}\,\ellstar}{\omega_{a}-\omega_{b}}\right)^{j}
    \psi^{\omega_{a}}_{0,k_{a}-j}(Z),
  \end{equation}
  where $b$ labels the other QH. The sum has a closed form: it is the inverse vortex factor written with the intra-LL lowering operator $\hat{b}_{a}$ of Eq.~\eqref{eq:dipole-ladder} centered at $\omega_{a}$, which acts as $\hat{b}_{a}\,\psi^{\omega_{a}}_{0,m}=\sqrt{m}\,\psi^{\omega_{a}}_{0,m-1}$,
    \begin{equation}\label{eq:hf-two-qh-orbital-closed}
      \xi_{a}
      =
      \left(1+\frac{\sqrt{2}\,\ellstar\,\hat{b}_{a}}{\omega_{a}-\omega_{b}}\right)^{-\alpha}
      \psi^{\omega_{a}}_{0,k_{a}},
    \end{equation}
  and expanding the bracket by the binomial series returns Eq.~\eqref{eq:hf-two-qh-orbital}. Because $\hat{b}_{a}$ annihilates the $k_{a}=0$ state, the series terminates after at most $k_{a}+1$ terms; unlike the QP case there is no truncation to make. In the LLL, lowering the angular momentum is a derivative rather than a multiplication, which is why the QH orbital has no form as simple as multiplication by $(Z-\omega_{b})^{-\alpha}$. The difference from Eq.~\eqref{eq:hf-two-qp-orbital} originates in the mixing between angular-momentum states: in the two-QP case, the modified orbital mixes in states of higher angular momentum, and in the two-QH case it mixes in states of lower angular momentum. The corresponding effective two-anyon wave function is
  \begin{equation}\label{eq:hf-two-qh-state}
    \Psi^{\mathrm{HF}}_{\mathrm{QH}}
    =
    e^{\imath\alpha\arg[\omega_{\rel}]}\,
    (\conj{Z}_{1}-\conj{Z}_{2})^{\alpha}
    \begin{vmatrix}
      \conj{\xi}_{1}(Z_{1}) & \conj{\xi}_{1}(Z_{2})\\
      \conj{\xi}_{2}(Z_{1}) & \conj{\xi}_{2}(Z_{2})
    \end{vmatrix},
  \end{equation}
with $\omega_{\rel}=(\omega_{1}-\omega_{2})/\sqrt{2}$.

\begin{figure}
  \includegraphics[width=\columnwidth]{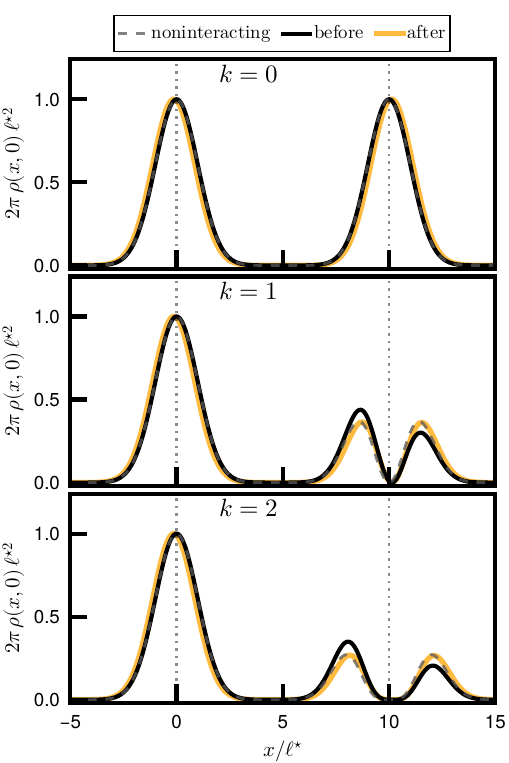}
  \caption{Same as Fig.~\ref{fig:hf-det-before-after} but for the two-QH determinant state of Eq.~\eqref{eq:hf-two-qh-state}: density profiles before (black) and after (orange) vortex attachment for $\alpha=2/3$, with one QH at the origin in the $k_{1}=0$ angular-momentum state and the second at $\omega=10\ellstar$ in the $k$ state. For this geometry the determinant coincides with the flux-dressed two-QH wave function to machine precision, so the curves also show the flux-dressed densities before and after vortex attachment. After vortex attachment, the $k\neq0$ shape distortion is removed; the outward displacement of the wavepackets decays as $1/d$. These are effective two-anyon densities and do not correspond to the microscopic two-QH density profiles.}
  \label{fig:hf-det-before-after-qh}
\end{figure}

This state has the same density as the effective two-anyon state corresponding to the flux-dressed two-QH wave function of Eq.~\eqref{eq:two-qh-corrected}. When either $k_{1}=0$ or $k_{2}=0$, the two coincide {exactly}, consistent with Eq.~\eqref{eq:two-qh-dressed-k}, in which the flux-dressed wave function can be rewritten as a single Slater determinant (due to the unentangled nature of its coefficients in the angular-momentum basis). Figure~\ref{fig:hf-det-before-after-qh} shows the density of the determinant state before and after vortex attachment for $k_{1}=0$. Before vortex attachment, the profiles are those of Fig.~\ref{fig:two-anyon-prevortex-qh}: for $k\neq0$, the lobe facing the other QH is enhanced. After vortex attachment, this distortion is removed, and the profile about each center is symmetric; the wavepackets are displaced outward by an amount that decays as $1/d$. We emphasize that these are densities of the effective two-anyon states: for QHs, the microscopic wave functions do not exhibit any density shift.

Figure~\ref{fig:hf-two-qh-density} shows the densities of the microscopic two-QH wave functions built with the modified hole orbitals (Sec.~\ref{sec:hf-many}), at $\nu=1/3$ and $\nu=2/5$: the shape distortion of the conventional two-QH profiles is removed, and the two-QH density is well described by the sum of the corresponding single-QH density profiles.

The braid statistics follows as for the QPs (Appendix~\ref{app:hf-braid}): in the large-separation limit it is $\alpha$, independent of $k_{1}$ and $k_{2}$. Again the removal of the deformation restores the shape independence.

\subsection{Arbitrary numbers of QPs and QHs, and the CF wave functions}
\label{sec:hf-many}

The determinant form generalizes immediately to any number $M$ of anyons, in contrast to the center-of-mass--relative-coordinate construction of the flux-dressed states, which is specific to $M=2$. The inverse vortex factor becomes a product over all other anyons. For $M$ QPs localized at $\omega_{1},\dots,\omega_{M}$ in the angular-momentum states $k_{1},\dots,k_{M}$, the modified orbital is
  \begin{equation}\label{eq:hf-multi-orbital-qp}
    \varphi_{a}(Z)
    \propto
    \prod_{b\neq a}
    \left(\frac{Z-\omega_{b}}{\ellstar}\right)^{-\alpha}
    \psi^{\omega_{a}}_{0,k_{a}}(Z),
  \end{equation}
  the product generalization of Eq.~\eqref{eq:hf-two-qp-orbital}. For $M$ QHs, the modified orbital is
  \begin{equation}\label{eq:hf-multi-orbital-qh}
    \xi_{a}
    =
    \prod_{b\neq a}
    \left(1+\frac{\sqrt{2}\,\ellstar\,\hat{b}_{a}}{\omega_{a}-\omega_{b}}\right)^{-\alpha}
    \psi^{\omega_{a}}_{0,k_{a}},
  \end{equation}
where $\hat{b}_{a}$ is the intra-LL lowering operator of Eq.~\eqref{eq:dipole-ladder} centered at $\omega_{a}$ (hatted to distinguish it from the anyon label $b$), $\hat{b}_{a}\,\psi^{\omega_{a}}_{0,m}=\sqrt{m}\,\psi^{\omega_{a}}_{0,m-1}$; each factor terminates after at most $k_{a}+1$ terms, as in Eq.~\eqref{eq:hf-two-qh-orbital}.

\begin{figure}
  \textbf{(a)}
  \begin{overpic}[width=0.9\columnwidth,percent]{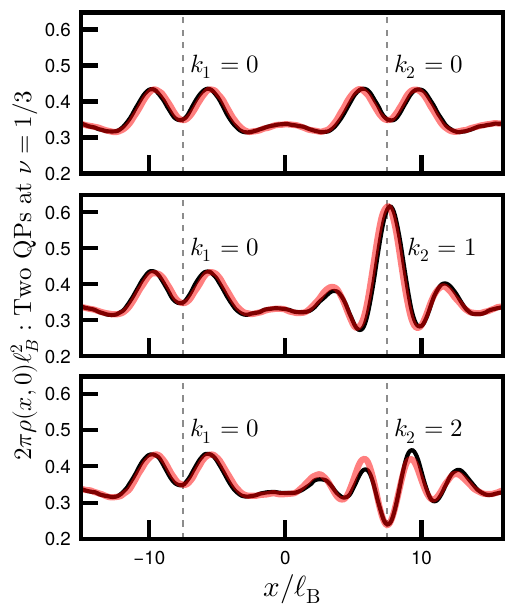}
    \put(-1,20){\color{white}\rule{7.8\unitlength}{68\unitlength}}
    \put(2.9,54){\rotatebox{90}{\makebox(0,0){\large $2\pi\rho(x,0)\lB^{2}$\,:\,Two QPs at $\nu=1/3$}}}
  \end{overpic}\\
  \textbf{(b)}
  \begin{overpic}[width=0.9\columnwidth,percent]{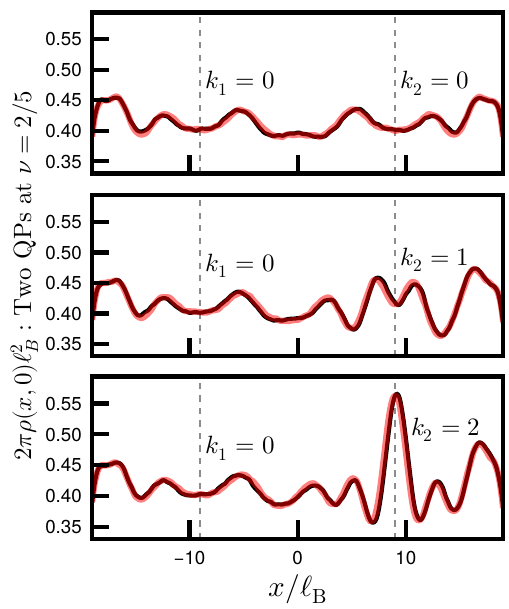}
    \put(-1,20){\color{white}\rule{7.8\unitlength}{68\unitlength}}
    \put(2.9,54){\rotatebox{90}{\makebox(0,0){\large $2\pi\rho(x,0)\lB^{2}$\,:\,Two QPs at $\nu=2/5$}}}
  \end{overpic}
  \caption{Density profiles $2\pi \rho(x,0)\lB^{2}$ for two QPs at (a) $\nu=1/3$ and (b) $\nu=2/5$. Black curves correspond to the conventional two-QP wave functions of Eq.~\eqref{eq:localized-two-qp-wavepacket}; semi-transparent red curves correspond to the HF determinant wave functions of Eq.~\eqref{eq:hf-jain-qp}. The shape specified by $k$ and the intended localization positions are shown in the panels, with the latter marked by vertical dashed lines. As for the flux-dressed wave functions, the displacement and shape distortion are removed: at these separations, the two-QP density is well described as the sum of the corresponding single-QP density profiles.}
  \label{fig:hf-two-qp-density}
\end{figure}

\begin{figure}
  \textbf{(a)}
  \begin{overpic}[width=0.9\columnwidth,percent]{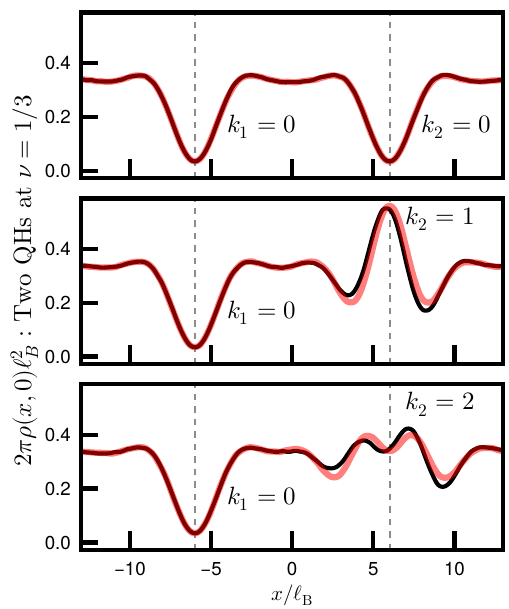}
    \put(-1,20){\color{white}\rule{7.8\unitlength}{68\unitlength}}
    \put(2.9,54){\rotatebox{90}{\makebox(0,0){\large $2\pi\rho(x,0)\lB^{2}$\,:\,Two QHs at $\nu=1/3$}}}
  \end{overpic}\\
  \textbf{(b)}
  \begin{overpic}[width=0.9\columnwidth,percent]{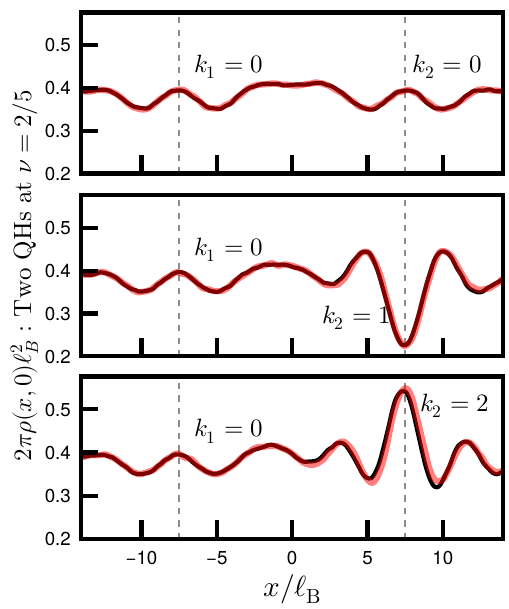}
    \put(-1,20){\color{white}\rule{7.8\unitlength}{68\unitlength}}
    \put(2.9,54){\rotatebox{90}{\makebox(0,0){\large $2\pi\rho(x,0)\lB^{2}$\,:\,Two QHs at $\nu=2/5$}}}
  \end{overpic}
  \caption{Density profiles $2\pi \rho(x,0)\lB^{2}$ for two QHs at (a) $\nu=1/3$ and (b) $\nu=2/5$. Black curves show the conventional two-QH wave functions of Eq.~\eqref{eq:localized-two-qh-wavepacket}; semi-transparent red curves show the HF determinant wave functions, Eq.~\eqref{eq:hf-jain-qp} with the removed orbitals replaced by the modified hole orbitals of Eq.~\eqref{eq:hf-jain-qh-orbital}. The shape specified by $k$ and the intended localization positions are indicated in the panels, with the latter marked by vertical dashed lines. For $k_{2}=0$ the HF and conventional constructions coincide, and the sampled densities agree within Monte Carlo error. For $k_{2}\neq0$ the HF wave function removes the shape distortion of the conventional profiles: at these separations, the two-QH density is well described as the sum of the corresponding single-QH density profiles.}
  \label{fig:hf-two-qh-density}
\end{figure}

\begin{figure}
  \includegraphics[width=\columnwidth]{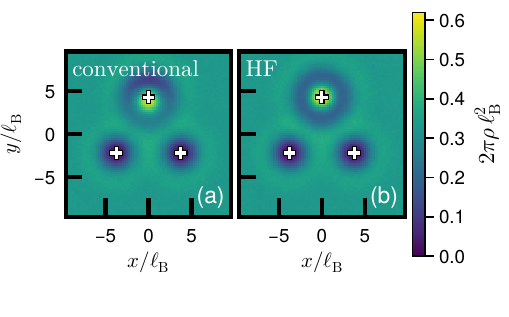}
  \caption{Density maps $2\pi\rho\,\lB^{2}$ for three QHs at $\nu=1/3$ ($N=64$), localized at the vertices of an equilateral triangle of side $d=7.5\lB$, with two QHs in the $k=0$ angular-momentum state and the one at the top vertex in the $k=1$ state; crosses mark the intended positions. (a) conventional wave function; (b) HF determinant wave function with the modified hole orbitals of Eq.~\eqref{eq:hf-jain-qh-orbital}. The $k=1$ QH is visibly distorted in (a), whereas in (b) it retains its circular shape to within the resolution of the plot.}
  \label{fig:hf-three-qh-density}
\end{figure}

The corresponding microscopic CF wave functions follow exactly as in Sec.~\ref{sec:localized-qh-qp-states}. For $M$ QPs localized at $\omega_{1},\dots,\omega_{M}$ in angular-momentum states $k_{1},\dots,k_{M}$ at $\nu=n/(2pn+1)$, with $\alpha=2p/(2pn+1)$,
  \begin{equation}\label{eq:hf-jain-qp}
    \Psi^{\{\QP{k_{a}},\omega_{a}\}}_{\frac{n}{2pn+1},\mathrm{HF}}
    =
    \LLL\,
    \Phi_{n}^{\mathrm{HF};\,\{\QP{k_{a}},\omega_{a}\}}\,
    \Phi_{1}^{2p},
  \end{equation}
  where $\Phi_{n}^{\mathrm{HF};\,\{\QP{k_{a}},\omega_{a}\}}$ is the $M$-QP generalization of the IQH determinant of Eq.~\eqref{eq:iqh-localized-two-electron-determinant} with each added-electron orbital $\psi^{\omega_{a}}_{n,k_{a}-n}$ replaced by its modified counterpart
  \begin{equation}\label{eq:hf-jain-qp-orbital}
    \varphi_{a}
    =
    \prod_{b\neq a}
    \left(1+\frac{\sqrt{2}\,\ellstar\,\hat{b}^{\dagger}_{a}}{\omega_{a}-\omega_{b}}\right)^{-\alpha}
    \psi^{\omega_{a}}_{n,k_{a}-n},
  \end{equation}
  with $\hat{b}^{\dagger}_{a}$ the raising counterpart of $\hat{b}_{a}$, $\hat{b}^{\dagger}_{a}\,\psi^{\omega_{a}}_{n,m}=\sqrt{m+n+1}\,\psi^{\omega_{a}}_{n,m+1}$. Outside the LLL the raising is no longer a multiplication by $z-\omega_{a}$, so the operator form is needed here.
  For two QPs, the Slater determinant is, explicitly,
  \begin{equation}
    \label{eq:hf-two-qp-microscopic-determinant}
    \begin{aligned}
      &\Phi_{n}^{\mathrm{HF};\,{\mathrm{QP}_{k_{1}}, \omega_{1}; \mathrm{QP}_{k_{2}}, \omega_{2}}}
      (\{z_i,\conj{z}_i\};B^{\star}) = \\
      &
      \begin{vmatrix}
        \eta_{0,0}(z_1) & \cdots & \eta_{0,0}(z_N)\\
        \eta_{0,1}(z_1) & \cdots & \eta_{0,1}(z_N)\\
        \vdots & \ddots & \vdots\\
        \eta_{n-1,M_{n-1}}(z_1) & \cdots & \eta_{n-1,M_{n-1}}(z_N)\\
        \varphi_{1}(z_1) & \cdots & \varphi_{1}(z_N)\\
        \varphi_{2}(z_1) & \cdots & \varphi_{2}(z_N)
      \end{vmatrix},
    \end{aligned}
  \end{equation}
  which differs from Eq.~\eqref{eq:iqh-localized-two-electron-determinant} only in the last two rows, with
  \begin{equation}\label{eq:hf-two-qp-microscopic-orbital}
    \varphi_{a}
    =
    \sum_{j\geq0}\binom{-\alpha}{j}
    \sqrt{\frac{(k_{a}+j)!}{k_{a}!}}
    \left(\frac{\sqrt{2}\,\ellstar}{\omega_{a}-\omega_{b}}\right)^{j}
    \psi^{\omega_{a}}_{n,k_{a}-n+j}.
  \end{equation}
  For $M$ QHs, each removed orbital of Eq.~\eqref{eq:iqh-localized-two-hole-determinant} is instead replaced by
  \begin{equation}\label{eq:hf-jain-qh-orbital}
    \xi_{a}
    =
    \prod_{b\neq a}
    \left(1+\frac{\sqrt{2}\,\ellstar\,\hat{b}_{a}}{\omega_{a}-\omega_{b}}\right)^{-\alpha}
    \psi^{\omega_{a}}_{n-1,k_{a}-(n-1)},
  \end{equation}
  a finite combination of at most $k_{a}+1$ hole orbitals. For two QHs, the Slater determinant is that of Eq.~\eqref{eq:iqh-localized-two-hole-determinant} with the two appended columns replaced by the expansion coefficients of the modified hole orbitals:
  \begin{equation}
    \label{eq:hf-two-qh-microscopic-determinant}
    \setlength{\arraycolsep}{3pt}
    \begin{aligned}
      &\Phi_{n}^{\mathrm{HF};\,{\mathrm{QH}_{k_{1}}, \omega_{1};\mathrm{QH}_{k_{2}}, \omega_{2}}}
      (\{z_i,\conj{z}_i\};B^{\star})= \\
      &\scalebox{0.95}{$\displaystyle{\setlength{\arraycolsep}{3pt}
          \begin{vmatrix}
            \eta_{0,0}(z_1) & \cdots & \eta_{0,0}(z_N) & 0 & 0\\
            \vdots & \ddots & \vdots & \vdots & \vdots\\
            \eta_{n-2,M_{n-2}}(z_1) & \cdots & \eta_{n-2,M_{n-2}}(z_N) & 0 & 0\\
            \eta_{n-1,-(n-1)}(z_1) & \cdots & \eta_{n-1,-(n-1)}(z_N) & \tilde{h}^{(1)}_{-(n-1)} & \tilde{h}^{(2)}_{-(n-1)}\\
            \vdots & \ddots & \vdots & \vdots & \vdots\\
            \eta_{n-1,M_{n-1}}(z_1) & \cdots & \eta_{n-1,M_{n-1}}(z_N) & \tilde{h}^{(1)}_{M_{n-1}} & \tilde{h}^{(2)}_{M_{n-1}}
      \end{vmatrix}}$},
    \end{aligned}
  \end{equation}
  with
  \begin{equation}\label{eq:hf-two-qh-microscopic-coefficients}
    \begin{split}
      \tilde{h}^{(a)}_{m}
      &=
      \sum_{j=0}^{k_{a}}\binom{-\alpha}{j}
      \sqrt{\frac{k_{a}!}{(k_{a}-j)!}}
      \left(\frac{\sqrt{2}\,\ellstar}{\omega_{a}-\omega_{b}}\right)^{j} \\
      &\quad\times
      \eta_{k_{a}-j,\,m+(n-1)-k_{a}+j}(\omega_{a});
    \end{split}
  \end{equation}
setting $\alpha=0$ retains only the $j=0$ term and recovers the conventional coefficients $h^{(a)}_{m}$ of Eq.~\eqref{eq:iqh-localized-two-hole-determinant}. The densities of these wave functions are shown in Figs.~\ref{fig:hf-two-qp-density} and \ref{fig:hf-two-qh-density} for two QPs and two QHs respectively at $\nu=1/3$ and $\nu=2/5$. Figure~\ref{fig:hf-three-qh-density} shows a three-QH example at $\nu=1/3$, with the QHs on an equilateral triangle, one placed in the $k=1$ angular-momentum state and the other two in $k=0$. As can be seen there, with the modified orbitals carrying the inverse vortex factor the $k=1$ QH shows almost no visible distortion, whereas in the conventional construction it is deformed by its two neighbors. Quantitatively, the leading angular ($m=1$) amplitude of the distortion falls from $0.040$ in units of $2\pi\rho\,\lB^{2}$ to a value consistent with zero at the Monte Carlo error, of order $10^{-3}$.

\section{Interference experiments}
\label{sec:interference}

Many ingenious interference experiments in recent years have observed fractional phase jumps~\cite{Nakamura20,Nakamura23,Kundu23,Kim24,Werkmeister25,Samuelson26,Ghosh25,Ghosh25A,Kim26}, which have been interpreted as signatures of the braid statistics of FQH QPs. In these experiments the interfering path runs predominantly along the two opposite edges of the FQH system, with two short tunneling segments coupling the edges and thereby closing the interference loop. However, we know that, unlike the incompressible bulk, the edges of the FQH system do not support QPs carrying fractionally quantized charge, due to the absence of a gap to protect these topological properties. How can one reconcile this fact with the experimental observations? 

One can argue that the tunneling regions effectively filter out the fractionally charged QPs, in the limit where the tunneling amplitude is small. Therefore, the entire interference process may be regarded as the coherent propagation of a fractionally charged QP around the loop, making the observed phase jumps a manifestation of the QP braid statistics~\cite{Chamon97,Rosenow07,Stern10A,Halperin11,Rosenow20,Feldman22,Wei24,Han16}; Mach--Zehnder geometries have been analyzed along similar lines~\cite{Law06,Feldman06,Feldman07,Wang10,Feldman08,Campagnano12,Batra23,Batra25}. Other proposed probes of the statistics include switching noise~\cite{Grosfeld06}, Coulomb blockade~\cite{Ilan08}, and the controlled localization of QPs inside the interferometer~\cite{Ofek10,Henzinger26}. Measurements that do not rely on a closed interference loop have also been carried out with anyon colliders and diluted anyon beams~\cite{Bartolomei20,Glidic23,Lee23}, building on theoretical proposals in which the statistics enters through braiding in the time domain~\cite{Rosenow16,Schiller23,Zhang25,Samal26,Rosenow25,Varada25}; we do not consider these here.

The CF framework suggests a way to understand the observations without assuming the existence of fractionally charged QPs at the edge. It only relies on the assumption that CFs exist at the edges of an FQH system. In other words, we assume that $2p$ vortices are bound to all electrons, including those at the edges, but we make no assumptions regarding the charge associated with an excited CF.

Let us recall that the concept of CFs is more generally valid than that of fractionally charged QPs. The formation of CFs is responsible for the FQHE at the Jain sequences, and incompressibility gives birth to fractionally charged QPs. While the existence of QPs with fractionally quantized charge is intimately tied to the incompressibility of the underlying state, the CFs can also exist in compressible states. A primary example is the CF metal at the half-filled LL~\cite{Halperin93,Halperin20,Shayegan20}. The model of QPs with fractionally quantized charge and statistics ceases to be meaningful when the QPs are close to one another and their density profiles have significant overlap, but CFs continue to provide an accurate description here as well. More pertinently to the present issue, the CFs can be perfectly well defined at the edges of an FQH state. There is indeed no conceptual problem in writing wave functions for CF systems with edges. Indeed, the CF theory has been extensively tested for FQH systems with edges, as those in the disk geometry, and shown to be quantitatively accurate~\cite{Jain95,Jeon04A,Jeon07}.

We note that while an isolated, excited CF in the interior of the FQH state carries a precisely quantized fractional charge relative to the uniform density incompressible state, that is no longer the case for excited CFs in compressible states or for CFs in regions without a gap. The reason is straightforward. The charge of a QP is equal to the charge of $2p$ vortices and an electron. A vortex has a charge equal to the filling factor in the bulk of the FQH state, which produces a fractionally quantized charge for the QP. However, near the edge, the charge of a vortex is not quantized. This is seen routinely in numerical evaluations of the charge of a vortex; see for example, Ref.~\cite{Kjonsberg97}, which evaluates the charge of the Laughlin QH, which is a vortex. Thus, while the CF, i.e. an electron with $2p$ vortices bound to it, is well defined at the edge, its charge is not quantized.

We attribute interference to a closed loop of a CF. The details of the loop are not important. The edge may have complex reconstructions~\cite{MacDonald90,Chamon94,Kane94,Wan02,Wan03,Bid10}, and the loop may involve ballistic as well as tunneling segments.  The only requirement is that the CF maintain its phase coherence around the closed loop. What charge is associated with it is irrelevant to the argument.

Let us now recall the derivation of braid statistics given in Eq.~\eqref{eq:braid_stat} with the help of Eqs.~\eqref{eq:Theta*} and \eqref{eq:B*}, which we reproduce here for convenience: 
\begin{equation}\label{eq:Theta*2}
  \begin{split}
    \Theta^*&=-\frac{2\pi B {A_{\cal L}}}{\phi_{0}}+2\pi\, 2p\int_{A_{\cal L}} d^2\vec{r}\, \rho(\vec{r})\\
    &\equiv -\frac{2\pi}{\phi_{0}}\int_{A_{\cal L}} d^2\vec{r}\, B^{\star}(\vec{r}) ,
  \end{split}
\end{equation}
\begin{equation} \label{eq:B*2}
  B^{\star}(\vec{r})=B-2p \rho(\vec{r})\phi_{0}
\end{equation}
\begin{equation}\label{eq:braid_stat2}
  \Delta \Theta^*=  2\pi \frac{2p}{2pn+1},
\end{equation}
where the QP executing the closed loop is an isolated CF in an excited $\Lambda$L. The remarkable aspect is that nowhere does the argument require that the CF executing the closed loop carry a fractionally quantized charge. The argument thus is also valid for a CF at the edge. The charge associated with the CF traversing the closed loop is irrelevant.

The interference experiments have observed oscillations whose period has been interpreted in terms of the fractionally quantized charge of the QP traversing the edge. To see how these may be understood without fractional charge at the edge, let us assume a spatially uniform filling $\nu=n/(2pn\pm 1)$, which corresponds to a uniform $B^{\star}=B/(2pn\pm 1)$. As $B$ is varied (with a corresponding variation in the density to stay at the fixed $\nu$), Eq.~\eqref{eq:Theta*2} implies that the resistance will have oscillations with period $\Delta(B^{\star}A)=\phi_{0}=hc/e$, or $\Delta(BA)=(2pn\pm 1)\phi_{0}=(2pn\pm 1)hc/e$, as observed experimentally. The period is thus interpreted in terms of the effective magnetic field $B^*=B/(2pn\pm 1)$ experienced by the CFs. 

Furthermore, the argument leading from Eq.~\eqref{eq:Theta*2} to the phase jumps in Eq.~\eqref{eq:braid_stat2} also remains valid. From this perspective, the phase jumps represent the fractional effective flux of the newly added QP in the interior of the FQH sample, measured by the looping CF.

In short, from this perspective, the periodic oscillations result from the effective magnetic field $B^*$ experienced by the looping CF, and the phase jumps directly measure the fractional flux of the QPs in the interior. In conjunction with the fractional charge measurements, these experiments provide a microscopic confirmation of the binding of $2p$ effective flux quanta, or vortices, to each electron.

We note that within the CF theory, the fractional flux of the QPs is an unavoidable consequence of two facts: (i) the excess charge associated with a QP is a fraction of an electron charge, and (ii) an effective flux $2p\phi_0$ is bound to each electron. However, the fractional flux does not follow in general for non-CF approaches. For example, Laughlin's QP at $\nu=1/3$ does not have a localized fractional flux associated with it, as evidenced by the fact that the Laughlin QPs, defined by the wave function proposed in Ref.~\cite{Laughlin83}, do not satisfy any well-defined braid statistics~\cite{Kjonsberg99}, even though they have well-defined fractional charge.

The challenges in measuring the braid statistics described in this article and summarized in the next section apply to these experiments as well, whether we view the object completing the loop as a CF or as a QP with a quantized fractional charge.  As noted above, in order for Eq.~\eqref{eq:braid_stat} to be valid, the addition of a new QP in the interior must neither alter the trajectory of the closed CF/QP loop nor distort the wave function of the CF/QP.  In the limit when the tunneling between the two edges is weak, the single particle states at the edges are invariant under the adiabatic addition of a flux quantum in the interior, as they do not enclose the flux. In general, the Coulomb potential emanating from the newly added QP will change the location of the closed loop at the Fermi energy and also distort the wave function of the CF/QP. However, in the experiments measuring fractional jumps, the Coulomb potential is effectively screened at long distances, and thus alters neither the location of the interfering path nor the shape of the CF/QP wave function.

\section{Discussion and Outlook}
\label{sec:discussion-outlook}

\label{sec:braid-statistics-challenges}

We have considered here the braid statistics of QPs contained entirely in the bulk of an FQH system, far from the edges.  We imagine a setup in which an STM tip produces a potential that creates and binds underneath it precisely one QP, which then follows the tip as it adiabatically executes a closed loop around another QP. Alternatively, one can create a closed tunneling loop by a judicious placement of impurities that serve as the sites for a QP~\cite{Gattu24}. We have evaluated the Berry phases in a computer experiment using the accurate and well-established theory of the QPs in terms of the CF theory~\cite{Kjonsberg97,Kjonsberg99,Kjonsberg99A,Jeon03,Jeon04,Tserkovnyak03,Nardin23A,Nardin23,Trung23,Gattu24,Bose24}.

For a reference, the Berry phase associated with a closed loop of an ``ideal anyon'' around another is robust, in that it is independent of the size or shape of the loop, and it directly yields the braid statistics. That is not the case for the FQH QPs. The Berry phase associated with a closed loop of an FQH QP around another is given by Eq.~\eqref{eq:berry-decomposition}, $$\Theta=\Theta_{\AB}+\Theta_{\rm braid},$$ where $\Theta_{\rm braid}$ on the right hand side is the braid statistics that we are after. To obtain it we must subtract $\Theta_{\AB}$, which we can do conveniently by determining the Berry phase associated with the identical closed loop of the FQH QP but without enclosing the other QP and subtracting it from $\Theta$. Two obstacles stand in the way: the uncertainty in $\Theta_{\AB}$, and changes in the shape of the QP between the two loops. Let us consider them one by one.

$\Theta_{\AB}$ is equal to the AB phase of a particle of charge $e^{\star}=-e/(2pn+1)$ moving in an external field $B$:
\begin{equation}
  \Theta_{\AB}=-2\pi \frac{B A}{(2pn+1)\phi_{0}}=-2\pi \frac{BA}{\phi_{0}^{\star}}
\end{equation}
where $\phi_{0}^{\star}=hc/|e^{\star}|$.
It is proportional to the area enclosed by the closed loop. For a meaningful determination of the braid statistics, the uncertainty in both $\Theta$ and $\Theta_{\AB}$  must be much less than the magnitude of the order-one contribution coming from statistics, given by $\Theta_{\rm braid}={4\pi p}/{(2pn+1)}$. For $\Theta_{\AB}$, taking the closed loop to be a circle of radius $R$, this implies that the error $\delta R$ in the location of the QP must satisfy:
\begin{equation}
  \delta R \ll 2p \left(\frac{\lB}{R}\right) \lB,
\end{equation}
which becomes arbitrarily small for large loops. Even a small uncertainty in $R$ can overwhelm, and thus wash out, the contribution from statistics. Note that $\delta R$ is in general much smaller than the size (diameter) of the QP, which is approximately given by~\cite{Jain07} $a=[8(2pn+2p+1)]^{1/2} \lB$.

The second obstacle is the shape of the QP. The shape of a QP is selected by the local potential, and, furthermore, it is deformed by the internal CF correlations (recall that the formation of CFs is what leads to the FQHE; the internal CF correlations are thus the correlations of the FQH liquid itself) due to the presence of other QPs. A change in the shape of the orbiting QP between the two loops, with and without the other QP enclosed, shifts the value of $\Theta_{\rm braid}$ extracted from their difference.  

In the ``conventional'' construction, the wave functions for states consisting of one or more QPs are obtained by composite-fermionizing wave functions of the corresponding {\it noninteracting} electron states. Here we find that correlated deformation produces the $k$-dependent statistics given in Eqs.~\eqref{eq:theta-two-qhs-k} and~\eqref{eq:theta-two-qps-k}. Our microscopic calculations show that the deformation decays as $1/d$ with the separation $d$ between the two QPs (Fig.~\ref{fig:two-qh-1-3-far}), becoming practically undetectable in the density at large separation, and yet it shifts the Berry phase of a closed loop by an amount comparable to, or even larger than, the statistics itself (Sec.~\ref{sec:shape-dependence-braid-statistics} explains why a $1/d$ deformation leaves a finite phase). This makes the Berry phase a sensitive function of the microscopic details of such an experiment.

We next demonstrate that it is possible to construct wave functions of states consisting of multiple QPs in such a manner that the shape of any given QP is unaffected by the presence of other QPs, i.e., the addition of a new QP leaves the already present QPs unaffected.  We present two explicit  constructions in Secs.~\ref{sec:flux-dressed-cf-wavefunctions} and~\ref{sec:hf-construction}. Each construction actually produces a single parameter ($\alpha'$) family of wave functions, and for the choice $\alpha'=2p/(2pn+1)$, the two-QP and two-QH densities become uncorrelated at large separation. With these wave functions a shape-independent statistics $\Delta\Theta=2\pi\alpha$ is recovered. Table~\ref{tab:constructions} summarizes these constructions. The two constructions differ in form: the first, referred to as flux-dressed states, produces a state that is a superposition of a thermodynamically large number of CF Slater determinants, whereas the second places the QPs / QHs in modified orbitals within a single Slater determinant, and generalizes to any number of QPs / QHs. Both, however, implement the same mechanism: prior to composite-fermionization, the localized QPs / QHs carry a compensating shift and deformation that cancels the shift and deformation produced by the vortex-attachment factor [Eq.~\eqref{eq:intro-vortex-cancellation}]. In both families, the exponent $\alpha'$ enters as a free parameter: nothing in the construction forces a particular value, and every value yields a legitimate wave function describing two localized QPs / QHs. Only when the parameter takes the value $\alpha'=2p/(2pn+1)$, i.e., the fractional flux bound to each QP / QH, does the cancellation take place (see Sec.~\ref{subsec:hf-two-qp}), leaving density profiles that are uncorrelated at large separation and producing the ``expected'' braid statistics, $\Delta\Theta/2\pi=\alpha$, independent of shape. We take this as evidence that the braid statistics of the FQH QPs remains meaningful in principle, although its extraction is delicate even in an idealized calculation with full microscopic control.

\begin{table*}
\caption{Summary of the three constructions of states consisting of two QHs / QPs. $\Delta\Theta$ is the Berry phase acquired in the large-separation limit in excess of the single-QH / single-QP Berry phases, for the general geometry in which both QHs / QPs rotate. For the wave functions obtained by composite-fermionizing the {\it noninteracting} electron wave functions (see Eqs.~\eqref{eq:localized-two-qh-wavepacket} and~\eqref{eq:localized-two-qp-wavepacket}), we find an angular-momentum- or shape-dependent answer $\Delta\Theta=\pm 2\pi\alpha\gamma$ with $\gamma\equiv1+k_{1}+k_{2}+(k_{1}-k_{2})\chi$ and $\chi$ given in Eq.~\eqref{eq:gamma-one-geometry-factor}; in the standard geometry, where one QH / QP is taken to be at the origin, we have $\chi=-1$ and $\gamma=1+2k_{2}$. For the flux-dressed wave functions in Eqs.~\eqref{eq:two-qh-corrected} and~\eqref{eq:two-qp-corrected}, and the single-determinant wave functions given in Eq.~\eqref{eq:hf-jain-qp}, we find a shape independent value  $\Delta\Theta=2\pi\alpha\equiv \Theta_{\rm braid}$. ``Shift'' and ``deformation'' refer to the correlated displacement and shape distortion of the density at large separation. The last two columns give the number of CF Slater determinants in the wave function and whether the construction extends to more than two QHs / QPs. 
}
  \label{tab:constructions}
  \begin{ruledtabular}
  \begin{tabular}{@{}lcccccc@{}}
    Construction & $\Delta\Theta^{\QHsym}/2\pi$ & $\Delta\Theta^{\QPsym}/2\pi$ & Shift & Deformation & determinants & generalizable? \\
    \hline
    Conventional (Secs.~\ref{sec:localized-qh-qp-states}, \ref{sec:shape-dependence-braid-statistics}) & $\alpha\gamma$ & $-\alpha\gamma$ & QPs only & for $k\neq0$ & one & yes \\
    Class I, flux-dressed (Sec.~\ref{sec:flux-dressed-cf-wavefunctions}) & $\alpha$ & $\alpha$ & none & none & thermodynamically many & no \\
    Class II, single determinant (Sec.~\ref{sec:hf-construction}) & $\alpha$ & $\alpha$ & none & none & one & yes \\
  \end{tabular}
  \end{ruledtabular}
\end{table*}

This raises another issue that deserves consideration. Replacing the  exponent $\alpha$ by a variational parameter $\alpha'$ in Eq.~\eqref{eq:qp-relative-coefficient-replacement}, or analogously in the inverse vortex factor of Eq.~\eqref{eq:hf-two-qp-orbital}, defines a one-parameter family of two-QH / two-QP wave functions, which are legitimate wave functions (for the single-determinant family, after the truncation described in Sec.~\ref{sec:hf-construction}) for arbitrary $\alpha'$ describing two QHs / QPs localized at $\omega_{1}$ and $\omega_{2}$ in prescribed angular-momentum orbitals. Two special points have been studied above: the conventional wave function ($\alpha'=0$), which yields the shape-dependent statistics $\Delta\Theta/2\pi=\pm\alpha(1+2k_{2})$ of Eq.~\eqref{eq:theta-standard-geometry}, and the flux-dressed wave function ($\alpha'=\alpha$), which yields the shape-independent value $\alpha$; varying $\alpha'$ interpolates continuously between them. These states are distinguished, at the level of the density, only by the correlated shift and shape deformation of Sec.~\ref{sec:shape-dependence-braid-statistics}, which decay as $1/d$. Consequently, at large separations the states with different $\alpha'$ in this range become practically indistinguishable to any density probe, no matter how accurate, while continuing to predict different braid statistics.
It appears to us that a superselection principle or an energetic criterion that sharply singles out the value $\alpha'=\alpha$ would be desirable for a sharply defined value of the braid statistics.  

How do these considerations enter into an experiment designed to measure the braid statistics? One approach for determining the braid statistics would be to determine the Berry phase twice -- once with and once without the other QP enclosed -- and take the difference. However, in this case it would be important to have exquisite control over the QP location to ensure that the two loops enclose the same area to extremely high accuracy, and that the QP shape also varies along the trajectory in exactly the same fashion in both experiments. Let us recall here the example of the $\nu=1/3$ QP.  Kjonsberg and Leinaas~\cite{Kjonsberg99A} evaluated the braid statistics for the Jain QP and found the statistics parameter to be $\Delta\Theta=-4\pi/3$  rather than the expected $\Theta_{\rm braid}=4\pi/3$.  This discrepancy was resolved in Refs.~\cite{Jeon03,Jeon04}, which noted that the insertion of another QP very slightly shifts the position of the QP going around in the loop; while the shift vanishes as the separation $R\rightarrow \infty$, the correction to the statistics remains finite, and accounting for it produces the expected braid statistics. That remains the case for the QPs of the other Jain FQH states.

The second approach is to measure the change in the Berry phase associated with a closed loop of a QP when another QP is inserted inside the loop. In this case, we need to make sure that the addition of the QP has no influence on either the trajectory or the shape of the QP executing the orbit. This was discussed in the context of the interference experiments in the preceding section.

Finally, we have considered here the possibility of measuring the braid statistics by going to its definition and considering braiding of the FQH QPs, but one might ask if it manifests through other consequences. Recall that the fermionic and bosonic statistics of particles manifest themselves in myriad ways. The fermionic statistics of electrons is the foundation on which the entire edifice of quantum chemistry and condensed matter physics is built. It is a necessary ingredient for the understanding of the spectra of atoms and molecules, as well as for the existence and properties of various many-body states of electrons such as metals, insulators, semiconductors, superconductors, and IQHE. Bosonic statistics is  responsible for the dramatic Bose-Einstein condensation. The fermionic statistics of CFs manifests through the appearance of the Jain sequences, the Fermi liquid metals at half fillings, paired FQH states, and various types of excited states. Analogous confirmations for the braid statistics of the FQH QPs appear to be more challenging because the notion of braid statistics is not applicable when the QPs are overlapping. For example, bound states of the QPs appear in STM experiments~\cite{Hu25,Gattu24A,Pu24}, and more recently, we have argued that the QPs often form bound molecules~\cite{Gattu25,Xu25,Wang26A,Li26}. However, the QPs are so strongly overlapping in these states that modeling them as anyons is not meaningful.  (The term ``molecular anyons'' in Ref.~\cite{Gattu25} does not mean a molecule of anyons, but rather a molecule that is an anyon under appropriate conditions.)

{\it Data availability.} The data that support the findings of this article are openly available~\cite{Gattu26}.

\begin{acknowledgments} We are grateful to Ajit C.~Balram, Yuval Gefen, T.~H.~Hansson, Kishore Iyer, Prashant Kumar, T.~Senthil, Steven H.~Simon, G.~J.~Sreejith, Sushanth Varada, and Ashvin Vishwanath for insightful discussions and feedback on the manuscript. 
  M.G. was supported in part by the National Science Foundation under Grant No.~DMR-2404619 and by a 2025/2026 Rising Researcher Grant from the Penn State Institute for Computational and Data Sciences (RRID:~SCR\_025154). The authors of this work also recognize the Institute for providing access to computational research infrastructure within the Roar Core Facility (RRID:~SCR\_026424). M.G. thanks the Lodha Theoretical Physics Institute for its hospitality during the summer of 2026. Anthropic's Claude (Fable) assisted with the preparation of figures, with the coordination and execution of the numerical calculations, and with critical review of the manuscript.
\end{acknowledgments}

\appendix

\section{Localized electron wavepackets in the lowest Landau level}
\label{app:localized-electron-wavepackets}

In this appendix, we derive the translated angular-momentum orbitals used in the localized-wavepacket construction in the main text. We begin with an electron restricted to the lowest Landau level (LLL) and then extend the result to higher Landau levels (LLs) using the Landau level (LL) raising operator.

In planar geometry under the symmetric gauge, the magnetic translation operator $\mathcal{T}_{\omega}$ is defined by
\begin{equation}
  \label{eq:appendix-magnetic-translation}
  \mathcal{T}_{\omega}\psi(z) = \exp\left(\frac{-\conj{\omega}z + \omega\conj{z}}{4\lB^{2}}\right)\psi(z+\omega),
\end{equation}
where $z=x-\imath y$ is the complex electron coordinate and $\omega=\omega_x-\imath\omega_y$ is the displacement, consistent with the convention of the main text. Suppose an electron, restricted to the $n^{\mathrm{th}}$ LL, is trapped in the orbital $\eta_{n,m}$ by a centrosymmetric potential centered at the origin. Equivalently, this implies that the orbital $\eta_{n,m}$ minimizes the matrix element:
\begin{equation}
  \label{eq:appendix-projected-potential-matrix-element}
  V_{n,m}=\int d\tau\,V(r)\abs{\eta_{n,m}(z)}^{2}.
\end{equation}
If this same localizing potential is translated to a point $\omega$, then the relevant eigenstates are the angular momentum orbitals centered at $\omega$ related to those at the origin via magnetic translation:
\begin{equation}
  \label{eq:appendix-localized-orbital-translation}
  \psi_{n, m}^{\omega}(z) = \mathcal{T}_{-\omega}\eta_{n, m}(z).
\end{equation}
From the definition of the magnetic translation operator,
\begin{equation}
  \label{eq:appendix-localized-orbital-explicit}
  \psi_{n, m}^{\omega}(z) = \exp\left(\frac{\conj{\omega}z-\omega\conj{z}}{4\lB^{2}}\right)\eta_{n, m}(z-\omega).
\end{equation}
Because $\mathcal{T}_{-\omega}$ commutes with the inter-Landau-level raising operator $a^{\dagger}$, we can simplify the problem by first deriving the expansion of $\psi_{n, m}$ in terms of $\eta_{n, m}$ explicitly in the LLL and subsequently obtaining the result in higher LLs through the action of the LL raising operator $a^{\dagger}$. That is, if the LLL expansion is:
\begin{equation}
  \label{eq:lll-translation-expansion}
  \mathcal{T}_{-\omega}\eta_{0,m}(z)=\sum_{\tilde M=0}^{\infty}c^{m}_{\tilde M}(\omega)\eta_{0,\tilde M}(z),
\end{equation}
then the corresponding expansion for the $n^{\mathrm{th}}$ LL takes the form
\begin{equation}
  \label{eq:nll-translation-expansion}
  \mathcal{T}_{-\omega}\eta_{n,m}(z)=\sum_{M=-n}^{\infty}c^{m+n}_{M+n}(\omega)\eta_{n,M}(z).
\end{equation}

Thus, we restrict our focus to the LLL. Setting the magnetic length $\lB=1$ for brevity, the LLL basis orbitals in the symmetric gauge are defined by
\begin{equation}\label{eq:lll-orbital}
  \eta_{0,m}(z) = \sqrt{\frac{1}{2\pi\,2^{m}\,m!}}\,z^{m}\,e^{-\abs{z}^{2}/4}.
\end{equation}
Applying this definition to our translated orbital gives
\begin{equation}
  \label{eq:lll-localized-orbital-explicit}
  \begin{split}
    &\psi_{0, m}^{\omega}(z) = \exp\left(\frac{\conj{\omega}z-\omega\conj{z}}{4}\right)\eta_{0, m}(z-\omega)\\
    & = \sqrt{\frac{1}{2\pi\,2^{m}\,m!}}\,(z-\omega)^{m}\exp\left[-\frac{\abs{z}^{2}+\abs{\omega}^{2}-2\conj{\omega}z}{4}\right].
  \end{split}
\end{equation}
The exponential is the standard coherent state in the LLL, which we can decompose as
\begin{equation}
  \label{eq:lll-coherent-state-resolution}
  \exp\left[-\frac{\abs{z}^{2}+\abs{\omega}^{2}-2\conj{\omega}z}{4}\right] = 2\pi \sum_{M=0}^{\infty}\conj{\eta}_{0, M}(\omega)\eta_{0, M}(z).
\end{equation}
Substituting this resolution in terms of the orbitals $\eta_{0, M}(z)$ and expanding the binomial $(z-\omega)^m$ gives
\begin{equation}
  \label{eq:lll-localized-orbital-double-sum}
  \begin{split}
    &\psi_{0, m}^{\omega}(z) = \sqrt{\frac{2\pi}{2^{m}\,m!}}\sum_{k=0}^{m}\sum_{M=0}^{\infty}\conj{\eta}_{0,M}(\omega) \eta_{0,M}(z) \\
    & \times (-1)^{k}\omega^{k}z^{m-k}\binom{m}{k} \\
    & = \sqrt{\frac{2\pi}{2^{m}\,m!}}\exp\left[-\frac{\abs{z}^{2}+\abs{\omega}^{2}}{4}\right]\\
    & \times \sum_{M=0}^{\infty}\sum_{k=0}^{m}\frac{(-1)^{k}}{2\pi \cdot 2^{M}M!}\conj{\omega}^{M}\omega^{k}z^{M+m-k}\frac{m!}{k!(m-k)!}.
  \end{split}
\end{equation}
To simplify this sum, we define $\tilde M = M+m-k$. Rearranging terms, we see that:
\begin{equation}
  \label{eq:lll-localized-orbital-shifted-sum}
  \begin{split}
    &\psi_{0, m}^{\omega}(z) = \sqrt{2\pi}\sum_{\tilde{M}=0}^{\infty}\sqrt{\frac{m!}{2\pi\,2^{\tilde{M}-m}\tilde{M}!}}{\conj{\omega}}^{\tilde{M}-m}e^{-\frac{\abs{\omega}^{2}}{4}}\eta_{0, \tilde{M}}(z)\\
    & \times \sum_{k=0}^{m}(-1)^{k}\left(\frac{\abs{\omega}^{2}}{2}\right)^{k}\frac{\tilde{M}!}{k!(m-k)!(\tilde{M}+k-m)!}.
  \end{split}
\end{equation}
The sum over $k$ is the series representation of the associated Laguerre polynomial:
\begin{equation}
  \label{eq:associated-laguerre-definition}
  L_{m}^{\tilde{M}-m}(x) = \sum_{k=0}^{m}(-1)^{k}\frac{\tilde{M}!}{k!(m-k)!(\tilde{M}+k-m)!}x^{k}.
\end{equation}
The definition of the higher-LL orbitals $\eta_{n,q}$ involves precisely these Laguerre polynomials:
\begin{equation}
  \label{eq:landau-orbital-unit-length}
  \eta_{n, q}(z)=\sqrt{\frac{n!}{2\pi\,2^{q}(n+q)!}}z^{q}L_{n}^{q}\left(\frac{\abs{z}^{2}}{2}\right)e^{-\frac{\abs{z}^{2}}{4}},
\end{equation}
where $q\geq -n$ specifies the azimuthal angular momentum. The summation therefore collapses to
\begin{equation}
  \label{eq:lll-localized-orbital-laguerre-expansion}
  \psi_{0, m}^{\omega}(z) = \sqrt{2\pi}\sum_{\tilde{M}=0}^{\infty}\conj{\eta}_{m, \tilde{M}-m}(\omega)\eta_{0, \tilde{M}}(z).
\end{equation}

Finally, acting with the raising operator $(a^{\dagger})^{n}/\sqrt{n!}$ on both sides gives the expansion in the $n^{\mathrm{th}}$ LL:
\begin{equation}
  \label{eq:nll-localized-orbital-expansion}
  \mathcal{T}_{-\omega}\eta_{n,m}(z) = \sqrt{2\pi}\sum_{M=-n}^{\infty}\conj{\eta}_{m+n,M-m}(\omega)\eta_{n,M}(z).
\end{equation}

\section{\texorpdfstring{Quasihole and quasiparticle wave functions at Jain fillings $\nu=n/(2n+1)$}{Quasihole and quasiparticle wave functions at Jain fillings}}
\label{app:jain-qh-qp-wavefunctions}
In this appendix, we provide a brief overview of the spherical geometry and single-particle physics within it. We then discuss the construction of composite fermion (CF) wave functions for the incompressible fractional quantum Hall (FQH) ground states at $\nu=n/(2n+1)$, along with the corresponding quasihole (QH) and quasiparticle (QP) wave functions, and the calculation of overlaps between them. We also introduce effective models for these calculations.

\subsection{Spherical geometry and planar limit}

In the spherical geometry~\cite{Haldane83}, electrons live on the surface of a two-dimensional sphere with a magnetic monopole of charge $Q$ at its center. The monopole produces a radial magnetic field of strength $B=2Q\phi_{0}/(4\pi R^{2})$, where $R$ is the radius of the sphere and $\phi_{0}=hc/e$ is the magnetic flux quantum. The single-particle eigenstates are the monopole harmonics $Y_{Q,l,m}(\theta,\phi)$, labeled by angular momentum $l$, quantized as $l=Q,Q+1,\dots$, and by the azimuthal quantum number $m=-l,-l+1,\dots,l$. A monopole harmonic of angular momentum $l$ has kinetic energy $\hbar\omega_{c}[l(l+1)-Q^{2}]/(2Q)$.

In the lowest Landau level ($l=Q$), the orbitals take a particularly simple form in terms of the spinor variables $u=\cos(\theta/2)\exp(\imath\phi/2)$, $v=\sin(\theta/2)\exp(-\imath\phi/2)$:
\begin{equation}
  \label{eq:lowest-ll-monopole-harmonic}
  Y_{Q,Q,m}(\theta,\phi) = \sqrt{\frac{2Q+1}{4\pi}\binom{2Q}{Q-m}}\,(-1)^{Q-m}\!u^{Q+m}\!v^{Q-m}\!.
\end{equation}
The higher-Landau-level monopole harmonics are multi-term polynomials in $u$, $\conj{u}$, $v$ and $\conj{v}$.

They can also be related to the Wigner-$D$ matrices,
\begin{equation}
  \label{eq:wigner-d-rotation-matrix}
  D^{l}_{m,m^{\prime}}(\mathcal{R}) = \bra{l,m}e^{-\imath\alpha J_{z}}e^{-\imath\beta J_{y}}e^{-\imath\gamma J_{z}}\ket{l,m^{\prime}},
\end{equation}
where $\mathcal{R}\equiv\mathcal{R}(\alpha,\beta,\gamma)$ denotes a rotation parametrized by the Euler angles $\alpha$--$\beta$--$\gamma$ about the $z$--$y$--$z$ axes, $J_{x},J_{y},J_{z}$ are the angular-momentum generators of rotations in three dimensions, and $\ket{l,m}$ are the ordinary spherical harmonics. In the gauge with Dirac strings at both poles,
\begin{equation}
  \label{eq:symmetric-monopole-gauge}
  \bm{A}(\theta,\phi) = -\frac{Q\phi_{0}}{2\pi R} \cot\theta\,\hat{\phi},
\end{equation}
the monopole harmonics take the form
\begin{equation}\label{eq:monopole-harmonics-wigner-D}
  Y_{Q,l,m}(\theta,\phi)=\sqrt{\frac{2l+1}{4\pi}}\,D^{l}_{-m,-Q}(\phi,\theta,0).
\end{equation}
This representation makes many properties of the monopole harmonics, such as their behavior under products and tensor decompositions, transparent.

\subsection{Composite-fermion ground state and single quasiholes}

We now turn to the construction of the CF wave functions. In the LLL, with interactions turned on, $N$ electrons facing a monopole of charge $Q$ each capture two vortices to form $N$ CFs facing a reduced monopole of charge $Q^{\star}=Q-(N-1)$. Facing this reduced monopole charge, CFs occupy their own LLs -- called $\Lambda$Ls -- separated by the CF cyclotron energy. When CFs occupy $n$ of these $\Lambda$Ls, we have an incompressible state of electrons at the Jain filling $\nu=n/(2n+1)$, with the corresponding many-body wave function written as:
\begin{equation}\label{eq:cf-gs-spherical-geometry}
  \begin{split}
    &\Psi_{\frac{n}{2n+1}}(\Omega_{1},\dots,\Omega_{N}) = \\
    & \LLL \phi_{n}(\Omega_{1},\dots,\Omega_{N};Q^{\star})J^{2}
  \end{split}
\end{equation}
Here, $\phi_{n}$ is a Slater determinant of $N$ electrons facing monopole strength $Q^{\star}=(N/n-n)/2$ and filling the first $n$ LLs, $J^{2}=\prod_{i<j}(u_{i}v_{j}-u_{j}v_{i})^{2}$ is the Jastrow factor that attaches two vortices to each electron in $\phi_{n}$ and converts them into CFs, and $\LLL$ is the projection operator into the LLL (implemented in practice via Jain--Kamilla projection). $\Omega_{i} \equiv (\theta_{i}, \phi_{i})$ is shorthand for the position of the $i^{\mathrm{th}}$ electron.

We first construct the quasihole wave functions. In the limit where any localizing or perturbing potential is much weaker than the CF cyclotron energy, the lowest-energy QH at $\nu=n/(2n+1)$ corresponds to a missing CF in the topmost ($n-1$)-th $\Lambda$L, i.e., from an orbital in the shell of angular momentum $L_{\mathrm{qh}}=Q^{\star}+n-1$. A general single-QH state is therefore a linear superposition
\begin{equation}
  \label{eq:single-qh-coefficient-state}
  \ket{c}\rangle \equiv \sum_{m}c_{m}\ket{m},
\end{equation}
where the doubled bracket indicates a coefficient-space state, distinguished from the microscopic many-body wave function. Here $\ket{m}$ denotes the state with a CF missing from the orbital of azimuthal quantum number $m$, corresponding to the unnormalized many-body wave function:
\begin{equation}
  \label{eq:cf-single-qh-spherical-state}
  \begin{split}
    &\ket{m} \equiv \Psi_{\frac{n}{2n+1}}^{\QHsym, m}(\Omega_{1},\dots,\Omega_{N})=\\
    &\LLL \phi_{n}^{\mathrm{+}, m}(\Omega_{1},\dots,\Omega_{N};Q^{\star})J^{2}
  \end{split}
\end{equation}
Here, $\phi_{n}^{\mathrm{+}, m}$ is the Slater determinant of $N$ electrons filling all orbitals of the first $n$ LLs at $Q^{\star}=(N/n-n)/2+(1/2n)$, except the orbital with azimuthal quantum number $m$ in the $(n-1)^{\mathrm{th}}$ LL.

Because the Jastrow factor $J^{2}$ preserves all total angular-momentum quantum numbers of the Slater determinant, each single-QH basis state $\ket{m}$ carries a definite total angular momentum $L=L_{\mathrm{qh}}$. This has an important consequence: the states $\ket{m}$ are connected by the ordinary angular momentum ladder operators, that is,
\begin{equation}
  \label{eq:qh-angular-momentum-ladder}
  L_{\pm}\ket{m} = \sqrt{L_{\mathrm{qh}}(L_{\mathrm{qh}}+1)-m(m\pm 1)}\ket{m \pm 1}.
\end{equation}
The overlap matrix $\overlap{m^{\prime}}{m}$ is therefore exactly proportional to the identity matrix~\cite{Gattu24},
\begin{equation}
  \label{eq:single-qh-overlap-normalization}
  \langle m' | m \rangle = \mathcal{N} \delta_{m,m'}.
\end{equation}
We can set the positive real normalization constant $\mathcal{N}$ to $1$. The inner product between any two single-QH states $\ket{f}\rangle$ and $\ket{g}\rangle$ then vastly simplifies to the inner product of their coefficient vectors:
\begin{equation}
  \label{eq:effective-single-qh-inner-product}
  \langle\langle f | g \rangle\rangle =\sum_{m}f^{\star}_{m}\,g_{m}.
\end{equation}

Consequently, for calculations that rely strictly on overlaps between single-QH states, such as evaluating the Berry phase for a single transported QH, the many-body problem maps onto the problem of a single particle at the effective monopole strength $-Q^{\star}$ in the angular-momentum shell $l=L_{\mathrm{qh}}$. The equivalent effective state is:
\begin{equation}
  \label{eq:effective-qh-sphere-wavefunction}
  \Psi^{\QHsym}_{\eff}(\Omega) = \sum_{m=-L_{\mathrm{qh}}}^{L_{\mathrm{qh}}}f_{m} \conj{Y}_{Q^{\star},\,L_{\mathrm{qh}},\,m}(\Omega).
\end{equation}
By employing the relationship between the monopole harmonics and $\eta_{n, m}$ (orbitals in the symmetric gauge in the planar geometry),
\begin{equation}
  Y_{Q,l,m}(\theta,\phi)\to \eta_{l-Q,Q-m}(z,\bar{z}).
\end{equation}
where $z=2R\tan(\theta/2)e^{-\imath\phi}$ with $R=\sqrt{Q}$ in units of the magnetic length $\lB$, we see that the effective QH wave function in the thermodynamic limit is
\begin{equation}
  \label{eq:effective-qh-planar-wavefunction}
  \Psi^{\QHsym}_{\eff}(z,\bar{z})=\sum_{m=-(n-1)}^{\infty}f_{Q^{\star}-m}\conj{\eta}_{n-1, m}(\frac{z}{\ellstar},\frac{\bar{z}}{\ellstar}).
\end{equation}

\subsection{Two-quasihole wave functions and overlaps}

Let us now turn to the construction of two-QH wave functions. A general two-QH state, $\ket{f}\rangle$,
\begin{equation}
  \label{eq:two-qh-coefficient-state}
  \ket{f}\rangle = \sum_{m_{1}<m_{2}} f_{m_{1},m_{2}} \ket{m_{1},m_{2}},
\end{equation}
is a superposition of the states $\ket{m_{1},m_{2}}$ defined by the unnormalized many-body wave function corresponding to missing CFs in azimuthal orbitals $m_{1}$ and $m_{2}$:
\begin{equation}\label{eq:two-qh-sphere}
  \begin{split}
    &\Psi_{\frac{n}{2n+1}}^{\QHsym, m_{1}; \QHsym, m_{2}}(\Omega_{1},\dots,\Omega_{N})=\\
    &\LLL \phi_{n}^{\mathrm{+}, m_{1}; \mathrm{+}, m_{2}}(\Omega_{1},\dots,\Omega_{N};Q^{\star})J^{2}.
  \end{split}
\end{equation}
Here, the underlying Slater determinant $\phi_{n}^{\mathrm{+}, m_{1}; \mathrm{+}, m_{2}}$ is the state of $N$ electrons filling the first $n$ LLs except for the $L_{z}=m_{1}$ and $L_{z}=m_{2}$ orbitals in the $(n-1)^{\mathrm{th}}$ LL at monopole strength $Q^{\star}=(N/n-n)/2+(1/n)$.

To compute overlaps between different two-QH states, we again use the fact that the Jastrow factor commutes with rotations. Rather than working in the $\ket{m_{1},m_{2}}$ basis, we decompose the underlying two-hole Slater determinants into total angular momentum eigenstates $\ket{L, M}$ using Clebsch--Gordan coefficients:
\begin{equation}
  \label{eq:two-qh-angular-momentum-decomposition}
  \ket{m_{1}, m_{2}} = \sum_{L, M} C^{L, M}_{L_{\mathrm{qh}}, -m_{1};L_{\mathrm{qh}}, -m_{2}}\ket{L, M}.
\end{equation}
Because we are dealing with holes, removing a CF from orbital $m_{i}$ contributes $-m_{i}$ to the total azimuthal quantum number, leading to an overall azimuthal quantum number $M = -(m_{1}+m_{2})$.

\begin{figure}
  \includegraphics[width=\columnwidth]{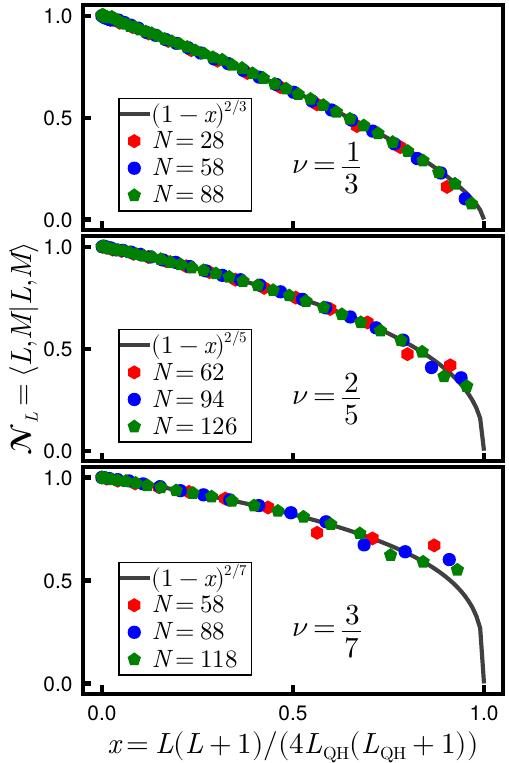}
  \caption{Norms of two-quasihole states at Jain fillings $\nu=1/3, 2/5,$ and $3/7$ (top to bottom). The states are obtained via composite-fermionization of two-hole states projected onto total angular momentum eigenstates $\ket{L, M}$ of the $\nu^{\star}=n$ integer quantum Hall states at effective monopole strength $Q^{\star}=(N/n-n)/2+1/n$. Each quasihole contributes angular momentum $L_{\mathrm{qh}}=Q^{\star}+n-1$, yielding a maximum allowed angular momentum of $2L_{\mathrm{qh}}-1$. Scatter points correspond to data for different system sizes $N$. Reference curves for each filling are based on a model of two anyons formed by attaching $\alpha=2/3, 2/5,$ and $2/7$ vortices to two electrons (holes).}
  \label{fig:two-qh-norms}
\end{figure}

Because total angular momentum $L$ is a conserved quantum number, the many-body states $\ket{L, M}$ are strictly orthogonal for distinct $(L, M)$ pairs:
\begin{equation}
  \label{eq:two-qh-overlap-normalization}
  \langle\langle L_{1}, M_{1} | L_{2}, M_{2} \rangle\rangle = \mathcal{N}_{L}\delta_{L_{1}, L_{2}}\delta_{M_{1}, M_{2}}.
\end{equation}
Crucially, within any specific $L$ sector, the norm $\mathcal{N}_{L}$ is independent of $M$ due to the underlying angular-momentum algebra. All information regarding the overlaps of arbitrary two-QH states is therefore entirely determined by the values of these norms $\mathcal{N}_{L}$. While we cannot extract these norms analytically from the microscopic CF wave functions, they can be efficiently evaluated using Monte Carlo integration. We plot the extracted norms in Fig.~\ref{fig:two-qh-norms}.

\begin{figure}
  \includegraphics[width=\columnwidth]{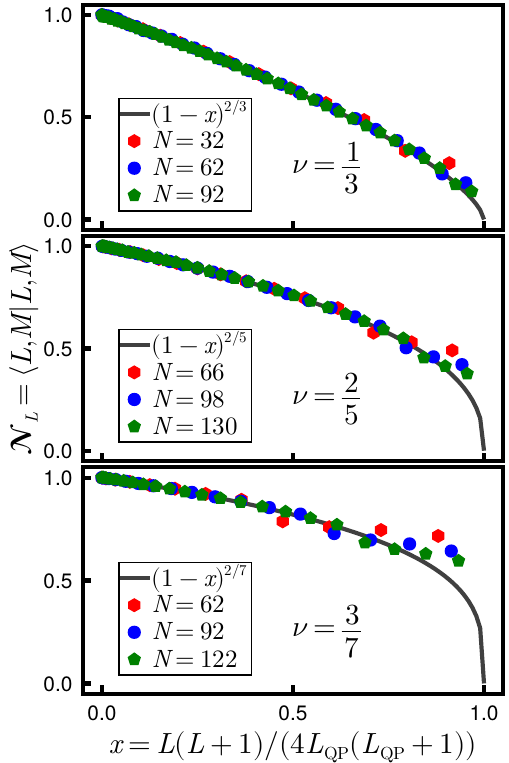}
  \caption{Norms of two-quasiparticle states at Jain fillings $\nu=1/3, 2/5,$ and $3/7$ (top to bottom). The states are obtained by composite-fermionizing two-electron states restricted to the $n^{\mathrm{th}}$ Landau level and projecting them onto total angular momentum eigenstates $\ket{L, M}$. For $N$ electrons, the effective monopole strength is $Q^{\star}=(N/n-n)/2-1/n$, and each QP carries angular momentum $L_{\mathrm{qp}}=Q^{\star}+n$, giving a maximum total angular momentum $2L_{\mathrm{qp}}-1$. Scatter points show data for different system sizes. Reference curves are based on a model of two anyons formed by attaching $\alpha=2/3, 2/5,$ and $2/7$ vortices to two electrons.}
  \label{fig:two-qp-norms}
\end{figure}

The quasiparticle construction is the particle analog of the two-QH construction above. For the $p=1$ sequence considered in this appendix, a single QP at $\nu=n/(2n+1)$ corresponds to an additional CF in the otherwise empty $n^{\mathrm{th}}$ $\Lambda$L. Thus, for overlap calculations involving single-QP states, the many-body problem maps onto a single (positively-charged) particle restricted to the $n^{\mathrm{th}}$ LL at the reduced magnetic field $B^{\star}$. The same logic applies to two QPs: we begin with two additional CFs in the $n^{\mathrm{th}}$ LL and decompose the corresponding two-particle states into total angular momentum eigenstates. For $N$ electrons, the effective monopole strength is $Q^{\star}=(N/n-n)/2-1/n$, and each QP carries angular momentum $L_{\mathrm{qp}}=Q^{\star}+n$. The resulting norms are shown in Fig.~\ref{fig:two-qp-norms}.

\subsection{Effective two-anyon overlap model}

The norm comparisons above show that, for well-separated QHs or QPs, both the two-QH and two-QP overlaps are reproduced by the same effective two-anyon model, with statistical  exponent $\alpha=2/(2n+1)$. Thus, for overlap calculations, a two-QH coefficient state $\ket{f}\rangle$ can be represented by

\begin{equation}
  \label{eq:effective-two-anyon-planar-state}
  \begin{split}
    &\vert f \rangle \rangle =\\
    &\sum_{m_{1}, m_{2}=0}^{\infty}f_{m_{1}-(n-1), m_{2}-(n-1)}\\
    &
      \begin{vmatrix}
        \conj{\eta}_{0,m_{1}}(\frac{Z_{1}}{\ellstar}, \frac{\bar{Z}_{1}}{\ellstar}) & \conj{\eta}_{0,m_{1}}(\frac{Z_{2}}{\ellstar}, \frac{\bar{Z}_{2}}{\ellstar}) \\
        \conj{\eta}_{0,m_{2}}(\frac{Z_{1}}{\ellstar}, \frac{\bar{Z}_{1}}{\ellstar}) & \conj{\eta}_{0,m_{2}}(\frac{Z_{2}}{\ellstar}, \frac{\bar{Z}_{2}}{\ellstar}) \\
    \end{vmatrix}\\
    &\times (\conj{Z}_{1}-\conj{Z}_{2})^{\alpha}
  \end{split}
\end{equation}
and the corresponding two-QP coefficient state by

\begin{equation}
  \label{eq:effective-two-anyon-planar-state-qp}
  \begin{split}
    &\vert f \rangle \rangle =\\
    &\sum_{m_{1}, m_{2}=0}^{\infty}f_{m_{1}-n, m_{2}-n}\\
    &
      \begin{vmatrix}
        \eta_{0,m_{1}}(\frac{Z_{1}}{\ellstar}, \frac{\bar{Z}_{1}}{\ellstar}) & \eta_{0,m_{1}}(\frac{Z_{2}}{\ellstar}, \frac{\bar{Z}_{2}}{\ellstar}) \\
        \eta_{0,m_{2}}(\frac{Z_{1}}{\ellstar}, \frac{\bar{Z}_{1}}{\ellstar}) & \eta_{0,m_{2}}(\frac{Z_{2}}{\ellstar}, \frac{\bar{Z}_{2}}{\ellstar}) \\
    \end{vmatrix}\\
    &\times (Z_{1}-Z_{2})^{\alpha}.
  \end{split}
\end{equation}

\section{Two anyons in spherical geometry}
\label{app:two-anyons-spherical-geometry}

In this appendix, we construct two-anyon states in the spherical geometry and derive an analytic expression for their norms as a function of total angular momentum. These norms serve as reference curves for the total-angular-momentum-resolved two-QH and two-QP norms, or normalization constants $\mathcal{N}_{L}$, shown in Figs.~\ref{fig:two-qh-norms} and \ref{fig:two-qp-norms}. The excellent agreement validates the use of an effective two-particle model for the many-body QH and QP wave functions at Jain fillings $\nu=n/(2n+1)$ for the purpose of calculating overlaps.

We build two-anyon wave functions in the spherical geometry by attaching vortices to states of noninteracting electrons at reduced monopole strength $q$. These anyons are fictitious particles introduced for the purposes of the calculation and should not be confused with the physical CFs.

Consider two electrons at monopole strength $q$ restricted to the $(l-q)^{\mathrm{th}}$ LL, i.e., the $l$-angular momentum shell. Any such state can be written as a linear superposition of basis states $\ket{m_{1}, m_{2}}$, following Eq.~\eqref{eq:monopole-harmonics-wigner-D}:
\begin{equation}
  \label{eq:spherical-two-particle-basis}
  \begin{split}
    &\langle \Omega_{1}, \Omega_{2} \vert m_{1}, m_{2} \rangle =\frac{1}{\sqrt{2}}\frac{2l+1}{4\pi} \times \\
    & [D^{l}_{-m_{1},-q}(\Omega_{1})D^{l}_{-m_{2},-q}(\Omega_{2})- \\
    &D^{l}_{-m_{2},-q}(\Omega_{1})D^{l}_{-m_{1},-q}(\Omega_{2})],
  \end{split}
\end{equation}
or equivalently in the total angular momentum basis $\ket{L, M}$, with $L$ restricted to $L=2l-1,2l-3,\dots$ by antisymmetry:
\begin{equation}
  \label{eq:two-particle-total-angular-momentum-state}
  \begin{split}
    & \vert L, M\rangle = \sum_{m_{1}, m_{2}}C^{LM}_{l,-m_{1};l,-m_{2}}\ket{m_{1}, m_{2}} \\
    & \equiv \sqrt{2}\frac{2l+1}{4\pi}\sum_{m_{1}, m_{2}}C^{LM}_{l,-m_{1};l,-m_{2}}\\
    &\hspace{2.5em}\times D^{l}_{-m_{1},-q}(\Omega_{1})D^{l}_{-m_{2},-q}(\Omega_{2}).
  \end{split}
\end{equation}
Here, we have used the relation $C^{L,M}_{l,m_{1};l,m_{2}}=(-1)^{2l-L}C^{L,M}_{l,m_{2};l,m_{1}}$ to condense the sum.

Next, we introduce anyonic behavior by attaching $\alpha$ vortices to each particle via the Jastrow-like factor $(u_{1}v_{2}-u_{2}v_{1})^{\alpha}$. We initially take $\alpha$ to be a positive integer, which enables the use of standard angular momentum recoupling identities~\cite{Varshalovich88}, and subsequently analytically continue to real $\alpha$ in a thermodynamic limit of large $l$ and $q$ (with finite $l-q$). The analytic continuation is applied only after the norm has been reduced to its thermodynamic-limit expression. The resulting two-anyon wave function takes the form
\begin{equation}
  \label{eq:spherical-two-anyon-state}
  \begin{split}
    &\Psi_{L, M}^{\alpha}(\Omega_{1}, \Omega_{2})=\\
    &[\sqrt{2}\frac{2l+1}{4\pi}\sum_{m_{1}, m_{2}}C^{LM}_{l,-m_{1};l,-m_{2}}D^{l}_{-m_{1},-q}(\Omega_{1})D^{l}_{-m_{2},-q}(\Omega_{2})] \\
    & \times (u_{1}v_{2}-u_{2}v_{1})^{\alpha}.
  \end{split}
\end{equation}
Due to the equivariance of the vortex attachment factor under magnetic translations, $\Psi_{L, M}^{\alpha}$ is an eigenstate of total angular momentum $(L, M)$ with the two particles at total monopole strength $Q=q+\alpha/2$. The initial two-electron state as written has norm $\sqrt{2}$, owing to the unrestricted sum over $(m_{1}, m_{2})$; this $L$-independent constant is dropped below. The composite two-anyon state is, moreover, not normalized. Determining its normalization constant is the central goal of this appendix, as it is required to evaluate wave function overlaps.

Following the equivariance arguments established previously, the norms $\mathcal{N}_{L}\equiv\|\Psi^{\alpha}_{L, M}\|^2$ depend exclusively on $L$ and $\alpha$, remaining independent of $M$.

For simplicity, we restrict our analysis to the $M=0$ sector. The same equivariance implies that the vortex attachment factor carries total angular momentum $L=M=0$, with each particle effectively at monopole strength $\alpha/2$. We may therefore expand it as (up to a real, $\alpha$-dependent normalization):
\begin{equation}
  \label{eq:jastrow-factor-spherical-expansion}
  \begin{split}
    &(u_{1}v_{2}-u_{2}v_{1})^{\alpha} \sim \\
    &\sum_{m=-\alpha/2}^{\alpha/2}C^{0,0}_{\alpha/2,-m;\alpha/2,m}D^{\alpha/2}_{-m,-\alpha/2}(\Omega_{1})D^{\alpha/2}_{m,-\alpha/2}(\Omega_{2}).
  \end{split}
\end{equation}
Thus, $\Psi_{L, 0}^{\alpha}$ can be written as:
\begin{equation}
  \label{eq:two-anyon-l-zero-expanded-state}
  \begin{split}
    &\Psi_{L, 0}^{\alpha} = \\
    &\left[\sqrt{2}\frac{2l+1}{4\pi}\sum_{m}C^{L,0}_{l,-m;l,m}D^{l}_{-m,-q}(\Omega_{1})D^{l}_{m,-q}(\Omega_{2})\right] \\
    & \times \sum_{\tilde{m}=-\alpha/2}^{\alpha/2}C^{0,0}_{\alpha/2,-\tilde{m};\alpha/2,\tilde{m}}D^{\alpha/2}_{-\tilde{m},-\alpha/2}(\Omega_{1})D^{\alpha/2}_{\tilde{m},-\alpha/2}(\Omega_{2}).
  \end{split}
\end{equation}
To combine the Wigner-$D$ matrix products, we use the following identity,
\begin{equation}
  \label{eq:wigner-d-product-clebsch-gordan}
  \begin{split}
    &D^{j_1}_{m_1, k_1}(R)D^{j_2}_{m_2, k_2}(R) =\\
    & \sum_{J} C^{J, m_1+m_2}_{j_1, m_1;j_2, m_2}C^{J, k_1+k_2}_{j_1, k_1;j_2, k_2}\\
    &\hspace{2.5em}\times D^J_{m_1+m_2,k_1+k_2}(R).
  \end{split}
\end{equation}
Applying this relation, and noting the relation between Wigner-$D$ matrices and monopole harmonics,
\begin{equation}
  \label{eq:wigner-d-to-monopole-harmonic}
  \begin{split}
    D^{j_1}_{-(m+\tilde{m}), -(q+\alpha/2)}(\Omega_1) &= \sqrt{\frac{4\pi}{2j_{1}+1}}Y_{q+\alpha/2,j_1, m+\tilde{m}}(\Omega_{1}) \\
    D^{j_2}_{(m+\tilde{m}), -(q+\alpha/2)}(\Omega_2) &= \sqrt{\frac{4\pi}{2j_{2}+1}}Y_{q+\alpha/2,j_2, -m-\tilde{m}}(\Omega_{2})
  \end{split}
\end{equation}
allows us to expand $\Psi^{\alpha}_{L, 0}$ over intermediate angular momenta $j_{1}$ and $j_{2}$ (which correspond to LL occupations $n_{i}=j_{i}-Q$ at the total monopole strength $Q=q+\alpha/2$):
\begin{equation}
  \label{eq:two-anyon-state-monopole-harmonic-expansion}
  \begin{split}
    &\Psi_{L, 0}^{\alpha} = \sum_{m, \tilde{m}, j_{1}, j_{2}}\sqrt{2}\frac{2l+1}{\sqrt{(2j_{1}+1)(2j_{2}+1)}} \times \\
    & C^{j_1, -(q+\alpha/2)}_{l, -q;\alpha/2, -\alpha/2}C^{j_2, -(q+\alpha/2)}_{l, -q;\alpha/2, -\alpha/2}\times \\
    &  C^{L,0}_{l,-m;\,l,m}  C^{0,0}_{\alpha/2,-\tilde{m};\alpha/2,\tilde{m}}C^{j_1, -(m+\tilde{m})}_{l, -m;\alpha/2, -\tilde{m}}C^{j_2, m+\tilde{m}}_{l, m;\alpha/2, \tilde{m}}\\
    &\times \left[Y_{q+\alpha/2,j_{1},m+\tilde{m}}(\Omega_{1}) Y_{q+\alpha/2,j_{2},-m-\tilde{m}}(\Omega_{2})\right].
  \end{split}
\end{equation}

We can simplify the above sum using the following identity from Ref.~\cite{Varshalovich88}:
\begin{equation}
  \label{eq:nine-j-recoupling-identity}
  \begin{split}
    &\sum_{\beta \gamma \epsilon \varphi}C^{a\alpha}_{b\beta;c\gamma}C^{d \delta}_{e \epsilon;f \varphi}C^{g\eta}_{e \epsilon;b \beta}C^{j \mu}_{f \varphi;c \gamma}=\\
    &=\sqrt{(2a+1)(2d+1)(2g+1)(2j+1)} \times \\
    &\sum_{k \kappa}C^{k\kappa}_{g \eta;j \mu}C^{k \kappa}_{d \delta;a \alpha}\underbrace{
      \begin{Bmatrix}
        c & b & a \\
        f & e & d \\
        j & g & k
    \end{Bmatrix}}_{9j-\text{symbol}}
  \end{split}
\end{equation}

Consider the expression:
\begin{equation}
  \label{eq:recoupling-coefficient-definition}
  \begin{split}
    &\mathcal{E}=\sum_{m+\tilde{m}=\text{constant}}\left(C^{L, 0}_{l,-m;l,m}C^{0,0}_{\alpha/2,-\tilde{m};\alpha/2,\tilde{m}}\right.\\
    &\left. C^{j_{1},-(m+\tilde{m})}_{l,-m;\alpha/2,-\tilde{m}}C^{j_{2},(m+\tilde{m})}_{l,m;\alpha/2,\tilde{m}}\right).
  \end{split}
\end{equation}
We can re-write this using the identity $C^{LM}_{j_1,m_1;j_2,m_2}=(-1)^{j_1+j_2-L}C^{LM}_{j_{2},m_{2};j_{1},m_{1}}$ and setting $\eta=-(m+\tilde{m})$ and $\mu=m+\tilde{m}$:
\begin{equation}
  \label{eq:recoupling-coefficient-nine-j}
  \begin{split}
    &\mathcal{E}=(-1)^{2l+\alpha-(j_{1}+j_{2})}\\
    &\sum_{\beta,\gamma,\epsilon,\delta}C^{L, 0}_{l,\beta;l,\gamma}C^{0,0}_{\alpha/2,\epsilon;\alpha/2,\delta}C^{j_{1},\eta}_{\alpha/2,\epsilon;l,\beta}C^{j_{2},\mu}_{\alpha/2,\delta;l,\gamma} \\
    &=(-1)^{2l+\alpha-(j_{1}+j_{2})}\sqrt{(2L+1)(2j_{1}+1)(2j_{2}+1)} \\
    & \sum_{k}C^{k0}_{j_{1},-(m+\tilde{m});j_{2},(m+\tilde{m})}C^{k0}_{0,0;L,0}
    \begin{Bmatrix}
      l & l & L \\
      \frac{\alpha}{2} & \frac{\alpha}{2} & 0 \\
      j_{2} & j_{1} & k
    \end{Bmatrix}
  \end{split}
\end{equation}
Now, noting that $C^{k0}_{0,0;L,0}=\delta_{k, L}$, and
\begin{equation}
  \label{eq:nine-j-reduction-identity}
  \begin{split}
    &
    \begin{Bmatrix}
      l & l & L \\
      \frac{\alpha}{2} & \frac{\alpha}{2} & 0 \\
      j_{2} & j_{1} & L
    \end{Bmatrix} =\\
    &\frac{(-1)^{l+L+\alpha/2+j_{2}}(-1)^{2l+2L+\alpha+j_{1}+j_{2}}}{\sqrt{(2L+1)(\alpha+1)}}
    \begin{Bmatrix}
      l & l & L \\
      j_1 & j_2 & \alpha/2
    \end{Bmatrix},
  \end{split}
\end{equation}
we find the following simplified expression for $\mathcal{E}$:
\begin{equation}
  \label{eq:recoupling-coefficient-reduced}
  \begin{split}
    &\mathcal{E}=(-1)^{l+\alpha/2+L+j_{2}}\sqrt{\frac{(2j_1+1)(2j_2+1)}{\alpha+1}} \times \\
    &C^{L0}_{j_{1},-(m+\tilde{m});j_{2},(m+\tilde{m})}
    \begin{Bmatrix}
      l & l & L \\
      j_1 & j_2 & \alpha/2
    \end{Bmatrix}
  \end{split}
\end{equation}
Therefore, in terms of the total angular momentum eigenstates,
\begin{equation}
  \label{eq:two-anyon-basis-j1-j2}
  \begin{split}
    &\Psi_{L,0}^{j_{1}, j_{2}}(\Omega_{1}, \Omega_{2})\\
    &=\sum_{m}C^{L,0}_{j_{1},-m;j_{2},m}Y_{q+\alpha/2,j_{1},m}(\Omega_{1})Y_{q+\alpha/2,j_{2},-m}(\Omega_{2})
  \end{split}
\end{equation}
The two-anyon wave function $\Psi^{\alpha}_{L, 0}$ can then be written as
\begin{equation}
  \label{eq:two-anyon-state-j1-j2-expansion}
  \begin{split}
    &\Psi^{\alpha}_{L, 0} = \frac{\sqrt{2}(2l+1)}{\sqrt{\alpha+1}}\sum_{j_{1}, j_{2}}(-1)^{l+\alpha/2+L+j_{2}}C^{j_{1},-(q+\alpha/2)}_{l,-q;\alpha/2,-\alpha/2} \times\\
    &C^{j_{2},-(q+\alpha/2)}_{l,-q;\alpha/2,-\alpha/2}
    \begin{Bmatrix}
      l & l & L \\
      j_{1} & j_{2} & \frac{\alpha}{2}
    \end{Bmatrix}\Psi_{L, 0}^{j_{1}, j_{2}}
  \end{split}
\end{equation}
With this simplified form, we can now calculate the norm of $\Psi_{L, 0}^{\alpha}$. The states $\Psi_{L,0}^{j_{1},j_{2}}$ are orthonormal for different $(j_{1}, j_{2})$ pairs, so:
\begin{equation}
  \label{eq:two-anyon-norm-nl}
  \begin{split}
    &\mathcal{N}_{L} \equiv \|\Psi_{L,0}^{\alpha}\|^{2}\\
    &= \sum_{j_1,j_2} \frac{2(2l+1)^2}{\alpha+1}\left(C^{j_{1},-(q+\alpha/2)}_{l,-q;\alpha/2,-\alpha/2}C^{j_{2},-(q+\alpha/2)}_{l,-q;\alpha/2,-\alpha/2}\right)^{2}\\
    &
    \begin{Bmatrix}
      l & l & L \\
      j_{1} & j_{2} & \frac{\alpha}{2}
    \end{Bmatrix}^2
  \end{split}
\end{equation}
We now obtain an expression for $\mathcal{N}_{L}$ in a thermodynamic limit, i.e., $l, q \gg \alpha/2$. Since $j_{1}, j_{2} \geq q+\alpha/2$, we have $j_{1}, j_{2} \gg \alpha/2$ as well. This allows us to make use of the following identities from Ref.~\cite{Varshalovich88}:
\begin{equation}
  \label{eq:clebsch-gordan-asymptotic}
  \begin{split}
    & \lim_{a,c\gg b}C^{c \gamma}_{a\alpha;b\beta} \to \delta_{\gamma-\alpha,\beta}d^{b}_{\beta (c-a)}(\vartheta)\\
  \end{split}
\end{equation}
with $\cos \vartheta = 2 \gamma / (2c+1)$ and
\begin{equation}
  \label{eq:six-j-asymptotic}
  \begin{split}
    &\lim_{a,b,c\gg f,m,n}
    \begin{Bmatrix}
      a & b & c \\
      b+m & a+n & f
    \end{Bmatrix} \\
    &\to \frac{(-1)^{a+b+c+f+m}}{\sqrt{(2a+1)(2b+1)}}d^{f}_{mn}(\Omega)
  \end{split}
\end{equation}
with
\begin{equation*}
  \cos \Omega = \frac{a(a+1)+b(b+1)-c(c+1)}{2\sqrt{a(a+1)b(b+1)}}.
\end{equation*}

These two asymptotic identities yield $\mathcal{N}_{L}$ in a thermodynamic limit. Dropping an $L$-independent overall prefactor, we obtain the remarkably simple parameter-free expression:
\begin{equation}
  \label{eq:two-anyon-norm-thermodynamic}
  \lim_{q \to \infty} \mathcal{N}_{L} = \left(1-\frac{L(L+1)}{4l(l+1)}\right)^{\alpha}.
\end{equation}

As previously shown in Fig.~\ref{fig:two-qh-norms} and Fig.~\ref{fig:two-qp-norms}, this asymptotic formula precisely matches the normalization constants extracted from the fully projected, microscopic many-body two-QH wave functions and two-QP wave functions at Jain fillings $\nu=1/3, 2/5,$ and $3/7$ (corresponding to vortex counts $\alpha=2/3, 2/5,$ and $2/7$, respectively).

This excellent agreement validates the use of our effective two-anyon model for evaluating the overlaps between generic two-QH or two-QP states when both QHs or QPs are well separated. Specifically, given two arbitrary many-body two-QH wave functions represented by the coefficients $f_{m_{1}, m_{2}}$ and $g_{m_{1}, m_{2}}$:
\begin{equation}
  \label{eq:two-qh-effective-coefficient-states}
  \begin{split}
    &\ket{f}\rangle \equiv \sum_{m_{1}, m_{2}}f_{m_{1}, m_{2}}\ket{m_{1}, m_{2}}\\
    &\ket{g}\rangle \equiv \sum_{m_{1}, m_{2}}g_{m_{1}, m_{2}}\ket{m_{1}, m_{2}}
  \end{split}
\end{equation}
The overlap $\langle\langle f | g \rangle\rangle$ can be computed from the overlap of the unnormalized state

\begin{equation}
  \label{eq:effective-two-anyon-state-f}
  \begin{split}
    &\ket{f} \rangle
      \equiv
      \sum_{m_{1},m_{2}=0}^{\infty}
      f_{m_{1}-(n-1),m_{2}-(n-1)}
    \\
    &\times
      \begin{vmatrix}
        \conj{\eta}_{0,m_{1}}(Z_{1}) &
        \conj{\eta}_{0,m_{1}}(Z_{2}) \\
        \conj{\eta}_{0,m_{2}}(Z_{1}) &
        \conj{\eta}_{0,m_{2}}(Z_{2})
      \end{vmatrix}
      (\bar{Z}_{1}-\bar{Z}_{2})^{\alpha}
  \end{split}
\end{equation}
and the unnormalized state

\begin{equation}
  \label{eq:effective-two-anyon-state-g}
  \begin{split}
    &\ket{g} \rangle
      \equiv
      \sum_{m_{1},m_{2}=0}^{\infty}
      g_{m_{1}-(n-1),m_{2}-(n-1)}
    \\
    &\times
      \begin{vmatrix}
        \conj{\eta}_{0,m_{1}}(Z_{1}) &
        \conj{\eta}_{0,m_{1}}(Z_{2}) \\
        \conj{\eta}_{0,m_{2}}(Z_{1}) &
        \conj{\eta}_{0,m_{2}}(Z_{2})
      \end{vmatrix}
      (\bar{Z}_{1}-\bar{Z}_{2})^{\alpha},
  \end{split}
\end{equation}
where $\alpha=2/(2n+1)$. The corresponding overlap will be accurate up to some real constant independent of $f$ and $g$.
An analogous statement holds for QPs, with the complex conjugation removed and with the coefficients shifted as $f_{m_{1}-n,m_{2}-n}$ and $g_{m_{1}-n,m_{2}-n}$. In these effective states, we place the anyons in the LLL. Since both effective particles are always restricted to a single LL, the LL index does not affect the overlap computation in the thermodynamic limit. This is also visible from the preceding derivation of the norms: the dominant contribution comes from $j_{1},j_{2}=q+\alpha/2$.

\section{Berry phase evaluation from the two-anyon model}
\label{app:berry-phase-two-anyon-model}
In this appendix, we use the two-anyon model to evaluate the Berry phase for the conventional two-QH and two-QP states, and for the corresponding states proposed in this work. We take the braid loop to be a rotation about the origin. Thus, if the two QHs or QPs are localized at $\omega_{1}$ and $\omega_{2}$, we evaluate the Berry phase $\Theta$  along $\omega_{1}(t)=\omega_{1}\exp(\imath \phi(t))$ and $\omega_{2}(t)=\omega_{2}\exp(\imath \phi(t))$. Throughout this appendix we set the effective magnetic length to $\ellstar=1$.

\subsection{Two QP states}
We first consider the effective state corresponding to $\Psi^{\QP{0}, \omega_{1}; \QP{0}, \omega_{2}}_{\frac{n}{2pn+1}}$ in Eq.~\eqref{eq:localized-two-qp-wavepacket}. This is the state with two QPs localized in the $k_{1}=k_{2}=0$ angular-momentum states about $\omega_{1}$ and $\omega_{2}$. Using Eq.~\eqref{eq:nll-localized-orbital-expansion}, we obtain

  \begin{equation}
    \begin{split}
      &\Psi^{\QP{0}, \omega_{1}; \QP{0}, \omega_{2}}_{\frac{n}{2pn+1}}(Z_{1}, \bar{Z}_{1}, Z_{2}, \bar{Z}_{2}) = \\
      & \left[
        \exp\left(\frac{\conj{\omega}_{1}Z_{1}
        +\conj{\omega}_{2}Z_{2}}{2}\right)
        -
        \exp\left(\frac{\conj{\omega}_{1}Z_{2}
        +\conj{\omega}_{2}Z_{1}}{2}\right)
      \right] \\
      &\times(Z_{1}-Z_{2})^{\alpha}
      \exp\left(
        -\frac{|Z_{1}|^{2}+|Z_{2}|^{2}}{4}
        -\frac{|\omega_{1}|^{2}+|\omega_{2}|^{2}}{4}
      \right).
    \end{split}
  \end{equation}

It is useful to rewrite this state in center-of-mass and relative coordinates,
\begin{equation}
  \begin{aligned}
    \omega_{\cm}&=\frac{\omega_{1}+\omega_{2}}{\sqrt{2}},
    &
    \omega_{\rel}&=\frac{\omega_{1}-\omega_{2}}{\sqrt{2}},
    \\
    Z_{\cm}&=\frac{Z_{1}+Z_{2}}{\sqrt{2}},
    &
    Z_{\rel}&=\frac{Z_{1}-Z_{2}}{\sqrt{2}}.
  \end{aligned}
\end{equation}
In these variables, the state factorizes into a center-of-mass coherent-state factor and a relative-coordinate factor:

  \begin{equation}
    \begin{split}
      &\Psi^{\QP{0}, \omega_{1}; \QP{0}, \omega_{2}}_{\frac{n}{2pn+1}}(Z_{\cm}, \bar{Z}_{\cm}, Z_{\rel}, \bar{Z}_{\rel}) =\\
      &\exp\left(
        \frac{\conj{\omega}_{\cm}Z_{\cm}}{2}
        -\frac{|Z_{\cm}|^{2}}{4}
        -\frac{|\omega_{\cm}|^{2}}{4}
      \right)\\
      &\times
      \exp\left(
        -\frac{|Z_{\rel}|^{2}}{4}
        -\frac{|\omega_{\rel}|^{2}}{4}
      \right)\\
      &\left[\exp(\frac{\conj{\omega}_{\rel}Z_{\rel}}{2})-\exp(-\frac{\conj{\omega}_{\rel}Z_{\rel}}{2})\right]\left(\sqrt{2}Z_{\rel}\right)^{\alpha}.
    \end{split}
  \end{equation}

Equivalently, expanding in orbitals $\eta_{0, M}(Z_{\cm})$ and $\eta_{0, m}(Z_{\rel})$, and dropping an overall real normalization constant independent of $\omega$, gives

  \begin{equation}
    \begin{split}
      &\Psi^{\QP{0}, \omega_{1}; \QP{0}, \omega_{2}}_{\frac{n}{2pn+1}}(Z_{\cm}, \bar{Z}_{\cm}, Z_{\rel}, \bar{Z}_{\rel}) = \\
      & \left[ \sum_{M=0}^{\infty}\conj{\eta}_{0, M}(\omega_{\cm})\eta_{0, M}(Z_{\cm}) \right] \\
      & \left[ \sum_{m=1,3,\dots}^{\infty}\conj{\eta}_{0, m}(\omega_{\rel})N_{\alpha, m}{\eta_{0, m+\alpha}(Z_{\rel})}\right].
    \end{split}
  \end{equation}

Here we used Eq.~\eqref{eq:relative-index-shift}, $\eta_{0, m}(Z_{\rel})Z_{\rel}^{\alpha}=N_{\alpha, m}\eta_{0, m+\alpha}(Z_{\rel})$, with $N_{\alpha, m}$ defined in Eq.~\eqref{eq:anyon-constant}, and absorbed the constant $2^{\alpha/2}$ into the overall normalization.

The same change of variables also relates the LL raising operators for $\omega_{1},\omega_{2}$ to those for $\omega_{\cm},\omega_{\rel}$. Applying this relation to general $k_{1},k_{2}$ gives

  \begin{equation}\label{eq:appendix-conventional-qp-k1-qp-k2-effective-state}
    \begin{split}
      &\Psi^{\QP{k_{1}}, \omega_{1}; \QP{k_{2}}, \omega_{2}}_{\frac{n}{2pn+1}}(Z_{\cm}, \bar{Z}_{\cm}, Z_{\rel}, \bar{Z}_{\rel}) = \\
      & \sum_{s=0}^{K}A_{s}(k_{1}, k_{2})\left[ \sum_{M=0}^{\infty}\conj{\eta}_{K-s, M+s-K}(\omega_{\cm})\eta_{0, M}(Z_{\cm}) \right] \\
      & \left[ \sum_{m=1,3,\dots}^{\infty}\conj{\eta}_{s, m-s}(\omega_{\rel})N_{\alpha, m}{\eta_{0, m+\alpha}(Z_{\rel})}\right].
    \end{split}
  \end{equation}

Here, $A_{s}(k_{1}, k_{2})$ is the coefficient defined in Eq.~\eqref{eq:cm-relative-binomial-coefficient} and $K=k_{1}+k_{2}$.

Equation~\eqref{eq:appendix-conventional-qp-k1-qp-k2-effective-state} is useful for calculating the Berry phase, since the center-of-mass and relative pieces can now be treated separately. For an infinitesimal step along the rotation loop, we denote by $\mathcal{O}_{\QPsym}(d\phi)$ the overlap between the state at $(\omega_{1},\omega_{2})$ and the state at $(\omega_{1}\exp(\imath d\phi),\omega_{2}\exp(\imath d\phi))$, and extract $\Theta$ from

\begin{equation}
  \begin{split}
    \frac{\Theta}{2\pi} =\lim_{d\phi \to 0}\frac{1}{\imath d\phi}
      \left[
        \frac{\mathcal{O}_{\QPsym}(d\phi)}{\mathcal{O}_{\QPsym}(0)}
        -1
      \right].
  \end{split}
\end{equation}

Using Eq.~\eqref{eq:appendix-conventional-qp-k1-qp-k2-effective-state}, this overlap is
\begin{equation}
  \begin{split}
    \mathcal{O}_{\QPsym}(d\phi)
    &= \sum_{s_1, s_2=0}^{K}A_{s_{1}}A_{s_{2}} \\
    &\times \left[ \sum_{M=0}^{\infty}
      \conj{\eta}_{K-s_1, M+s_1-K}(\omega_{\cm}e^{\imath d\phi})
      \right. \\
      &\qquad\left.
    \eta_{K-s_2, M+s_2-K}(\omega_{\cm}) \right] \\
    &\times \left[ \sum_{m=1,3,\dots}^{\infty}
      \conj{\eta}_{s_1, m-s_1}(\omega_{\rel}e^{\imath d\phi})
      \right. \\
      &\qquad\left.
    {\eta_{s_2, m-s_2}(\omega_{\rel})}N_{\alpha, m}^{2}\right].
  \end{split}
\end{equation}

The center-of-mass sum is fixed by Eq.~\eqref{eq:ll1-ll2-overlap}. In the limit $d\phi\to 0$,
\begin{equation}
  \begin{split}
    &2\pi \lim_{d\phi \to 0}\left[\sum_{M=0}^{\infty}\conj{\eta}_{K-s_1, M+s_1-K}(\omega_{\cm}(1+\imath\, d\phi)) \right.\\
    & \left.\eta_{K-s_2, M+s_2-K}(\omega_{\cm}) \right]\\
    &\to \delta_{s_{1}, s_{2}} - \imath d\phi \left( \delta_{s_{1}, s_{2}}\frac{\abs{\omega_{\cm}}^{2}}{2} \right. \\
    &\left.+\delta_{s_{1}, s_{2}+1}\frac{\omega_{\cm}\sqrt{(K-s_{2})}}{\sqrt{2}}+\delta_{s_{1}, s_{2}-1}\frac{\conj{\omega}_{\cm}\sqrt{K-s_{1}}}{\sqrt{2}}\right)
  \end{split}
\end{equation}
We next evaluate the relative-coordinate sum. For the infinitesimal overlaps needed below, the antisymmetrization of the two fermions does not change the leading contribution: the two wavepackets are centered at $\omega$ and $\omega^{\prime}$, which are infinitesimally separated, while the exchange term involves the overlap between $\omega$ and $-\omega^{\prime}$. That exchange term is relevant only when the separation is small enough that the two overlaps are comparable. We therefore work with
\begin{equation}
  \begin{split}
    &\mathcal{I}_{s_{1}, s_{2}} = \sum_{m=0}^{\infty}\conj{\eta}_{s_{1}, m-s_{1}}(\omega){\eta}_{s_{2}, m-s_{2}}(\omega^{\prime})N_{\alpha, m}^{2}\\
    & = \frac{\exp\left(-\frac{\abs{\omega}^{2}}{4}-\frac{\abs{\omega^{\prime}}^{2}}{4}\right)}{2\pi \sqrt{2^{s_{1}+s_{2}}s_{1}!s_{2}!}} \times \\
    &\int dx dy \; [r^{2\alpha}(Z-\omega)^{s_{1}}(\conj{Z}-\conj{\omega}^{\prime})^{s_{2}}] \\
    & \times \exp\left(\frac{\conj{\omega}Z}{2}+\frac{\conj{Z}\omega^{\prime}}{2}-\frac{\abs{Z}^{2}}{2}\right)
  \end{split}
\end{equation}
Completing the square in the Gaussian gives
\begin{equation}
  \begin{split}
    &\int dx dy \; [r^{2\alpha}(Z-\omega)^{s_{1}}(\conj{Z}-\conj{\omega}^{\prime})^{s_{2}}] \\
    &\times \exp\left(\frac{\conj{\omega}Z}{2}+\frac{\conj{Z}\omega^{\prime}}{2}-\frac{\abs{Z}^{2}}{2}\right) = \\
    & \exp(\frac{\conj{\omega}\omega^{\prime}}{2}) \int dx dy [Z+(\omega^{\prime}-\omega)]^{s_{1}}[\conj{Z}-(\conj{\omega}^{\prime}-\conj{\omega})]^{s_{2}} \\
    & \times (\conj{Z}+\conj{\omega})^{\alpha}(Z+\omega^{\prime})^{\alpha}\exp(-\frac{Z\conj{Z}}{2})
  \end{split}
\end{equation}
We now take $\omega \to \omega_{\rel}+\imath\omega_{\rel}d\phi$ and $\omega^{\prime}\to\omega_{\rel}$, and keep only terms through first order in $d\phi$. This gives
\begin{equation}
  \begin{split}
    &\mathcal{I}_{s_{1}, s_{2}} = \mathcal{I}^{0}_{s_{1}, s_{2}} - \imath d\phi \left(\frac{\abs{\omega_{\rel}}^{2}}{2}\mathcal{I}^{0}_{s_{1}, s_{2}}+\frac{\sqrt{s_{1}}\omega_{\rel}}{\sqrt{2}}\mathcal{I}^{0}_{s_{1}-1, s_{2}} + \right. \\
    &\left.+\frac{\sqrt{s_{2}}\conj{\omega}_{\rel}}{\sqrt{2}}\mathcal{I}^{0}_{s_{1}, s_{2}-1} + \conj{\omega}_{\rel}\mathcal{I}^{\prime}_{s_{1}, s_{2}}\right)
  \end{split}
\end{equation}
Here,
\begin{equation}
  \begin{split}
    \mathcal{I}^{0}_{s_{1}, s_{2}} &= \frac{1}{2\pi \sqrt{2^{s_{1}+s_{2}}s_{1}!s_{2}!}} \\
    &\quad\times\int dx dy\, |Z+\omega_{\rel}|^{2\alpha}Z^{s_{1}}\conj{Z}^{s_{2}}e^{-\frac{Z\conj{Z}}{2}}\\
    \mathcal{I}^{\prime}_{s_{1}, s_{2}} &= \frac{1}{2\pi \sqrt{2^{s_{1}+s_{2}}s_{1}!s_{2}!}} \\
    &\quad\times\int dx dy\, \frac{|Z+\omega_{\rel}|^{2\alpha}Z^{s_{1}}\conj{Z}^{s_{2}}}{\conj{Z}+\conj{\omega}_{\rel}}e^{-\frac{Z\conj{Z}}{2}}\\
  \end{split}
\end{equation}

We now evaluate $\mathcal{I}^{0}_{s_{1}, s_{2}}$ and $\mathcal{I}^{\prime}_{s_{1}, s_{2}}$ for large $|\omega_{\rel}|$. To this end, let us introduce:
\begin{equation}
  \begin{split}
    &\mathcal{J}_{s_{1}, s_{2}, \alpha_{1}, \alpha_{2}} = \\
    &\frac{1}{2\pi}\int dx dy Z^{s_{1}}\conj{Z}^{s_{2}}(Z+\omega_{\rel})^{\alpha_{1}}(\conj{Z}+\conj{\omega}_{\rel})^{\alpha_{2}}  e^{-\frac{Z\conj{Z}}{2}}
  \end{split}
\end{equation}

Expanding $(Z+\omega_{\rel})^{\alpha_{1}}$ and $(\conj{Z}+\conj{\omega}_{\rel})^{\alpha_{2}}$ as infinite series in powers of $Z$ and $\conj{Z}$, respectively, and then evaluating the integral gives the asymptotic series
\begin{equation}
  \begin{split}
    &\mathcal{J}_{s_{1}, s_{2}, \alpha_{1}, \alpha_{2}} = \\
    &\frac{2^{s_{1}}\omega_{\rel}^{\alpha_{1}}\conj{\omega}_{\rel}^{\alpha_{2}}}{\conj{\omega}_{\rel}^{s_{1}-s_{2}}}\sum_{m = 0}^{\infty}\binom{\alpha_{1}}{m}\binom{\alpha_{2}}{m+s_{1}-s_{2}}\\
    &\times \frac{2^{m}\Gamma(m+s_{1}+1)}{|\omega_{\rel}|^{2m}}
  \end{split}
\end{equation}

This should be viewed as an asymptotic expansion rather than a convergent representation of the integral; the expansion is reliable only for $m \ll |\omega_{\rel}|^2/2$. Since only the leading term is needed below, this subtlety does not affect the calculation of the Berry phase. Thus, for $s_{1} \geq s_{2}$,
\begin{equation}
  \begin{split}
    &\mathcal{I}^{0}_{s_{1}, s_{2}} \sim \sqrt{\frac{s_{1}!}{s_{2}!}}\frac{|\omega_{\rel}|^{2\alpha}}{\left(\frac{\conj{\omega}_{\rel}}{\sqrt{2}}\right)^{s_{1}-s_{2}}}\binom{\alpha}{s_{1}-s_{2}} \\
    &\mathcal{I}^{\prime}_{s_{1}, s_{2}} \sim \sqrt{\frac{s_{1}!}{s_{2}!}}\frac{|\omega_{\rel}|^{2\alpha}}{\conj{\omega}_{\rel}\left(\frac{\conj{\omega}_{\rel}}{\sqrt{2}}\right)^{s_{1}-s_{2}}}\binom{\alpha-1}{s_{1}-s_{2}} \\
  \end{split}
\end{equation}
while for $s_{1}<s_{2}$,
\begin{equation}
  \begin{split}
    &\mathcal{I}^{0}_{s_{1}, s_{2}} \sim \sqrt{\frac{s_{2}!}{s_{1}!}}\frac{|\omega_{\rel}|^{2\alpha}}{\left(\frac{{\omega}_{\rel}}{\sqrt{2}}\right)^{s_{2}-s_{1}}}\binom{\alpha}{s_{2}-s_{1}} \\
    &\mathcal{I}^{\prime}_{s_{1}, s_{2}} \sim \sqrt{\frac{s_{2}!}{s_{1}!}}\frac{|\omega_{\rel}|^{2\alpha}}{\conj{\omega}_{\rel}\left(\frac{{\omega}_{\rel}}{\sqrt{2}}\right)^{s_{2}-s_{1}}}\binom{\alpha-1}{s_{2}-s_{1}} \\
  \end{split}
\end{equation}

Combining the center-of-mass contribution with the relative-coordinate expansion gives

\begin{equation}
  \begin{split}
    &\frac{\Theta}{2\pi} = -\frac{|\omega_{\cm}|^{2}}{2} - \frac{|\omega_{\rel}|^{2}}{2} \\
    &-\frac{\sum_{s=0}^{K}A_{s}^{2}\left[\sqrt{2}\conj{\omega}_{\rel}\mathcal{I}^{\prime}_{s, s} + \sqrt{{s}}\left( \omega_{\rel}\mathcal{I}^{0}_{s-1, s} + \conj{\omega}_{\rel}\mathcal{I}^{0}_{s, s-1}\right)\right]}{\sqrt{2}\sum_{s=0}^{K}A_{s}^{2}\mathcal{I}^{0}_{s, s}} \\
    &-\frac{\sum_{s=0}^{K}\sqrt{K-s}A_{s}A_{s+1}\left({\omega}_{\cm}\mathcal{I}^{0}_{s,s+1}+\conj{\omega}_{\cm}\mathcal{I}^{0}_{s+1,s}\right)}{\sqrt{2}\sum_{s=0}^{K}A_{s}^{2}\mathcal{I}^{0}_{s, s}}
  \end{split}
\end{equation}

Substituting the leading order asymptotic form for $\mathcal{I}^{0}$ and $\mathcal{I}^{\prime}$ into this expression gives

\begin{equation}
  \begin{split}
    &\frac{\Theta}{2\pi} = -\frac{|\omega_{\cm}|^{2}}{2} - \frac{|\omega_{\rel}|^{2}}{2}-\alpha\frac{\sum_{s=0}^{K}A_{s}^{2}(2s+1)}{\sum_{s=0}^{K}A_{s}^{2}} \\
    &-2\alpha\Re \left(\frac{\omega_{\cm}}{\omega_{\rel}}\right)\frac{\sum_{s=0}^{K-1}\sqrt{(K-s)(s+1)}A_{s}A_{s+1}}{\sum_{s=0}^{K}A_{s}^{2}}
  \end{split}
\end{equation}
The remaining coefficients are independent of $\omega_{\cm}$ and $\omega_{\rel}$ which we now evaluate. We first make use of the symmetry
\begin{equation}
  \begin{split}
    &A_{s}=(-1)^{k_{2}} A_{K-s},\\
    &\frac{\sum_{s=0}^{K}A_{s}^{2}(2s+1)}{\sum_{s=0}^{K}A_{s}^{2}}
    =
    \frac{\sum_{s=0}^{K}A_{s}^{2}(2(K-s)+1)}{\sum_{s=0}^{K}A_{s}^{2}}.
  \end{split}
\end{equation}
Therefore,
\begin{equation}
  \frac{\sum_{s=0}^{K}A_{s}^{2}(2s+1)}{\sum_{s=0}^{K}A_{s}^{2}}
  =
  K+1.
\end{equation}
For the second coefficient, let
\begin{equation}
  B_s=[x^s](1+x)^{k_{1}}(1-x)^{k_{2}}.
\end{equation}
Then
\begin{equation}
  \begin{split}
    &A_{s+1}(k_{1}, k_{2}) = \sqrt{\frac{s+1}{K-s}}A_{s}(k_{1}, k_{2}) \\
    &\times \frac{B_{s+1}}{B_{s}}.
  \end{split}
\end{equation}
The polynomial coefficients satisfy the recurrence relation
\begin{equation}
  B_{s+1}(s+1) = (k_{1}-k_{2})B_{s}-(k_{1}+k_{2}-s+1)B_{s-1}
\end{equation}
with $B_{-1}=B_{K+1}=0$. Using this recurrence in the definition of $A_s$ gives
\begin{equation}
  \begin{split}
    &\sqrt{(K-s)(s+1)}A_{s}A_{s+1}
    =\frac{(s+1)!(K-s)!}{k_{1}!k_{2}!2^K}B_{s}B_{s+1} \\
    & = \frac{(K-s)!s!}{k_{1}!k_{2}!2^K} B_{s}
    \left( (k_{1}-k_{2})B_{s}-(K+1-s)B_{s-1} \right) \\
    & = (k_{1}-k_{2})A_s^2-\sqrt{(K+1-s)s}A_sA_{s-1},
  \end{split}
\end{equation}
Summing this identity from $s=0$ to $K$, and using $A_{-1}=A_{K+1}=0$, yields
\begin{equation}
  \begin{split}
    &\sum_{s=0}^{K-1}A_{s}A_{s+1}\sqrt{(K-s)(s+1)} \\
    &=(k_{1}-k_{2})\sum_{s=0}^{K}A_s^2
    -\sum_{s=1}^{K}A_{s}A_{s-1}\sqrt{(K+1-s)s}.
  \end{split}
\end{equation}
The last sum is the same as the first after shifting the summation index. Therefore,
\begin{equation}
  \sum_{s=0}^{K-1}A_{s}A_{s+1}\sqrt{(K-s)(s+1)}
  =
  \frac{k_{1}-k_{2}}{2}\sum_{s=0}^{K}A_{s}^{2}.
\end{equation}
Hence,

\begin{equation}
  \begin{split}
    &\frac{\Theta}{2\pi} = -\frac{|\omega_{\cm}|^{2}}{2}-\frac{|\omega_{\rel}|^{2}}{2}-\alpha(1+k_{1}+k_{2}) \\
    &-\alpha(k_{1}-k_{2})\Re\left( \frac{\omega_{\cm}}{\omega_{\rel}} \right).
  \end{split}
\end{equation}

For our proposed state, the only change is in the relative-coordinate sum. It is simplest to begin with $s_{1}=s_{2}=0$ and then obtain the general $s_{1},s_{2}$ result using ladder operators. For $s_{1}=s_{2}=0$,
\begin{equation}
  \begin{split}
    &2\pi \sum_{m=0}^{\infty}\conj{\eta}_{0, m+\alpha}(\omega^{\prime})\eta_{0, m+\alpha}(\omega) \\
    & = \sum_{m=0}^{\infty}\frac{1}{\Gamma(m+\alpha+1)}\left( \frac{\conj{\omega}^{\prime}\omega}{2} \right)^{m+\alpha}\\
    &=\exp\left(-\frac{\abs{\omega}^{2}}{4}-\frac{\abs{\omega^{\prime}}^{2}}{4}\right) \times \left[\exp\left( \frac{\conj{\omega}^{\prime}\omega}{2} \right)- \right.\\
    &\left.\sum_{k=0}^{\infty}\frac{1}{\Gamma(\alpha-k)}\left(\frac{\conj{\omega}^{\prime}\omega}{2}\right)^{\alpha-1-k}\right]
  \end{split}
\end{equation}
In the limit $\omega^{\prime} \to \omega$, the second term in brackets is subleading as $|\omega|\to\infty$ (exponentially less relevant). Consequently, the relative-coordinate sum reduces to that of a single electron, up to the overall phase $\exp(\imath\alpha(\arg[\omega^{\prime}]-\arg[\omega]))$.

Hence,

\begin{equation}
  \begin{split}
    &\frac{\Theta}{2\pi} = \alpha-\frac{|\omega_{\cm}|^{2}}{2}-\frac{|\omega_{\rel}|^{2}}{2}
  \end{split}
\end{equation}

\subsection{Two QHs}
For the conventional two localized QH wave functions, the corresponding effective state is the complex conjugate of the two-QP effective state. Consequently, the Berry phase $\Theta$ is:

\begin{equation}
  \begin{split}
    \frac{\Theta}{2\pi}
    &=
      \frac{|\omega_{\cm}|^2}{2}
      +
    \frac{|\omega_{\rel}|^2}{2} \\
    &\quad+
      \alpha\left[
        1+k_{1}+k_{2}
        +(k_{1}-k_{2})\Re\left(\frac{\omega_{\cm}}{\omega_{\rel}}\right)
      \right].
  \end{split}
\end{equation}

For our proposed wave functions, following the center-of-mass and relative-coordinate decomposition from the preceding subsection, the infinitesimal overlap is
\begin{equation}
  \begin{split}
    \mathcal{O}_{\QHsym}(d\phi)
    &= \sum_{s_1, s_2=0}^{K}A_{s_{1}}A_{s_{2}} \\
    &\times \left[ \sum_{M=0}^{\infty}
      {\eta}_{K-s_1, M+s_1-K}(\omega_{\cm}e^{-\imath d\phi})
      \right. \\
      &\qquad\left.
    \conj{\eta}_{K-s_2, M+s_2-K}(\omega_{\cm}) \right] \\
    &\times \left[ \sum_{m=1,3,\dots}^{\infty}
      {\eta}_{s_1, m-\alpha-s_1}(\omega_{\rel}e^{-\imath d\phi})
      \right. \\
      &\qquad\left.
      \conj{\eta}_{s_2, m-\alpha-s_2}(\omega_{\rel}) \right. \\
    & \left. N^2_{\alpha, m}N^2_{-\alpha, m}\right] e^{\imath \alpha d\phi}.
  \end{split}
\end{equation}
The sum over relative coordinates in the large-$m$ limit reduces to the corresponding sum for the proposed QP wave functions, up to complex conjugation and $\alpha \to -\alpha$. This follows because the product $N_{\alpha, m}^{2}N_{-\alpha, m}^{2} \to 1$ as $m \to \infty$, and in the $|\omega_{\rel}| \to \infty$ limit the support is concentrated over large-$m$ orbitals.
Therefore, from the previous subsection, we have

\begin{equation}
  \frac{\Theta}{2\pi} = \frac{|\omega_{\cm}|^{2}}{2}+\frac{|\omega_{\rel}|^{2}}{2} + \alpha.
\end{equation}

\section{Braid statistics of the single-Slater-determinant wave functions}
\label{app:hf-braid}

In this appendix we show numerically that the single-Slater-determinant wave functions of Eqs.~\eqref{eq:hf-two-qp-state} and \eqref{eq:hf-two-qh-state}, built from the modified orbitals of Eqs.~\eqref{eq:hf-two-qp-orbital} and \eqref{eq:hf-two-qh-orbital}, give a braid statistics $\Delta\Theta/(2\pi)$ that approaches $\alpha$ at large separations, irrespective of the shapes of the two anyons.

We proceed as in Appendix~\ref{app:berry-phase-two-anyon-model}. The two anyons are placed at $\omega_{1}=d/2$ and $\omega_{2}=-d/2$ and rotated together about the origin, $\omega_{a}\to\omega_{a}e^{\imath\phi}$. We compute the infinitesimal overlap $\mathcal{O}(d\phi)$ between the state at $\phi=0$ and at $\phi=d\phi$, extract $\Theta$  from Eq.~\eqref{eq:berry-phase-overlap-definition}, and subtract the two single-particle Aharonov--Bohm phases as in Eq.~\eqref{eq:braid-statistics-large-separations} to obtain $\Delta\Theta$. Applied to the conventional states the same procedure returns $\Delta\Theta/(2\pi)=\mp\alpha(1+k_{1}+k_{2})$ for QPs and QHs, and applied to the flux-dressed states of Sec.~\ref{sec:flux-dressed-cf-wavefunctions} it returns $\alpha$.

\begin{figure*}
  \includegraphics[width=\textwidth]{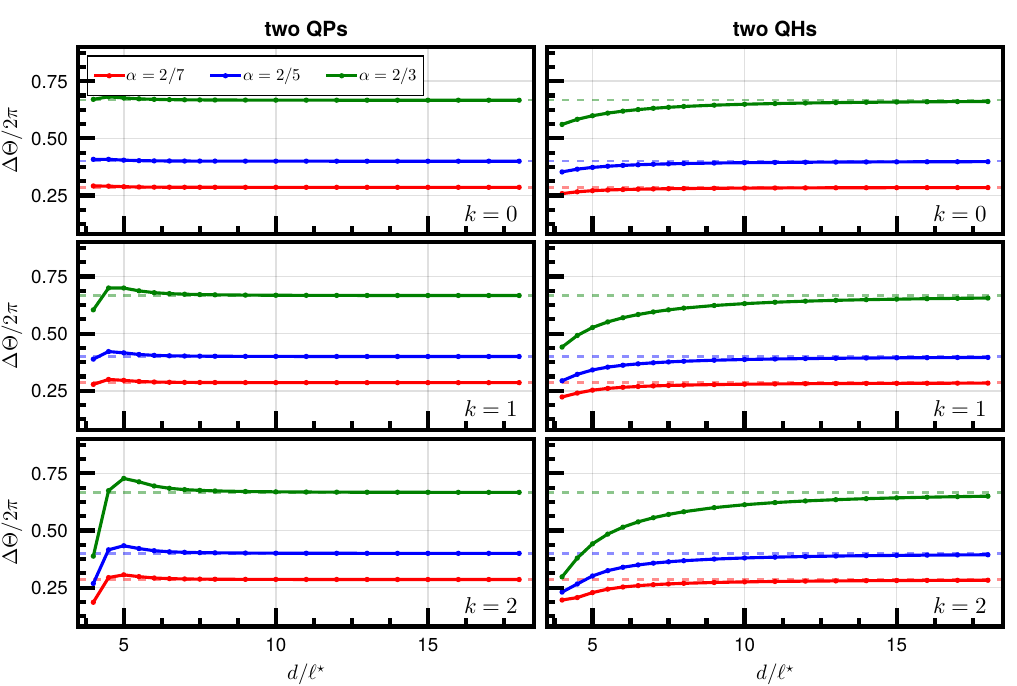}
  \caption{Braid statistics $\Delta\Theta/2\pi$ (more precisely, the contribution to the Berry phase in excess of the Aharonov--Bohm phase) of the single-Slater-determinant states as a function of the separation $d$ between the two anyons, for two quasiparticles (QPs, left, Eq.~\eqref{eq:hf-two-qp-state}) and two quasiholes (QHs, right, Eq.~\eqref{eq:hf-two-qh-state}), at $\alpha=2/7,\,2/5,\,2/3$. One anyon is in the $k_{1}=0$ angular-momentum state and the other in the $k_{2}=k$ state, with $k=0,1,2$ from top to bottom. Dashed horizontal lines mark the expected value $\Delta\Theta/2\pi=\alpha$. The QH curves converge more slowly than the QP curves, as $1/d^{2}$.}
  \label{fig:hf-braid-statistics}
\end{figure*}

Figure~\ref{fig:hf-braid-statistics} shows the result for the three statistical exponents $\alpha=2/7,\,2/5,\,2/3$, corresponding to $\nu=3/7,\,2/5,\,1/3$, and for one anyon in the $k_{1}=0$ state and the other in $k_{2}=k$ with $k=0,1,2$. In every case $\Delta\Theta/(2\pi)$ approaches $\alpha$ as the separation grows, independently of $k$.

\section{Asymptotic nature of the modified-orbital expansion}
\label{app:dressing-asymptotics}

In this appendix, we analyze the modified orbitals proposed in Sec.~\ref{sec:hf-construction}; specifically their series expansion in terms of the localized angular-momentum states about $\omega_{a}$. For simplicity, we set $\omega_{a}=0$ and write $\omega$ for the localization center of the other anyon, so that $d=\abs{\omega}$ is the separation between the two. We also set $\ellstar=1$. The proposed orbital is therefore
\begin{equation}\label{eq:app-modified-orbital}
  \varphi(Z)\propto\frac{1}{(Z-\omega)^{\alpha}}\,\psi_{0,k}(Z),
\end{equation}
with $\psi_{0,k}$ the localized wavepacket of Eq.~\eqref{eq:intro-lll-packet} centered at the origin.

\textit{Normalizability.} The factor $(Z-\omega)^{-\alpha}$ has a branch point at $\omega$, so $\varphi$ is not an LLL state, i.e., not a holomorphic function times a Gaussian. It is nevertheless a normalizable function. For $k=0$, a direct evaluation of the norm gives
\begin{equation}\label{eq:app-dressed-norm}
  \Vert\varphi\Vert^{2}
  =
  \frac{\Gamma(1-\alpha)}{2^{\alpha}}\;
  {}_{1}F_{1}\!\left(\alpha;1;-\frac{d^{2}}{2}\right),
\end{equation}
where ${}_{1}F_{1}$ is Kummer's confluent hypergeometric function. The norm is finite if and only if $\alpha<1$. Since $\alpha=2p/(2pn+1)<1$ at filling $\nu=n/(2pn+1)$, the modified orbitals are normalizable at every physical value of $\alpha$.

\textit{Asymptotic series.} The orbital of Eq.~\eqref{eq:app-modified-orbital} is not a state within the LLL. We can convert it into one by expanding $(Z-\omega)^{-\alpha}$ about the origin with the binomial series and truncating the result. This has to be done with care, because the series so obtained is asymptotic rather than convergent: its error stops decreasing once a certain number of terms is included, and grows thereafter. The expansion gives the series in the second line of Eq.~\eqref{eq:hf-two-qp-orbital}, namely a superposition of the angular-momentum states $\psi_{0,k+j}$ about the origin,
\begin{equation}\label{eq:app-series-coefficients}
  \varphi\propto\sum_{j\geq0}c_{j}\,\psi_{0,k+j},
  \qquad
  \abs{c_{j}}
  =
  \left|\binom{-\alpha}{j}\right|
  \sqrt{\frac{2^{j}(k+j)!}{k!}}\;
  \frac{1}{d^{\,j}}.
\end{equation}
The ratio of successive terms is
\begin{equation}\label{eq:app-term-ratio}
  \left|\frac{c_{j+1}}{c_{j}}\right|
  =
  \frac{\alpha+j}{j+1}\,\frac{\sqrt{2(k+j+1)}}{d}
  \;\xrightarrow[\;j\gg k\;]{}\;
  \frac{\sqrt{2j}}{d},
\end{equation}
which is smaller than one for $j<d^{2}/2$ and larger than one for $j>d^{2}/2$. The terms therefore decrease in magnitude up to
\begin{equation}\label{eq:app-optimal-truncation}
  j^{\star}\approx\frac{d^{2}}{2},
\end{equation}
and grow thereafter. We therefore truncate the series at $j^{\star}$. All numerical results reported in this work use this truncation. The corresponding series for the QH orbitals, Eq.~\eqref{eq:hf-two-qh-orbital}, terminates after $k+1$ terms, so no truncation is required there.

A caveat is in order regarding the criterion itself. What enters a physical quantity is not the coefficient $c_{j}$ alone but the product $c_{j}\psi_{0,k+j}$, so in principle one ought to weigh each coefficient together with the orbital it multiplies in order to decide where the series is best truncated. Doing so term by term is tedious. Moreover, in the setting we are ultimately interested in, that of the microscopic FQH wave functions, it cannot be done at all: there the orbitals are the Jain--Kamilla projected orbitals~\cite{Jain97}, which are inherently many-body objects and cannot be examined one at a time. We therefore base the truncation on the coefficients alone, which is what gives $j^{\star}\approx d^{2}/2$.

  \section{Berry phase and density relation}
  \label{app:berry-density}

  In this appendix, we review the argument by Nardin {\it et al.}~\cite{Nardin23A} connecting the Berry phase to the density profile of the many-body state.

  Ref.~\cite{Nardin23A} notes that for an FQH state in the presence of a radially symmetric confining potential, and with QHs or QPs in the presence of centrosymmetric localizing potentials centered at $\omega_{1}, \dots, \omega_{k}$, the following holds: suppose we have $N$ electrons and the ground state wave function can be parametrically written as $\Psi(z_1, \dots, z_N; \omega_{1}, \dots, \omega_{k})$. Since the Hamiltonian is invariant under a combined rotation of the electron coordinates $z_{1}, \dots, z_{N}$ and the impurity coordinates $\omega_{1}, \dots, \omega_{k}$, the Berry phase $\Theta$ acquired by the wave function under closed braid loops with $\omega_{i}(t) = |\omega_{i}(0)|\exp(\imath \phi(t))$ from $\phi(0)=0$ to $\phi(T)=2\pi$ can be written as:
  \begin{equation}
    \Theta / 2\pi = -\bra{\Psi}\sum_{i}L_{z_{i}}\ket{\Psi}
  \end{equation}
  up to some integer. Here, $\sum_{i}L_{z_{i}}$ is the total angular momentum operator.

  For wave functions confined to the lowest Landau level (a consequence of the conformal invariance of the polynomial part), the expectation value of the total angular momentum can be related to the expectation value of the second moment of the density:
  \begin{equation}
    \bra{\Psi}\sum_{i}L_{z_{i}}\ket{\Psi} = \bra{\Psi}\left[ \sum_{i}\left( \frac{r^{2}_{i}}{2\lB^{2}}-1 \right) \right]\ket{\Psi}
  \end{equation}
  up to some $N$-dependent integer. The operator on the right-hand side has a very elegant interpretation: it measures the Berry phase acquired by a classical electron fluid whose density equals the density of the many-body wave function when it is taken around a loop about the origin.

  This leads to the following important consequence. Consider two different two-QH states, $\Psi^{(1)}$ and $\Psi^{(2)}$, in which the QHs are localized by centrosymmetric potentials. The difference between the Berry phases acquired by such wave functions under a rotation about the origin of the potentials can be related precisely to the densities $\rho_{1}$ and $\rho_{2}$ of the states $\Psi^{(1)}$ and $\Psi^{(2)}$ as:
  \begin{equation}\label{eq:berry-phase-density-difference}
    \frac{\Theta^{(1)}-\Theta^{(2)}}{2\pi} = -\int d^{2}r [\rho_{1}(\bm{r})-\rho_{2}(\bm{r})]\left(\frac{r^{2}}{2\lB^{2}}-1\right).
  \end{equation}
  Thus, any difference between the Berry phase acquired by two different QH wave functions is tied exactly to their densities. Assuming that the QHs are indeed localized (i.e., within the bulk, the density deviates from $\nu=n/(2pn+1)$ only in some compact regions with exponential localization), and the density at the edge is tied only to the number of QHs in the system (the screening hypothesis), this tells us that the Berry phase difference is tied strictly to the difference in the density of the two QHs in $\Psi^{(1)}$ and $\Psi^{(2)}$.

  Given that the two-QH wave functions with the QHs localized in the $m=-(n-1)$ orbital at $\nu=n/(2pn+1)$ yield the expected Berry phase (an Aharonov-Bohm contribution plus the statistical phase), any deviation from this phase, such as when one of the QHs is localized in an orbital different from the $m=-(n-1)$ orbital, must be a result of a density distortion.

  More precisely, the spin-statistics relation derived in Ref.~\cite{Nardin23A} rests on the assumption that when two QHs or QPs are sufficiently far apart, their density profiles can be treated independently. That is, if the density profile of the full state on top of the ground state is $\delta \rho(x,y;\omega_{1}, \omega_{2})$ and the density profile of a single QH/QP is $\delta \rho (x, y; \omega)$, it was assumed that $\delta \rho(x,y;\omega_{1}, \omega_{2}) = \delta \rho(x, y; \omega_{1}) + \delta \rho(x, y;\omega_{2})$. For the conventional CF wave functions, obtained by composite-fermionizing the solution of the corresponding IQH problem, this assumption fails: the two-QH / two-QP densities acquire a correlated deformation, and the shape-dependent braid statistics follows from it. The flux-dressed wave functions of Sec.~\ref{sec:flux-dressed-cf-wavefunctions} carry no such correlated deformation, and correspondingly recover the expected, shape-independent statistics.

\bibliography{references}

\end{document}